\documentclass{WileyMSP-template}
\usepackage{amssymb,latexsym,amsmath}
\usepackage{bm}
\usepackage{tensor}
\usepackage{graphicx,xcolor}
\usepackage{float}
\usepackage[titletoc, title]{appendix}
\usepackage[caption=false,lofdepth,lotdepth]{subfig}
\usepackage{soul}
\usepackage[normalem]{ulem}
\usepackage[version=4]{mhchem} 
\usepackage{textcomp}
\usepackage{mathtools}
\usepackage{float}
\usepackage{hyperref}
\hypersetup{%
	pdfpagemode=FullScreen,
	pdfstartpage=1,
	pdfmenubar=true,
	pdftoolbar=true,
	colorlinks = true,
	linkcolor=blue,
	citecolor=blue,
	urlcolor=blue,
	bookmarksopen=false
}
\usepackage{ragged2e}

\begin{document}

\pagestyle{fancy}
\rhead{\includegraphics[width=2.5cm]{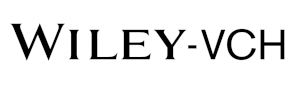}}

\title{Supercurrents and tunneling in massive many-vortex necklaces and star-lattices}

\maketitle

\author{Alice Bellettini*} 
\author{Vittorio Penna}

\begin{affiliations}
Dr. A. Bellettini\\
Department of Applied Science and Technology, Politecnico di Torino, 10129 Torino, Italy\\
alice.bellettini@polito.it 

Prof. V. Penna\\
Department of Applied Science and Technology, Politecnico di Torino, 10129 Torino, Italy

\end{affiliations}

\keywords{Bose-Einstein condensate, Mixture, Vortex, Gross-Pitaevskii equation, tunneling, Bosonic Josephson Junction}

\begin{abstract}
\justifying
Recently, cold atoms mixtures have attracted broad interest due to their novel properties and exotic quantum effects with respect to single-component systems.
In this paper the focus is on massive many-vortex states and their dynamics. 
Vortex configurations characterized by the same discrete rotational symmetry
are investigated when confined within topologically nonequivalent geometries, and the relative stability properties at varying number of vortices and infilling mass are highlighted.
It is numerically shown how massive many-vortex systems, in a mixture of Bose-Einstein condensates, can host the bosonic tunneling of the infilling component both in a disordered way, with tunneling events involving two or more close vortices, or in an almost-periodic way when the vortices are organized in persisting necklaces or star-lattices.
The purpose is to 
explore a variety of situations involving the interplay between the highly-nonlinear vortex dynamics and the inter-vortex atomic transfer, and 
so to better understand the conditions for the onset of Josephson supercurrents in rotating systems, or to reveal phenomena that could be of interest for a future application e.g. in the context of atomtronics. 
\end{abstract}

\section{Introduction}
\justifying

Binary mixtures of atomic gases \cite{Myatt1997, Modugno2002} offer intriguing opportunities for the study of macroscopic quantum phenomena and the interplay of nonlinear interactions \cite{KASAMATSU2005, Kuopanportti2015, Pitaevskii2016} and the intervortex boson exchange \cite{Bellettini2024}. In such systems, the dynamics of massive vortices \cite{McGee2001, Bandyopadhyay2017, Gallemi2018, Richaud2020}, i.e. quantum vortices \cite{Onsager1949, Fiszdon1991, Madison2000, Anderson2000, Raman2001} endowed with an inertial effect or an effective ``vortex mass'', was shown to present exotic features in comparison to that of standard massless vortices \cite{Ragazzo1994, Law2010, Ruban2022, Bellettini2024_CO, Bellettini2024}. Massive vortices are excited states of an (repulsive) \textit{immiscible} binary mixture \cite{Ao1998, Trippenbach2000}, where a majority component (hereafter ``$a$") hosts the quantum vortices while a minority component (``$b$") accumulates at the vortex cores. In this case, the $a$-density profile at the vortex wells constitute an effective external potential for the component $b$, which then obeys the physics of many particles within locally harmonic potentials. If the barrier between two wells is weak enough, e.g. if the vortices are close enough, $b$ undergoes a quantum tunneling process.

In a recent work, we showed that massive-vortex systems can in fact support phenomena such as the solitonic tunneling of the infilling component \cite{Bellettini2024} among different vortex wells, where the latter act as effective trapping potentials for the tunneling component. In this context, 
the physics of Bose-Hubbard (BH) models naturally
arises, so that a fascinating relation between the physics of Josephson supercurrents and the
quantum-vortex dynamics emerges. 
While in Ref. \cite{Bellettini2024} 
we examined the pivotal case of two massive vortices, whose internal dynamics makes up a single Bosonic Josephson Junction (BJJ), we address here as a follow-up 
the obvious question on the phenomenology of many-vortex systems.

In general, this paper tries to shed some light on a the dramatically complex dynamics, where the behavior of the system is controlled by the interplay of the highly-nonlinear dynamics of quantum vortices combined with a simultaneous boson transfer caused by the tunneling effect. 
While the dynamics of vortices with constant vortex masses was shown to be well captured e.g. in Ref. \cite{Bellettini2024_CO} by applying point-like (PL) vortex models \cite{Richaud2020,Kim2004}, 
when the constant-mass condition breaks down the tunneling of the infilling component actively triggers a parallel dynamics,
affecting the distribution of the vortex masses and in turn the evolution of the interacting vortices.
In this paper, we analyze many-vortex configurations in different geometries and arranged in different configurations. Specifically, we consider the relatively simple case of vortex necklaces and star-lattices, which depending on the choice of the model parameters and of the initial conditions can develop both stable and unstable behaviors. 
The simple structure of such states
is exploited to explore and analyze the interplay between the parallel dynamics of the boson transfer and of the vortices.

Among many aspects,
we try to highlight the situations where point-like models for the few- to many-vortex dynamics break down and tunneling phenomena take place in the field dynamics. These include the fast swapping of the vortex-cores population and the emergence of metastable bosonic Josephson currents.
In some cases, we also show how the vortex dynamics supports the absorption of the vortices at a boundary, accompanied by the formation of a background supercurrent. 
We leave aside here the characterization of three- and four-vortex systems, deserving a dedicated analysis, to investigate more general phenomena involving many-vortex interactions.

Experimentally, vortex necklaces were recently observed for example in Ref. \cite{Hernndez-Rajkov2024}, for the case of fermionic superfluids, where a connection with a shear-flow instability mechanism was established, whereas vortex lattices were produced by Abo-Shaeer \textit{et al.} \cite{Abo-Shaeer2001}, and Schweikhard \textit{et al.} \cite{Schweikhard2004} in early studies.
On the other hand, atomic Josephson currents \cite{Smerzi1997, Raghavan1999} in optical necklace-shaped lattices were observed and studied by the group of Roati in Ref. \cite{Pezze2024} (2024), where their stability applicability for the realization of atomtronic circuits was proven. Earlier on, atomic supercurrents had already been observed in one-dimensional arrays of Josephson junctions \cite{Cataliotti2001, Anker2005}, while the realization of a single Bosonic Josephson Junction was due to Albiez \textit{et al.} in 2005 \cite{Albiez2005}.

Our work prompts the modeling of various interesting phenomena in terms of rotating Bose-Hubbard models \cite{Milburn1997, Kasamatsu2008, Arwas2015, Richaud2017, Penna2017, Capuzzi2025}
and a future application e.g in the context of atomtronics \cite{Amico2017, Pezze2024}.
Furthermore, 
the study of matter waves in vortex star-lattices, whose stability properties, among other things, were studied by Campbell \textit{ 
et al.} \cite{Campbell1979} and by Kim \textit{et al.} in Ref. \cite{Kim2004}, 
offers
an interesting parallelism with
the study of the discretized nonlinear Schr{\"o}dinger equation performed
by Jason \textit{et al.} in Ref.
\cite{Jason2016}.
Moreover, within the literature, the case of tunneling bosons within a lattice was investigated by Caracanhas \textit{et al.} \cite{Caracanhas2015} and Chaviguri \textit{et al.} \cite{Chaviguri2018}
in the context of impurity physics.
Finally, as already mentioned, the rise of atomic $b$-supercurrents sustained by the vortex lattices, and the combination with the $a$-superflow generated by the vortices, 
suggests some connections with the field of atomtronics \cite{Amico2017}.

Our work is structured in the following way. 
In Section \ref{sec:PL}, we present the point-like models' solutions describing a special type of rotating massive-vortex lattice and a rotating massive-vortex necklace in a two-dimensional (2D) disk an in a 2D annulus.
Here we also illustrate the 
small-oscillation analysis which gives information on a fixed-point solution stability character. 
In Section \ref{sec:GPE}, we proceed with  
introducing the Gross-Pitaevskii equations describing the mixture at a mean-field level (see Section \ref{sec:GPE_intro}), we individuate different stability properties of vortex necklaces at varying physical parameters,
and we examine some tunneling phenomena occurring in the massive vortex necklaces. These include tunneling events occurring after the necklace's disruption, accompanying the subsequent chaotic vortex dynamics (see Section \ref{sec:necklaces}), and
atomic Josephson currents of the infilling component throughout the vortex necklace. 
In general, we study massive vortex necklaces that are confined within two different geometries, i.e. a disk and an annulus, and we individuate some metastable behaviors associated to their dynamics.
In Section \ref{sec:peak_propagation}, we then focus on the initial transient dynamics of a single peak in the infilling component.
We then proceed with the study of 
bosonic supercurrents in a stable vortex necklace.
In this case,
presented in Section \ref{sec:current_in_necklace}, we find instances of Bosonic-Josephson-Junction-like dynamical regimes. Further on, in Section \ref{sec:collapse}, we illustrate the vortex absorption, in an annular geometry, at the inner boundary. In our example, the resulting background superflow are accompanied by the rearrangement of the massive vortices into (ordered) fewer-vortex necklaces.
In general, throughout the whole Section \ref{sec:GPE}, we present some qualitative comparisons with the point-like model results on the necklaces stability properties. Such results are promising for a future more detailed comparison. 
In Section \ref{sec:lattices}, we then move on to the study of massive-vortex star-lattices, and observe
the inset of radial atomic currents
at varying physical parameters. Here, we also observe different stability properties of the lattices at varying number of vortices and infilling atoms.
Finally,
in Section \ref{sec:conclusions}, we conclude presenting our final considerations and outlooks.

\section{Point-like models for massive vortices}
\label{sec:PL}

The point-like Lagrangian describing $N_v$ massive vortices in a two-dimensional (2D) disk or annulus is \cite{Richaud2021}, in polar coordinates $(r_j \;\theta_j )\coloneq(r_j(t)\;\theta_j(t))$, $j,k=1,...,N_v$,

\begin{align}
\mathcal{L}=& \sum_j 
 \frac{\kappa_j m_a n_a}{2}\left(  R^2  -r_j ^2\right) \dot{\theta}_j + \frac{M_c}{2}  \sum_j \left( r_j ^2 \dot{\theta}_j ^2 + \dot{r}_j ^2 \right)
- \sum_j \phi_j  - \sum_{k}\sum_{j\neq k}
V_{jk},
\label{eq:Lagr}
\end{align}

\medskip
\noindent
where $m_a$ and $m_b$ are the atomic masses of the components $a$ and $b$, $\kappa_j=N_j h/m_a$ is the circulation quantum related to the $j$th vortex charge $N_j=\pm 1$,
$R$ stands for the disk radius or for the annulus' external radius respectively for the two cases, and
$M_c=m_b N_b/N_v$ is a single vortex' core-mass, common to all the vortices. This Lagrangian is obtained via a time-dependent variational formalism taking as an \textit{Ansatz} for $\psi_a$ a field with uniform density and phase singularities in correspondence of the vortices' centres, and as an \textit{Ansatz} for $\psi_b$ a superposition of Gaussians centered as well at the vortices' centres (see Refs. \cite{Kim2004} and \cite{Richaud2021}). The boundaries are modeled via the method of the virtual-charges, consisting in the introduction of imaginary vortices outside of the physical system, so to guarantee that the boundary is the locus of points with zero net perpendicular flow. In light of this, the \textit{Ansatz} for $\psi_a$ contains phase singularities in correspondence of any real or virtual vortices.

In the special case of a 2D disk, 
the potential energy that includes the boundary contribution is

\begin{equation}
    \phi_j = \frac{\kappa_j^2 \, m_a \, n_a}{4\pi} \ln\left(1 - \frac{r_j^2}{R^2} \right) ,
\end{equation}

\medskip
\noindent
while the vortices' interaction energy is

\begin{equation}
    V_{jk} = \frac{\kappa_j \kappa_k \, m_a \, n_a}{8\pi} \ln\left( 
\frac{R^2 - 2   r_j  r_k  \cos \theta_{jk} + r_j ^2 r_k ^2/R^2 }
{ r_j ^2 - 2 r_j  r_k \cos\theta_{jk}  + r_k ^2  } 
\right),
\end{equation}

\medskip
\noindent
with
$\theta_{jk} =\theta_j  - \theta_k$.
The Euler-Lagrangian equations relative to Lagrangian \eqref{eq:Lagr} 

\begin{equation}
M_c \, \ddot{r}_j  = 
- \kappa_j m_a n_a \, r_j  \, \dot{\theta}_j 
+ M_c\, r_j  \, \dot{\theta}_j ^2 
- \frac{\partial \phi_j}{\partial r_j}-\sum_k \frac{\partial V_{jk}}{\partial r_j}  ,
\label{eq:eq_rho}
\end{equation}

\medskip
\noindent
and

\begin{equation}
M_c \, r_j ^2 \, \ddot{\theta}_j  = 
 \kappa_j m_a n_a \, r_j  \, \dot{r}_j  - 
2M_c \, r_j  \, \dot{\theta}_j  \, \dot{r}_j - \sum_k\frac{\partial V_{jk}}{\partial \theta_j} ,
\label{eq:eq_theta}
\end{equation}
\medskip
\noindent
feature an electromagnetic analogy \cite{Richaud2021} with the Lorentz equations for a massive charge.

On the other hand, for an annular geometry, the point-like Lagrangian and the equations of motion are respectively Lagrangian \eqref{eq:Lagr} and Equation \eqref{eq:eq_rho} and \eqref{eq:eq_theta},
with the potential energy terms changing, according to Ref. 
\cite{Caldara2023}, as follows

\begin{equation}
\phi_j=\frac{\kappa_j^2 m_a n_a}{4\pi} \left[
\ln\left(
-\frac{2 i \, \vartheta_1\left(-i \ln\left( \frac{r_j }{R} \right), q\right)}
{\vartheta_1'(q)}
\right)
+   \ln \left(\frac{r_j }{R} \right)
\right],
\end{equation}

\begin{equation}
\begin{split}
V_{jk}&=\frac{\kappa_j \kappa_k m_a n_a}{4\pi}\;\text{Re}\left[ \ln\left(
\frac{
\vartheta_1\left(
-\frac{i}{2} \ln\left(\frac{r_j r_k }{R^2}\right) + \frac{1}{2} \theta_{jk} , q
\right)
}{
\vartheta_1\left(
-\frac{i}{2} \ln\left(\frac{r_j }{r_k }\right) + \frac{1}{2} \theta_{jk} , q
\right)
}
\right)\right].
\end{split}
\end{equation}

\medskip
\noindent
The Lagrangian \eqref{eq:Lagr} is rotation invariant,
so that the system features two conserved quantities: the energy (see the Hamiltonian \eqref{eq:Ham}) and the angular momentum, featuring only one non-zero component, that is the $z$-coordinate $L_3$.
We have that, for both geometries,  

$$ L_3 = \sum_j \frac{\partial\mathcal{L}}{\partial \dot\theta_j } = \sum_j \frac{\kappa_j m_a n_a}{2}(R^2-r_j ^2)+\sum_j M_c\, r_j ^2\dot \theta_j  \;\;\;j=1,...,N_v.$$

In such PL models, the vortex masses are typically constant parameters, condition which breaks down
if $\psi_b$ undergoes a quantum tunneling between different vortex wells in a non-uniform way. Hence, the possibility of a time-dependent vortex mass is legitimate, and we studied it in Ref. \cite{Bellettini2024} for the simple system of two vortices, finding out the realization of a Bosonic Josephson Junction. In Section \ref{sec:necklaces} and \ref{sec:lattices}, we investigate such intriguing tunneling effects
in more-vortex systems, representing the fields' dynamics 
via the simulation of Gross-Pitaevskii equations.

\subsection{Rotationally symmetric orbits}
\label{sec:rot_symm_orbits}

Both the disk and the annular geometry admit solutions for the equations of motion \eqref{eq:eq_rho}-\eqref{eq:eq_theta} that feature a discrete rotational symmetry at any time $t$, and are fixed points in a frame of reference rotating at angular frequency $\Omega$ (see also Ref. \cite{Campbell1979} for studies on massless vortices). 
Such solutions can describe vortex necklaces in both geometries, and be of the type

\begin{equation}
    r_j(t)= r_0,\;\;\;\theta_j(t)=\frac{2\pi}{N_v} j+\Omega t,
    \label{eq:necklace_sol}    
\end{equation}

\medskip
\noindent
with $N_j=+1$, i.e. $\kappa_j=\kappa$ (equal vorticities), and 
where the index $j=1,..,N_v$ labels the vortex, and $N_v\geq 2$. Also, the rotationally symmetric solutions can, in the case of the 2D disk, represent a vortex lattice, where one vortex is centered in the origin and the other $N_v-1$, with $N_v\geq 3$, vortices organize in a necklace or ``vortex crown'' around it. 
We will refer to this type of lattice as to a star-lattice.
In this case, we have

\begin{equation}
    x_1(t)=0,\;\;\;y_1(t)=0,\;\;\; r_{j\neq 0}(t)= r_0,\;\;\;\theta_{j\neq 0}(t)=\frac{2\pi}{N_v} j+\Omega t,
    \label{eq:lattice_sol}
\end{equation}

\medskip
\noindent
with $j=2,..,N_v$, and where $x_1(t)= r_1(t)\cos\theta_1(t)$ and $y_1(t)= r_1(t)\sin\theta_1(t)$, namely we represent the first vortex in Cartesian coordinates \cite{Kim2004}, due to the indeterminate character of the angular coordinate in the origin.
Also in this case we consider vortices with identical vorticities, meaning that all the configurations of the necklace- and star-lattice-type that we will consider in this paper will, at least in the case of the disk, always feature a global anti-clockwise precession.

Note that the value of $\Omega$ for the given geometry and number of vortices is found by plugging the \textit{Ans{\"a}tze} \eqref{eq:necklace_sol} and \eqref{eq:lattice_sol} in the equations of motion \eqref{eq:eq_rho} and \eqref{eq:eq_theta}. Due to the presence of a non-zero vortex mass, $\Omega$ features two branches, as a function of $r_0$, where however the greater one is most often unstable. 
Also, the annular geometry admit the rotation inversion, i.e. $\Omega$ can be both positive or negative, whereas in the case of the disk $\Omega$ is always positive (anti-clockwise precession) 
if the vortices' charges are positive. In other words, in a disk the direction precession only depends on the vortex charge, whereas in an annulus $\Omega$ also depends on the radial coordinate of the vortices.

\subsubsection{Small oscillations around a fixed point}
\label{sec:small_oscill}

To describe the stability properties of the fixed point solutions described in Section \ref{sec:rot_symm_orbits} with respect to small perturbations, is convenient to switch to the Hamiltonian formalism. 
The Hamiltonian corresponding to Lagrangian 
\eqref{eq:Lagr} reads

\begin{equation}
\mathcal{H}=\frac{1}{ 2 M_c} \sum_j \bigg[
p_{r_j}^2 + \frac{p_{\theta_j}^2}{ r_j^2}+ \kappa_j m_a n_a\left(1-\frac{R^2}{r_j^2}\right) p_{\theta_j} +\frac{\kappa_j^2 m_a^2 n_a^2 }{4 } \left(r_j- \frac{R^2}{r_j}\right)^2 \bigg] +\sum_j \phi_j  +\sum_{k}\sum_{j\neq k} V_{jk},
\label{eq:Ham}
\end{equation}

\medskip
\noindent
where the indices 
run along the same interval of
vortex labels, $j,k=1,...,N_v$, and the canonical moment are defined by

\begin{equation}
    p_{r_j} = M_c\, \dot r_j
,\;\;\;p_{\theta_j} =M_c\, r_j^2\, \dot \theta_j +\frac{\kappa_j m_a n_a}{2}\left( R^2-r_j^2 \right) .
\end{equation}

\bigskip
\noindent
The Hamilton equations corresponding to the Hamiltonian \eqref{eq:Ham} are found from the 
standard Poisson Brackets

\begin{equation}
    \{O,P\}=\sum_j \left[ \frac{\partial O}{\partial r_j}\frac{\partial P}{\partial p_{r_j}}  +\frac{\partial O}{\partial \theta_j}  \frac{\partial P}{\partial p_{\theta_j}}  -\frac{\partial P}{\partial r_j}  \frac{\partial O}{\partial p_{r_j}}  -\frac{\partial P}{\partial  \theta_j}\frac{\partial O}{\partial p_{\theta_j}}     \right].
\end{equation}
\medskip
\noindent
When switching to a rotating frame of reference, with rotation frequency $\Omega$, the solutions described in Section \ref{sec:rot_symm_orbits} become fixed points of the dynamical system. 
The Hamiltonian of the rotating, primed system and relevant primed coordinates
are
$$\mathcal{H}'=\mathcal{H}-\Omega L_3
, \qquad
r_j'= r_j,\;\;\; \theta_j'=\theta_j-\Omega t,
$$
respectively,
with $r_j$ and $\theta_j$ the coordinates in the reference frame at rest.
\medskip
\noindent
To find the small oscillations properties of our fixed-point solutions we linearize the equations of motion, in the primed system, around the fixed points (see, for example, Ref.\cite{Bellettini2024_CO}). 
Let us call the fixed point in the rotating frame $\tilde{\bm z}'$, and consider perturbed points 
\begin{equation}
\bm z' =  \tilde{\bm z}' +\bm \xi =(r_1',\;...,\; r_{N_v}',\;p_{r_1}',\;...,\;p_{ r_{N_v}}',\;\theta_1',\;...,\;\theta_{N_v}',\;p_{\theta_1}',\;...,\;p_{\theta_{N_v}}')^T,
\label{eq:vector}
\end{equation}
where, for brevity, we removed the prime index from the rotating-frame canonical variables.
We then get the linear
dynamical system 

\begin{equation}
    \dot{\bm{\xi}}=J\,\bm{\xi},
    \label{eq:dyn_sys}
\end{equation}
\medskip
\noindent
with
\medskip

\noindent
\begin{equation}
J=E\,\mathbb{H}, 
\qquad
    E=
\begin{pmatrix}
0_{N_v } & I_{N_v} & 0_{N_v } & 0_{N_v } \\
-I_{N_v} & 0_{N_v } & 0_{N_v } & 0_{N_v } \\
0_{N_v } & 0_{N_v } & 0_{N_v } & I_{N_v} \\
0_{N_v } & 0_{N_v } & -I_{N_v} & 0_{N_v }
\end{pmatrix},
\qquad
h_{ij}=\frac{\partial\mathcal{H}'}{\partial z_i'\partial z_j'}\bigg|_{\tilde{\bm z}' }.
\end{equation}
\bigskip

\begin{figure}
\centering
\begin{minipage}{0.49\textwidth}
  \centering
  \includegraphics[width=1\linewidth]{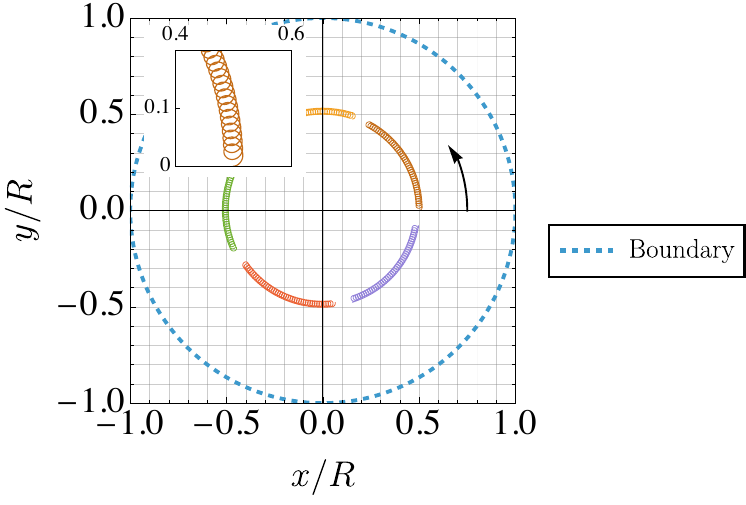}
  \caption{Example of a stable fixed-point trajectory of a five-vortex necklace in a disk. Upon a small perturbation of the massive vortices' velocities, the system stays close to the fixed point during its evolution. The inset shows a zoom of a vortex' trajectory, while the arrow indicates the sense of the vortices' average precession. We have: $N_a=10^5$, $N_b=10^3$ and $\Omega\simeq 2.36$ $rad/s$, while the evolution time for the plotted dynamics is $0.45$ $s$.}
\label{fig:ex_disk}
\end{minipage}
\hfill
\begin{minipage}{0.49\textwidth}
  \centering
  \includegraphics[width=1\linewidth]{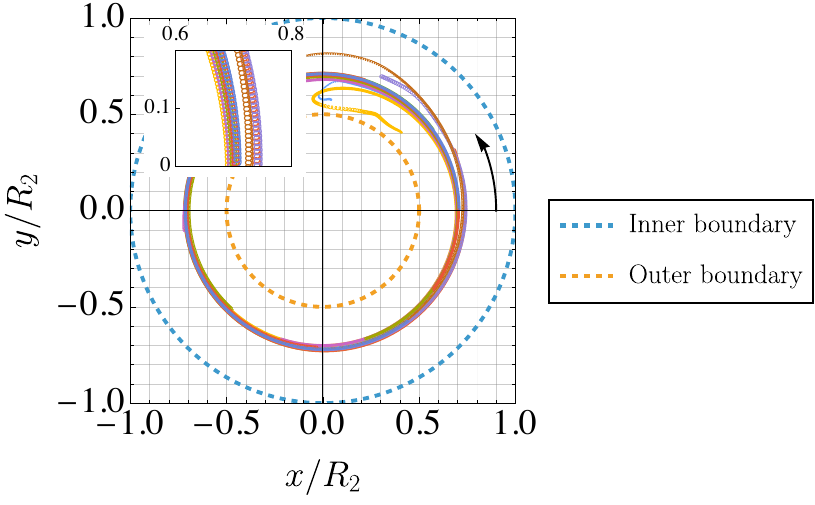}
  \caption{Example of an unstable fixed-point solution: upon a small perturbation, the system evolves further from the fixed-point. In this case the system is made up of ten massive vortices, in an annular geometry (with $q=0.5$). The arrow indicates the direction of the vortex-necklace's precession in the fixed-point and the inset is a zoom in of some vortex trajectories, as in Figure \ref{fig:ex_disk}, and the microscopic parameters also the same of Figure \ref{fig:ex_disk}. The plot's evolution time is of $2$ $s$, 
  and $\Omega\simeq 2.64$.}  
\label{fig:unstable}
\end{minipage}
\end{figure}

\noindent
In the above, $E$ is the standard symplectic matrix, and 
$\mathbb{H}$ the symmetric Hessian matrix associated with Hamiltonian 
$\mathcal{H}'$ whose elements $h_{ij}$ are computed in the fixed point $\bm z'$.
The parameter $z_i'$ is the $i$th component of vector (\ref{eq:vector}).
The fixed point is 
stable if all the eigenvalues of $J$ are pure imaginary,
whereas, if at least one eigenvalue has a nonzero real part, the fixed point is unstable.
These situations are illustrated in
\textbf{Figure \ref{fig:ex_disk}} and \textbf{Figure \ref{fig:unstable}}, respectively for a disk and an annular geometry. In the former,
the trajectories are obtained by perturbing a stable fixed point, corresponding to a necklace of five massive vortices, whereas the latter
exhibits vortex trajectories relevant to an unstable fixed point. The small-amplitude oscillations that are visible in the insets are a signature of the vortex mass \cite{Bellettini2023}.

\begin{figure}
\centering
\begin{minipage}{0.49\textwidth}
  \centering
  \includegraphics[width=1\linewidth]{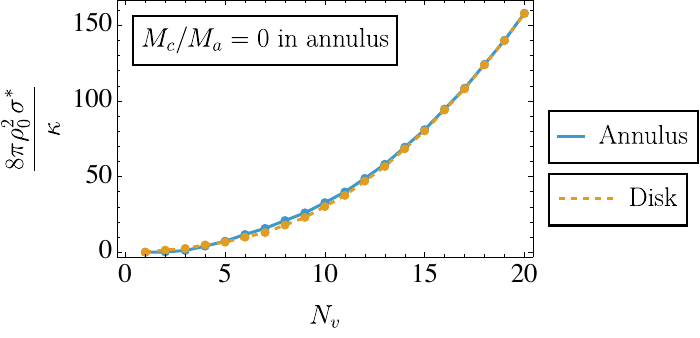}
\end{minipage}%
\hfill
\begin{minipage}{0.49\textwidth}
  \centering
  \includegraphics[width=1\linewidth]{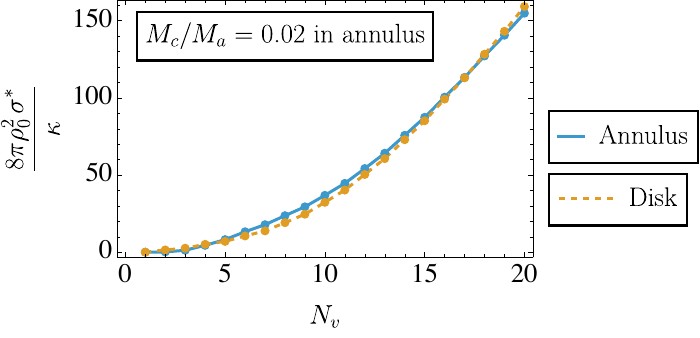}
\end{minipage}
\begin{minipage}{0.49\textwidth}
  \centering
  \includegraphics[width=1\linewidth]{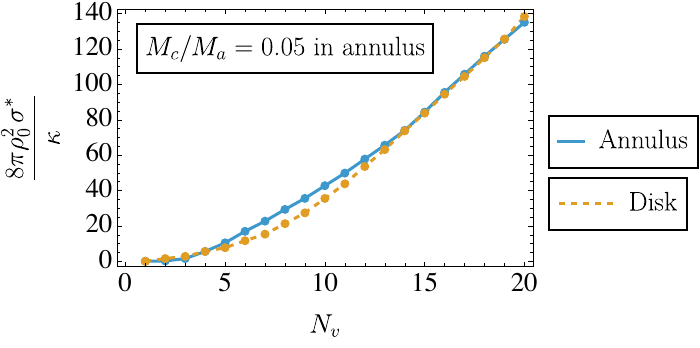}
\end{minipage}%
\hfill
\begin{minipage}{0.49\textwidth}
  \centering
  \includegraphics[width=1\linewidth]{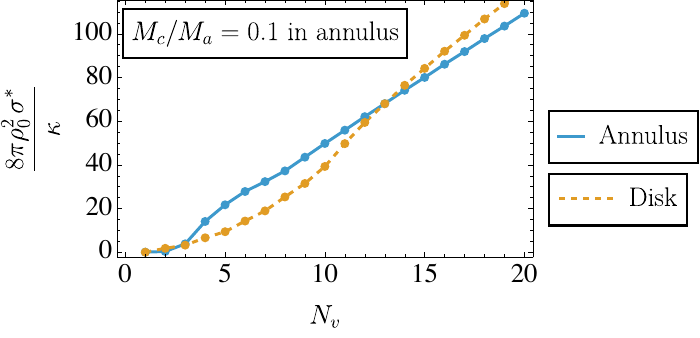}
\end{minipage}
\caption{Comparison of the maximum instability growth rate $\sigma^*$, normalized to dimensionless units, as a function of the number $N_v$ of massive vortices in a necklace for two different trap geometries: a disk and an annulus with $q=0.5$. The four panels are relevant to different values of the vortices core mass $M_c$, which is taken, for each panel, to be the same (in absolute value) for the annulus and the disk. We take also equal mean-value $a$-density for the two traps, and the disk radius is identical to the external radius of the annulus. }
\label{fig:sigma_star}
\end{figure}

In \textbf{Figure \ref{fig:sigma_star}}, we plot the maximum among the real parts of the eigenvalues of the system \eqref{eq:dyn_sys}, which is a measure for the maximum instability growth rate 
of the configuration, at varying number of vortices in the necklace (we take example from Ref. \cite{Caldara2024} and their study of necklaces in annular geometries). We call 
such maximal real part $\sigma^*$.
We compare necklaces in a disk with necklaces in a planar annulus with an external radius equal to the disk size and with equal average density $N_a/S$, with $S$ the surface of respectively the disk or the annulus. Also, we consider vortices with the same number of infilling atoms for both geometries, and examine four different values of the core mass $M_c$.
Interestingly, we see that accordingly to the PL 
model, for our annular geometry at $q=0.5$, at large vortex masses the necklace in our disk is ``less unstable'' for low $N_v$, whereas after a threshold value of $N_v$ the annular geometry supports necklaces featuring lower maximum instability growth rate $\sigma^*$. 
The difference in the values of $\sigma^*$ for the two geometries grows smaller at smaller core masses, while the threshold value of $N_v$ increases, so that the disk geometry leads to a slightly lower $\sigma^*$ for more values of $N_v$.  We remark that in our case there are neither background flows in the annulus, nor central vortices in the disk.

\section{Vortex necklaces in planar disk and annulus}
\label{sec:GPE}

After presenting the standard PL models for representing, at a simplified level, the motion of massive quantum vortices, let us now move on to the actual field dynamics as predicted by the solution of the time-dependent Gross-Pitaevskii equations (GPEs) \cite{Gross1961, Pitaevskii1961}.
Their solution well describes the properties of the mixture of the two ultracold quantum gases.
In the current section, we group a collection of phenomena involving vortex necklaces, and show how the dynamics predicted by the GPEs can feature tunneling phenomena that lead to vortex mass exchanges and atomic Josephson supercurrents.

In the following, when referring to a vortex ``necklace'' in the GPEs solution, we mean a closed
sequence of vortices that does not develop any changes in the order of the vortices over time.
Such a configuration can be the same as in the context of the PL model (see Equation 
\eqref{eq:necklace_sol}), or its deformation. We run here our numerical simulations up to the final time of $10$ $s$, so that when we talk about a ``stable'' configuration in the context of the GPEs study, we mean that within our simulation time of $10$ $s$ this is not destroyed.

\subsection{Gross-Pitaevskii equations' simulation}
\label{sec:GPE_intro}

In the following sections we employ numerical simulations to solve the following coupled 2D Gross-Pitaevskii equations for the mixture of BECs,

\begin{equation}
\begin{aligned}
    i\hbar\, \frac{\partial \psi_a(\bm{r}, t)}{\partial t} &=  -\frac{\hbar^2}{2m_a} \nabla^2 \psi_a(\bm{r}, t)+ \left(V_{ \text{ext}}(\bm{r}) + \frac{g_{ab}}{L_z} |\psi_b(\bm{r}, t)|^2 \right) \psi_a(\bm{r}, t)    + \frac{g_{a}}{L_z}  |\psi_a(\bm{r}, t)|^2 \psi_a(\bm{r}, t), \\
    i\hbar\, \frac{\partial \psi_b(\bm{r}, t)}{\partial t} &= -\frac{\hbar^2}{2m_b} \nabla^2 \psi_b(\bm{r}, t) + \left( V_{ \text{ext}}(\bm{r}) + \frac{g_{ab}}{L_z}  |\psi_a(\bm{r}, t)|^2 \right) \psi_b(\bm{r}, t)   + \frac{g_{b}}{L_z}  |\psi_b(\bm{r}, t)|^2 \psi_b(\bm{r}, t),
\end{aligned}
\label{eq:GPEs}
\end{equation}

\medskip
\noindent
where $\psi_j\coloneq\psi_j(\bm{r}, t)$, $j=a,b$ is the order parameter of the component $j$ and $\bm r = (x\;y)^T$, $m_j$ is its atomic mass, $V_{ \text{ext}}(\bm{r}) $ is the external potential 
, and $g_j = 4 \pi \hbar^2 a_j/m_j$, with $a_j$ the $s$-wave scattering length of the component $j$, while $g_{ab} = 2\pi \hbar^2 a_{ab} / m_r$ and 
$1/m_r = 1/m_a + 1/m_b$, with $a_{ab}$ the $s$-wave scattering length describing the interaction between an atom of $a$ and an atom of $b$. Finally, $L_z$ is the thickness of the mixture in the $z$-direction, where the dynamics is frozen. The rescaling of the interaction parameters $g_j$ and $g_{ab}$ is due to the dimensional reduction of the GPEs to two dimensions.
The GPE for $\psi_b$ highlights how $g_{ab}/L_z |\psi_a(\bm{r}, t)|^2 $ is an effective potential for the component $b$. 

In the present work, we always consider a mixture of 
\ce{^{87}  Rb}, as the ``$a$'' majority component and \ce{^{41}  K}
as the ``$b$'' minority component, with a number of $a$-atoms in the trap $N_a=10^5$, and a varying number $N_b$ of $b$-atoms in the trap.
An example of such a binary mixture was realized 
by Burchianti \textit{et al.} \cite{Burchianti2018}.
The external potential $V_{ \text{ext}}(\bm{r}) $, common to the two components, is here a hard-wall potential with either the shape of a 2D disk or of a planar annulus. We take here a disk
of radius $R=50$ $\mu m$ and an annulus of external radius $R$, and variable internal radius $R_1$. The thickness $L_z$ is $2$ $\mu m$, and the intra- and inter-species scattering lengths are, if not explicitly indicated,
$a_a=99a_0$, $a_b=65 a_0$, and $a_{ab}=163 a_0$, with $a_0$ the Bohr radius. We have that $g_{ab}/\sqrt{g_a g_b}\simeq 2.2$, in agreement with the immiscibility condition required for the massive vortices.
Note that the system of equations \eqref{eq:GPEs} only features contact interactions, and that the number of atoms in the two components is conserved and given by

$$\int d^2 r\, |\psi_j|^2=N_j,\;\;\; j=a,b.$$

\medskip
\noindent
Observe, in the upper relation, the atom-number-density meaning of $|\psi_j|^2$.

Our code for simulating the GPEs is based on an imaginary time algorithm to produce the input configuration, and on the Runge-Kutta method of the fourth order for solving our nonlinear time-dependent equations.
Our spacial grid is of $256\times 256 $ points.
The trial state used in the imaginary-time algorithm for producing the desired initial configuration of the real-time evolution contains Gaussian $b$-peaks at the vortex sites, to model the infilling masses, and phase singularities in the phase field $\theta_a$, with $\psi_a=|\psi_a|\,e^{i\theta_a}$, to model the quantum vortices.

\begin{figure}
\centering
\begin{minipage}{0.33\textwidth}
  \centering
  \includegraphics[width=1\linewidth]{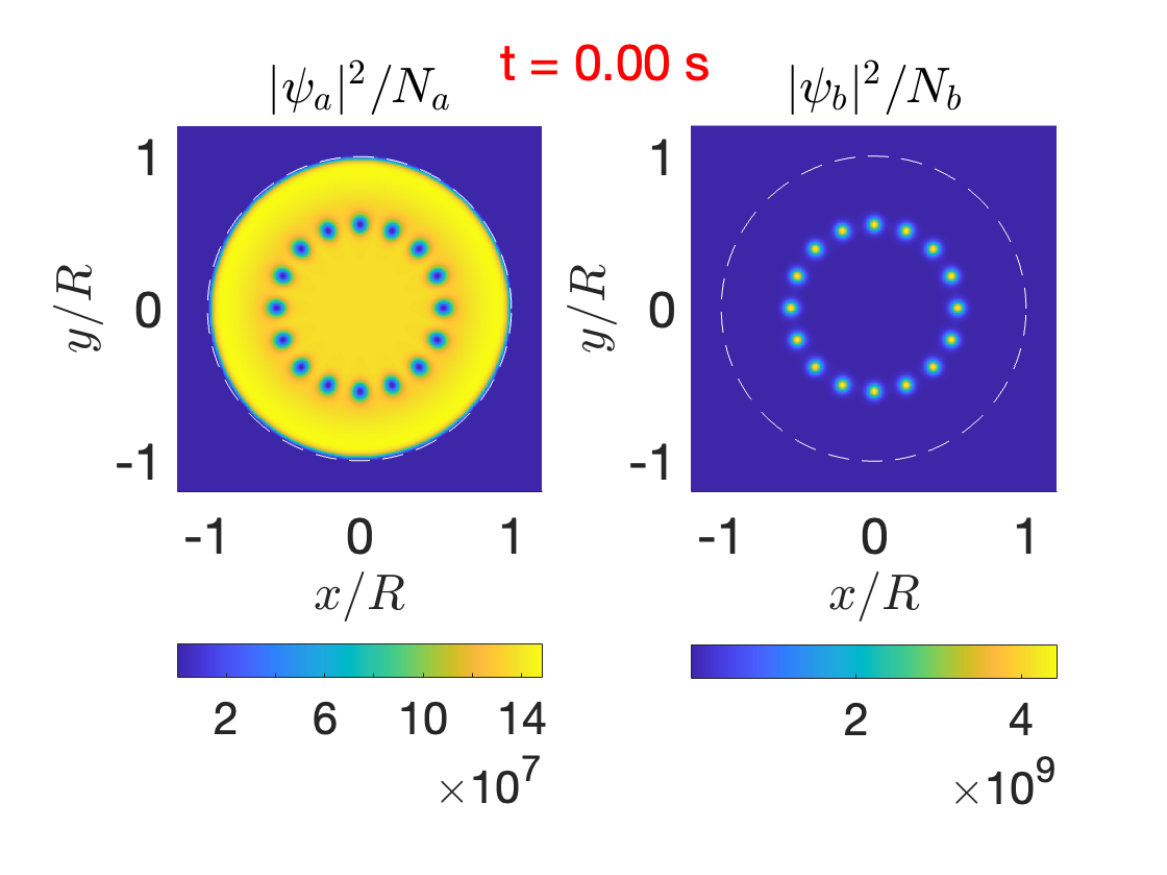}
\end{minipage}
\begin{minipage}{0.33\textwidth}
  \centering
  \includegraphics[width=1\linewidth]{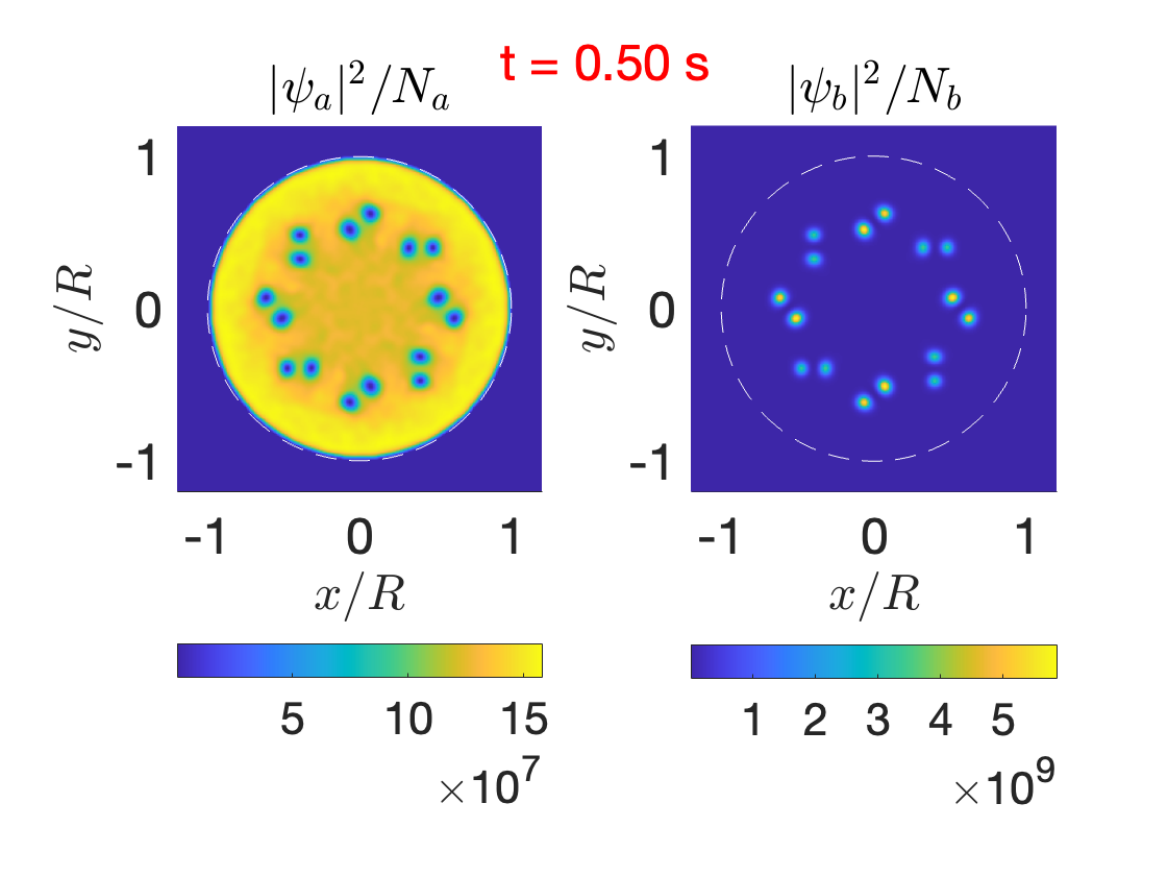}
\end{minipage}%
\begin{minipage}{0.33\textwidth}
  \centering
  \includegraphics[width=1\linewidth]{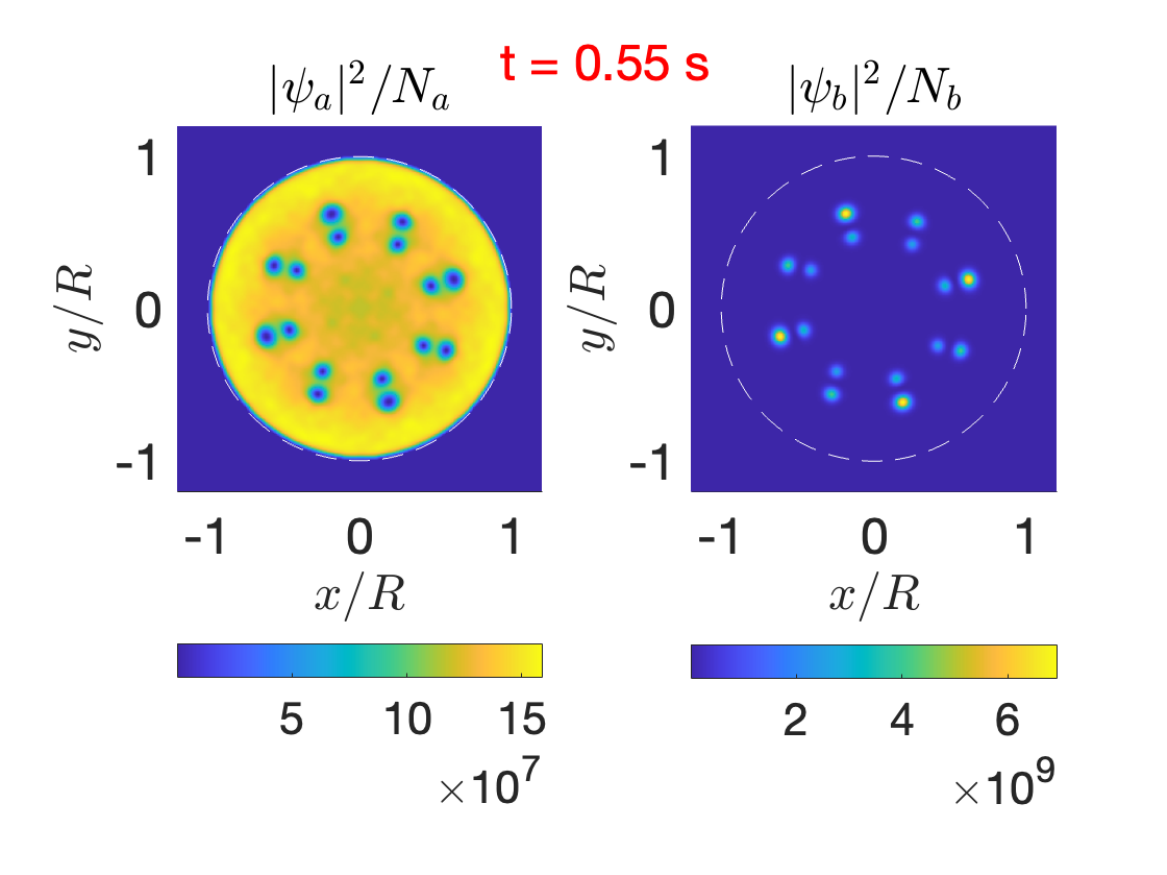}
\end{minipage}
\begin{minipage}{0.33\textwidth}
  \centering
  \includegraphics[width=1\linewidth]{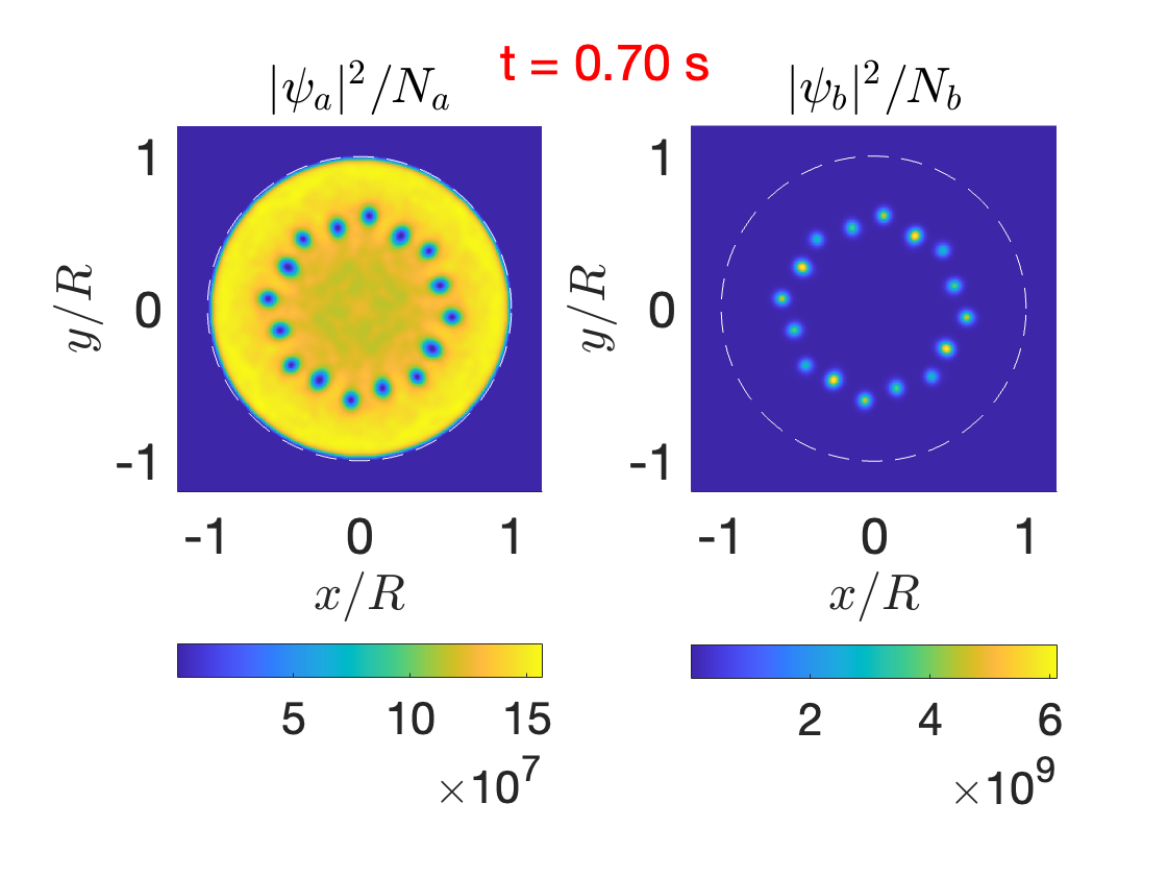}
\end{minipage}
\begin{minipage}{0.33\textwidth}
  \centering
  \includegraphics[width=1\linewidth]{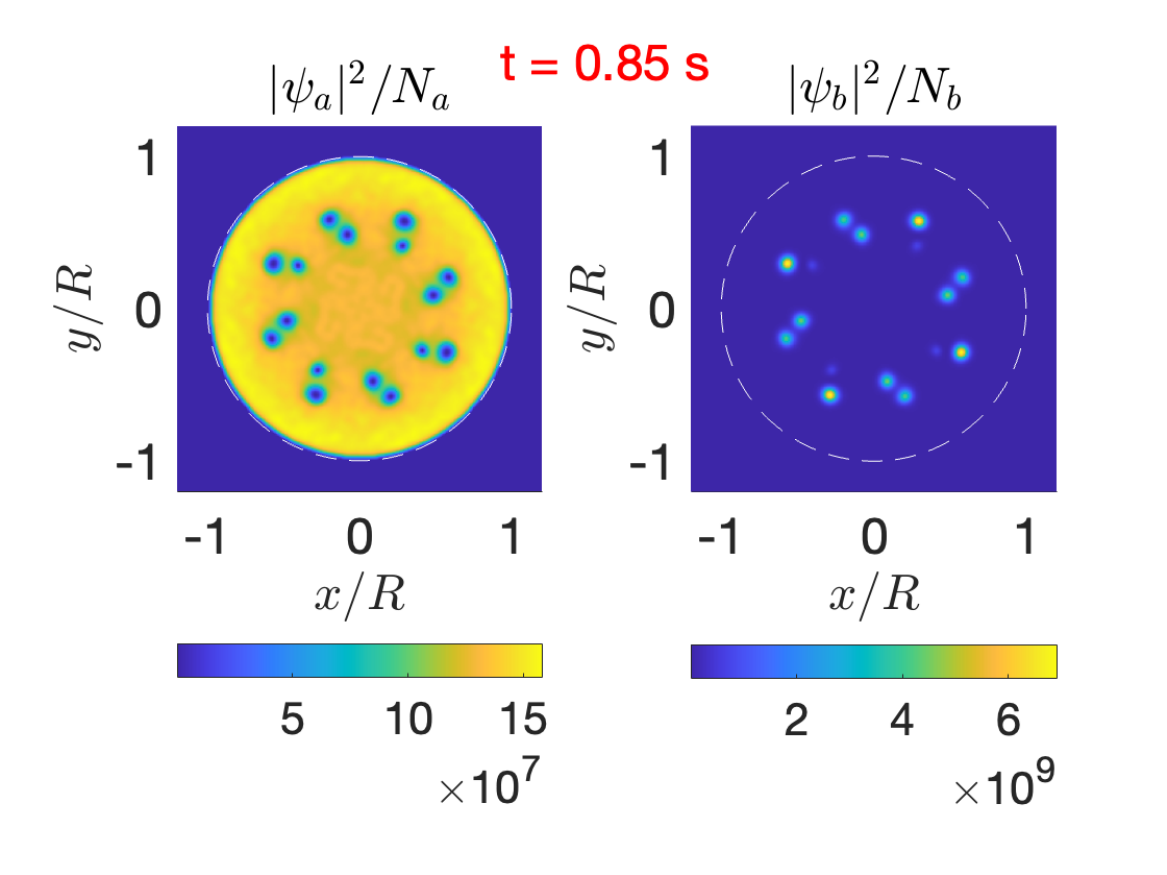}
\end{minipage}%
\begin{minipage}{0.33\textwidth}
  \centering
  \includegraphics[width=1\linewidth]{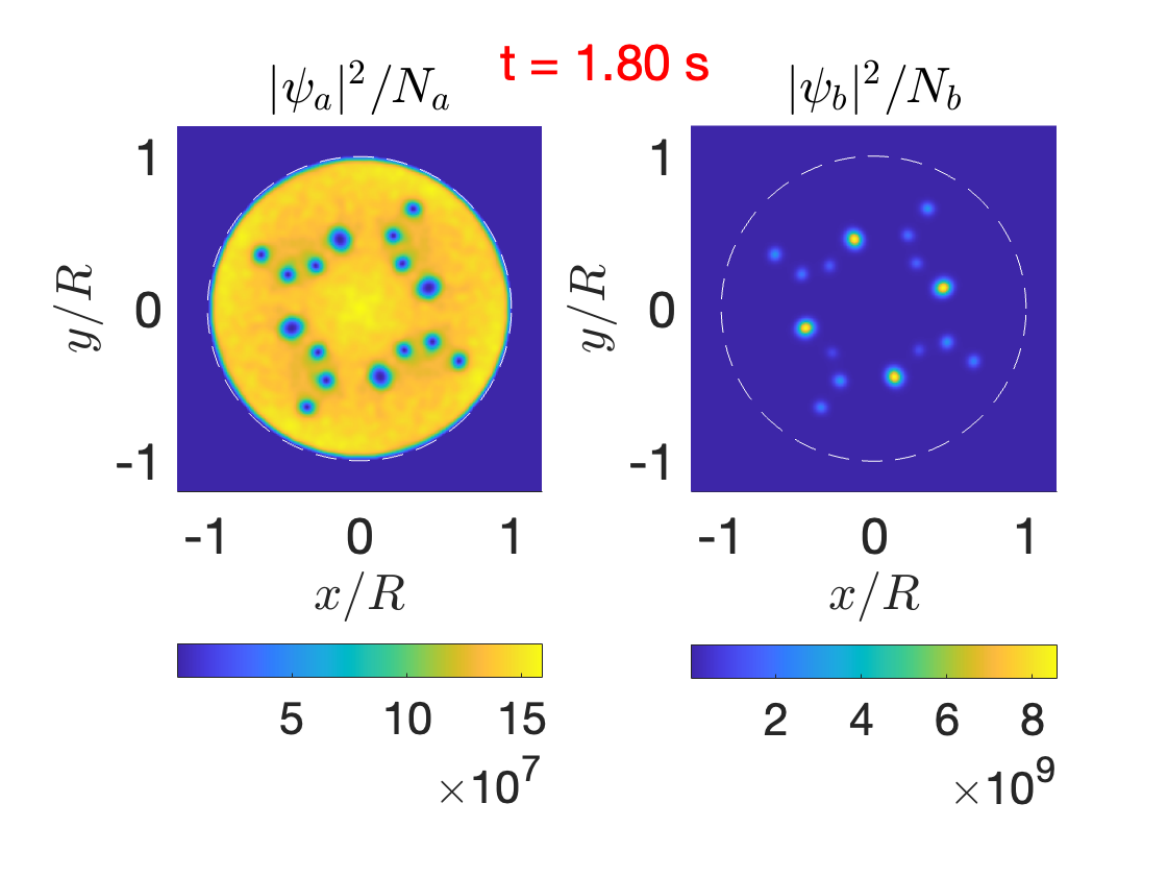}
\end{minipage}
\begin{minipage}{0.33\textwidth}
  \centering
  \includegraphics[width=1\linewidth]{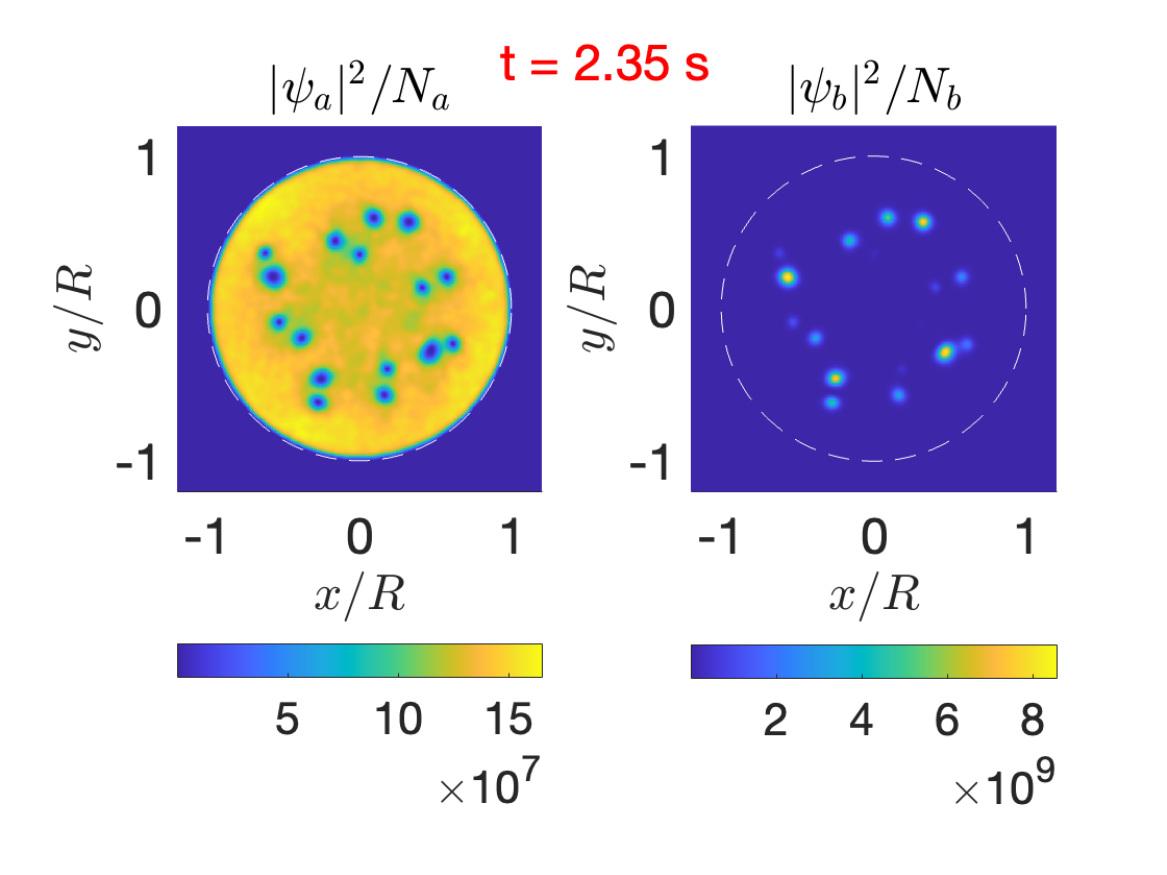}
\end{minipage}
\begin{minipage}{0.33\textwidth}
  \centering
  \includegraphics[width=1\linewidth]{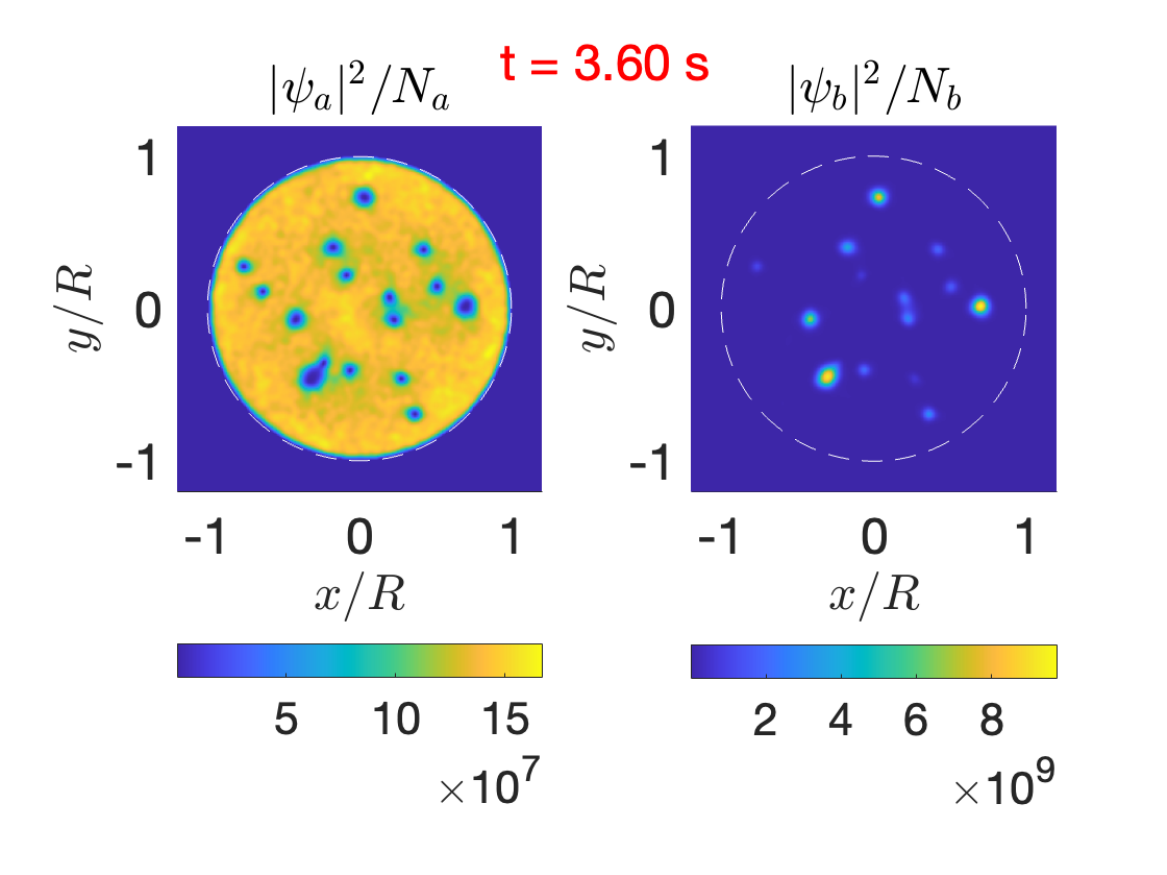}
\end{minipage}
\caption{Snapshots of the time evolution, at different times $t$, of the density fields $|\psi_a|^2$ and $|\psi_b|^2$, where the former hosts $16$ quantum-vortex wells initially arranged in a necklace, while the latter is the vortices' infilling-component undergoing quantum tunneling. In this system, the condensates are trapped in disk, and $N_b=10^3$.}
\label{fig:neckl_disk_16v}
\end{figure}

\subsection{Metastable behaviors in unstable massive vortex necklaces}
\label{sec:necklaces}

As a first example of massive vortex necklaces we examine some cases of unstable configurations, to characterize their destruction mechanisms and the role of the mass exchanges between the vortices, identifying some metastable dynamical features.

\begin{figure}
\centering
\begin{minipage}{0.33\textwidth}
  \centering
  \includegraphics[width=1\linewidth]{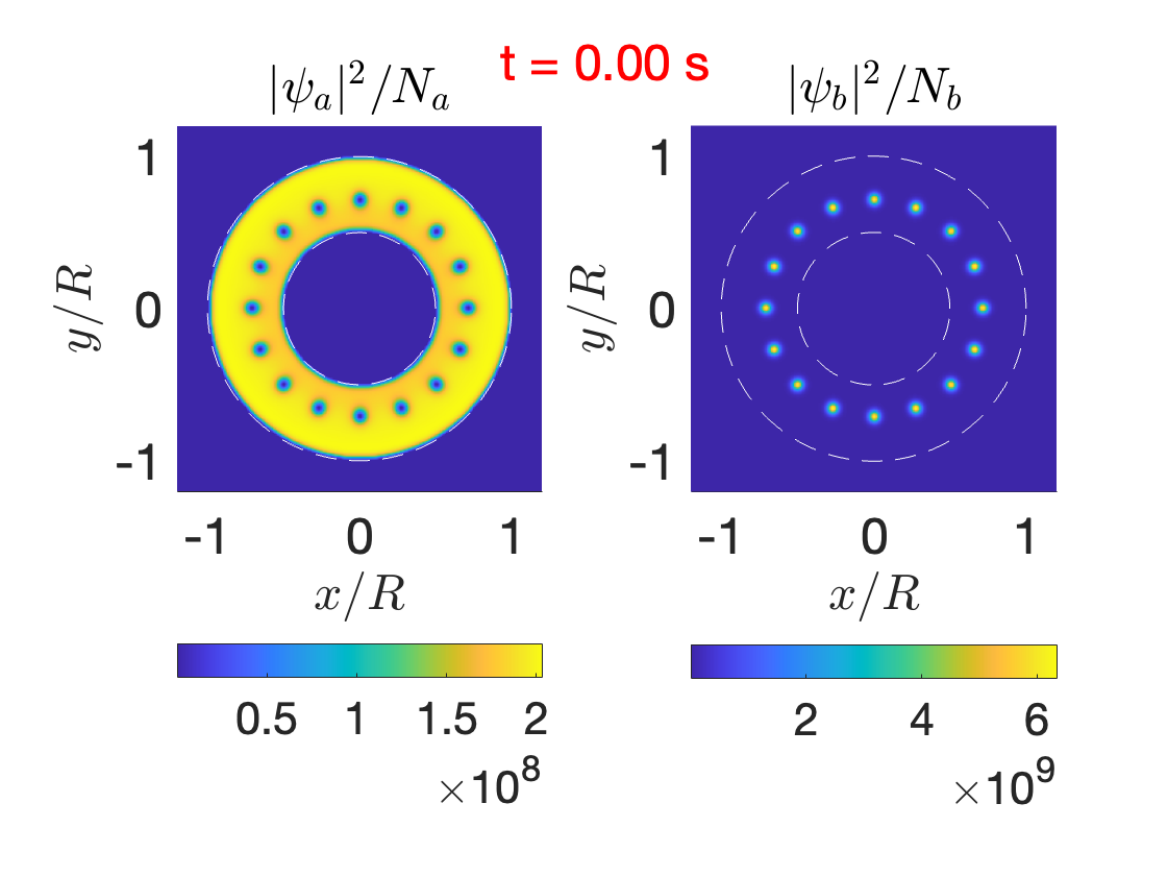}
\end{minipage}
\begin{minipage}{0.33\textwidth}
  \centering
  \includegraphics[width=1\linewidth]{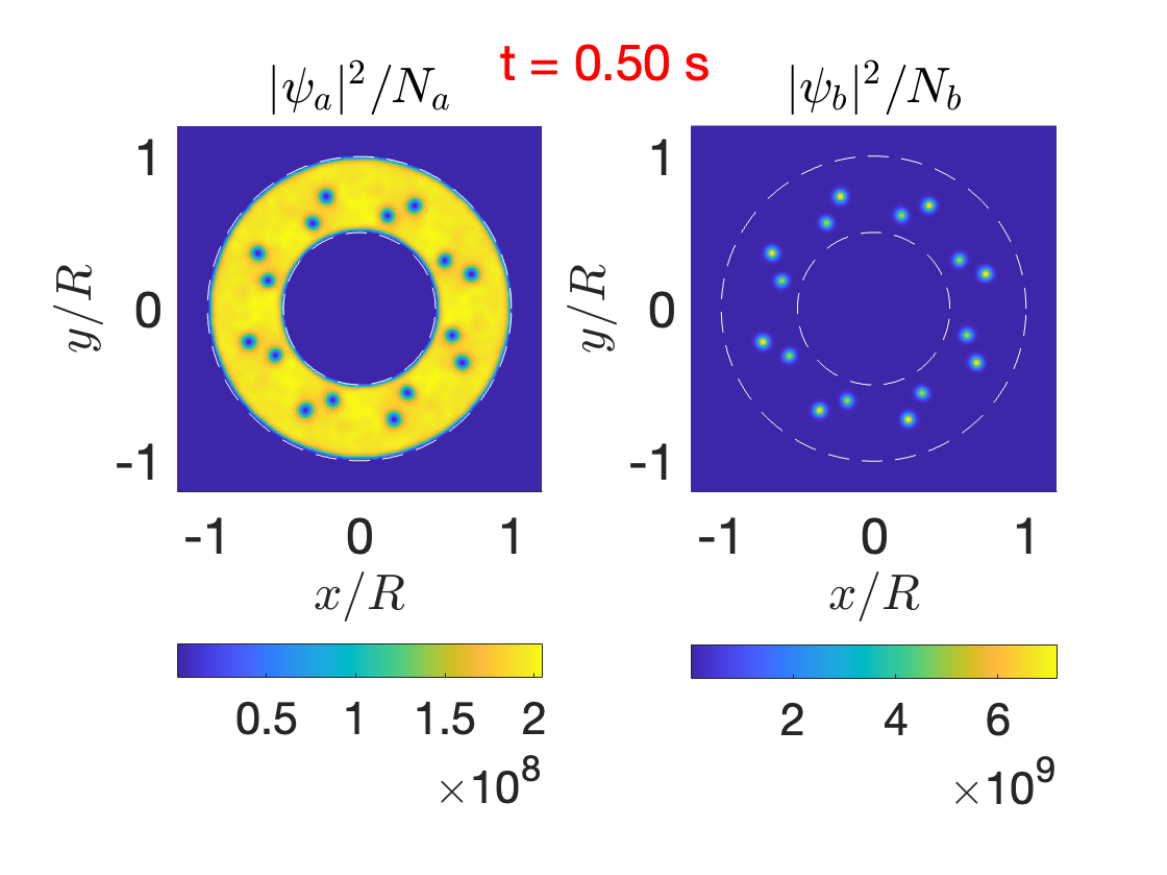}
\end{minipage}%
\begin{minipage}{0.33\textwidth}
  \centering
  \includegraphics[width=1\linewidth]{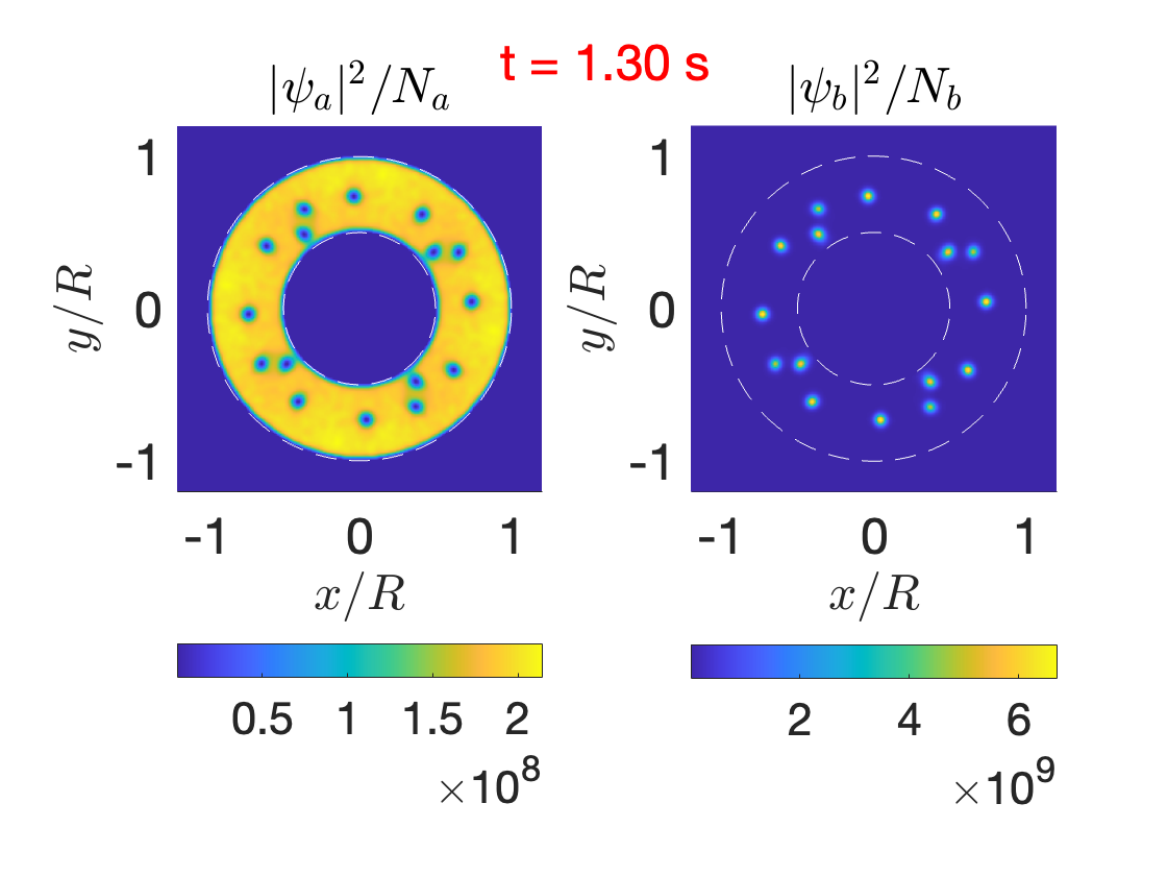}
\end{minipage}
\begin{minipage}{0.33\textwidth}
  \centering
  \includegraphics[width=1\linewidth]{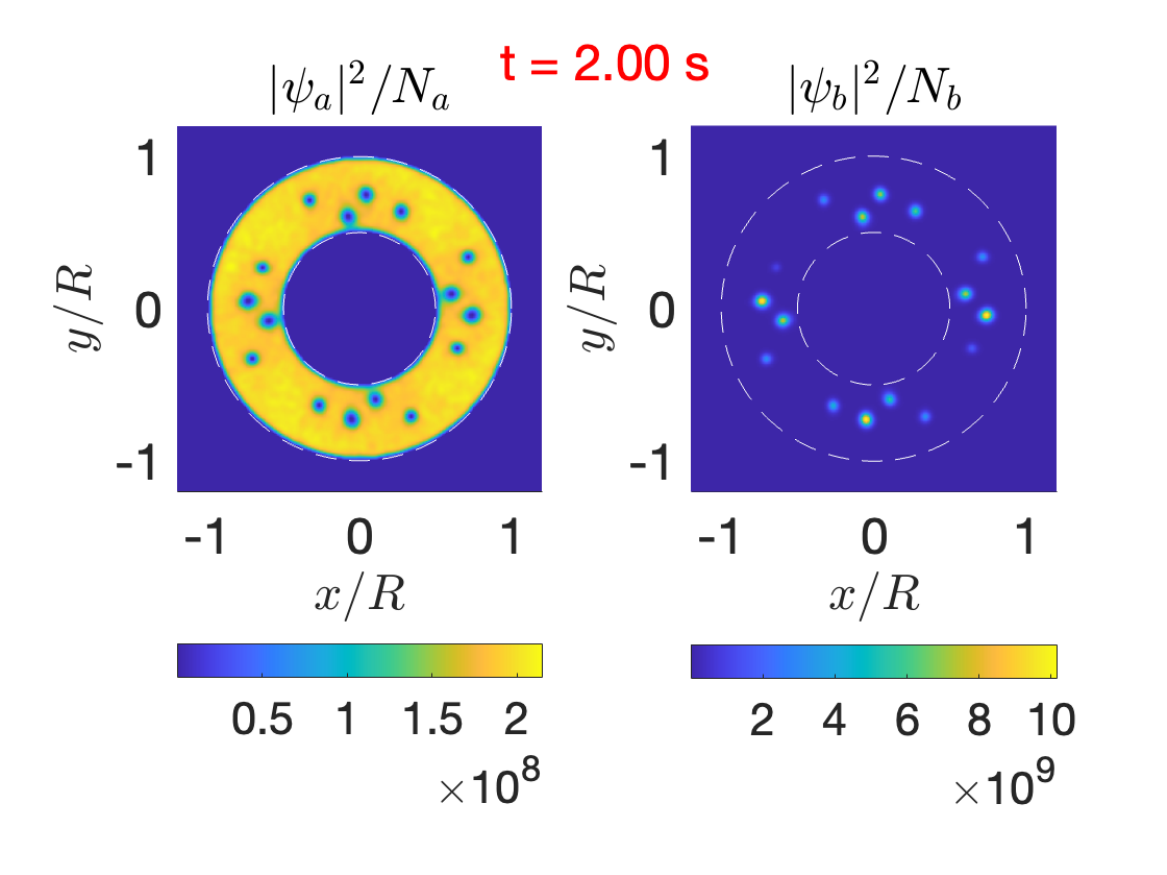}
\end{minipage}
\begin{minipage}{0.33\textwidth}
  \centering
  \includegraphics[width=1\linewidth]{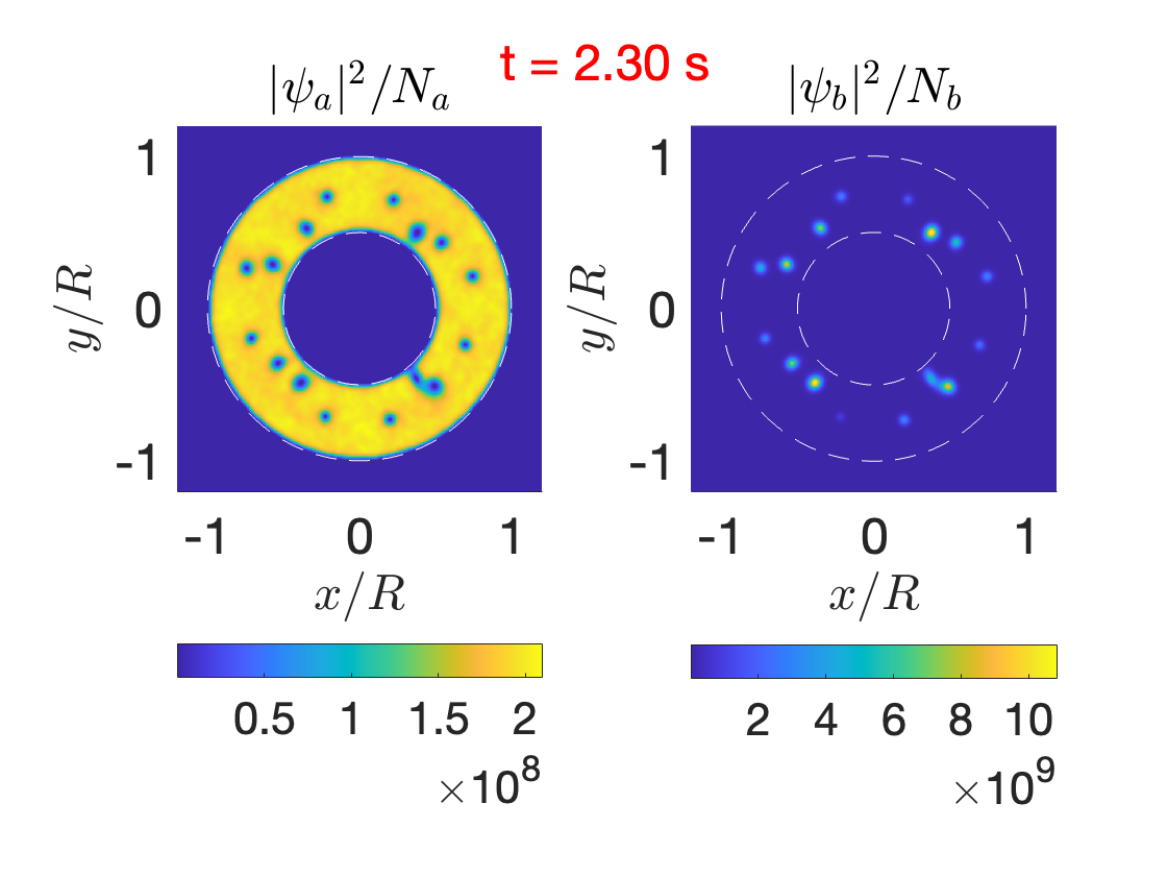}
\end{minipage}%
\begin{minipage}{0.33\textwidth}
  \centering
  \includegraphics[width=1\linewidth]{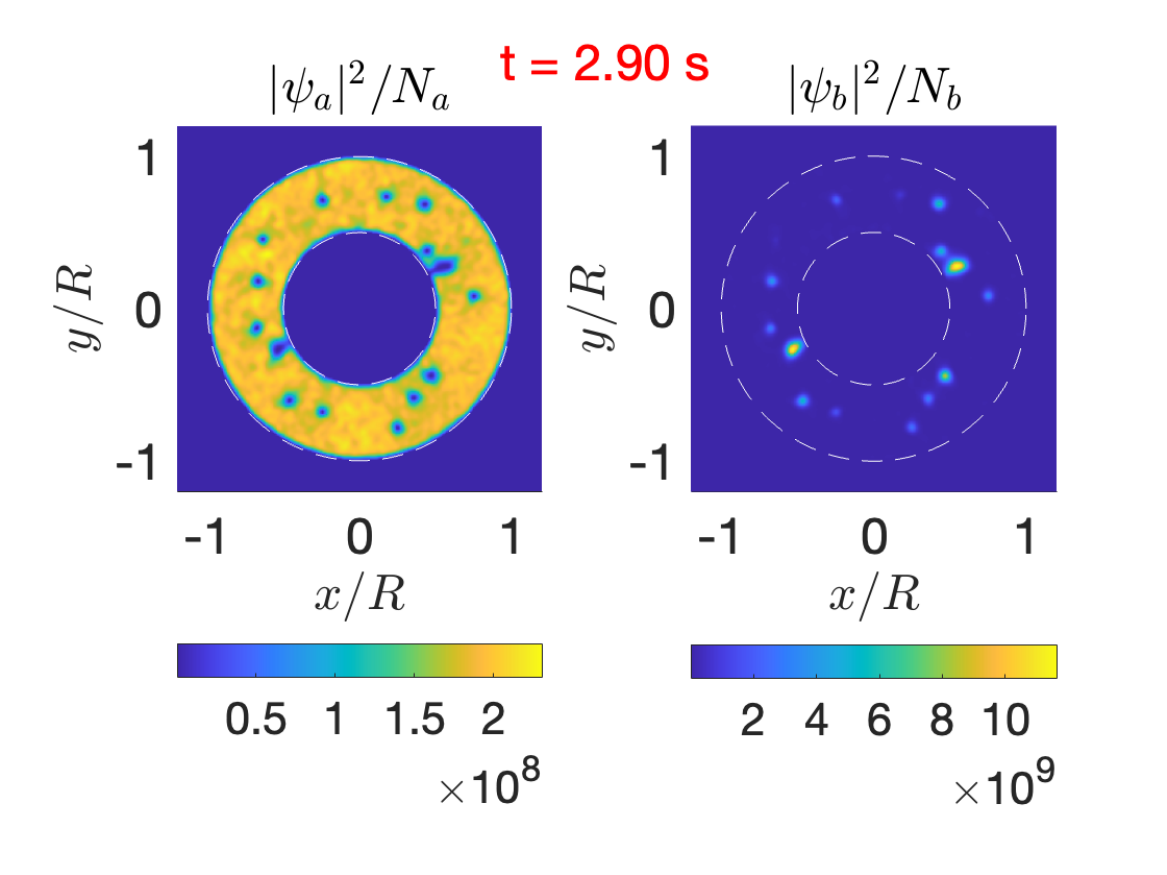}
\end{minipage}
\begin{minipage}{0.33\textwidth}
  \centering
  \includegraphics[width=1\linewidth]{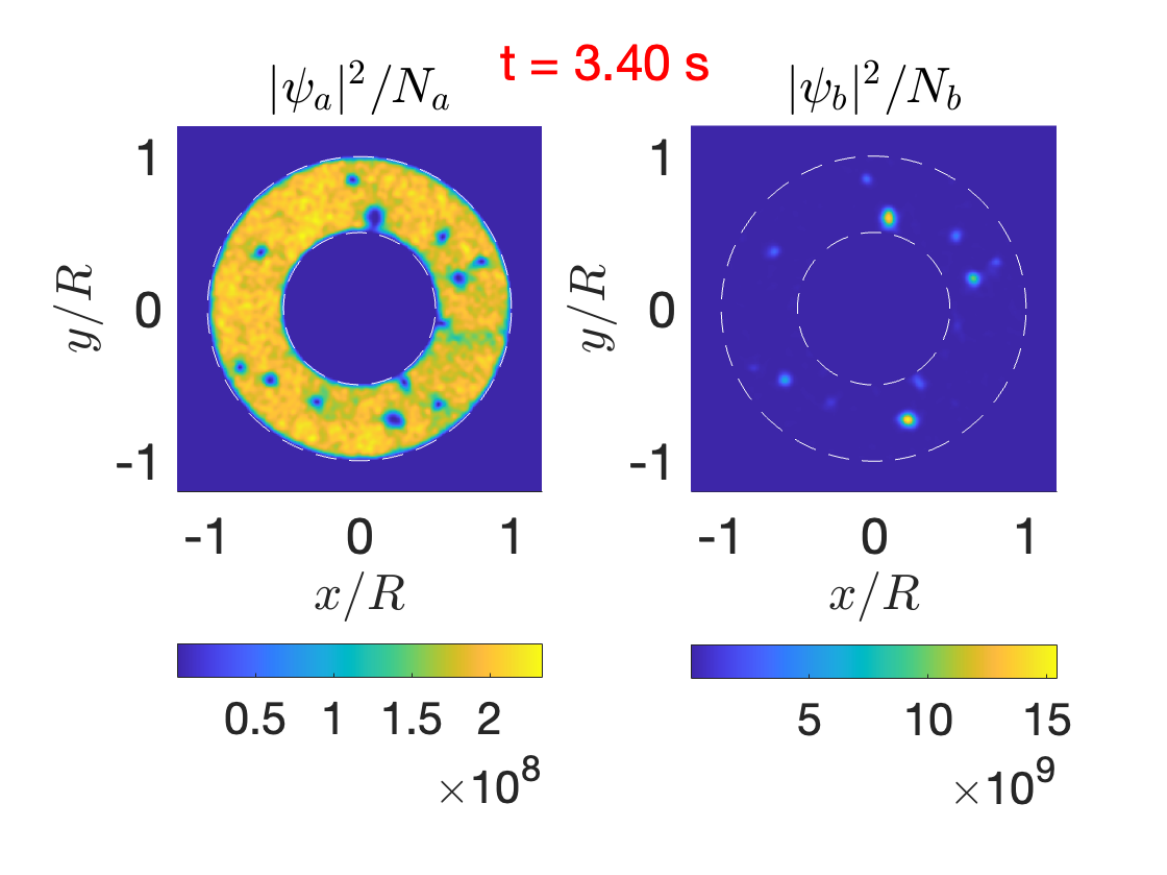}
\end{minipage}
\begin{minipage}{0.33\textwidth}
  \centering
  \includegraphics[width=0.8\linewidth]{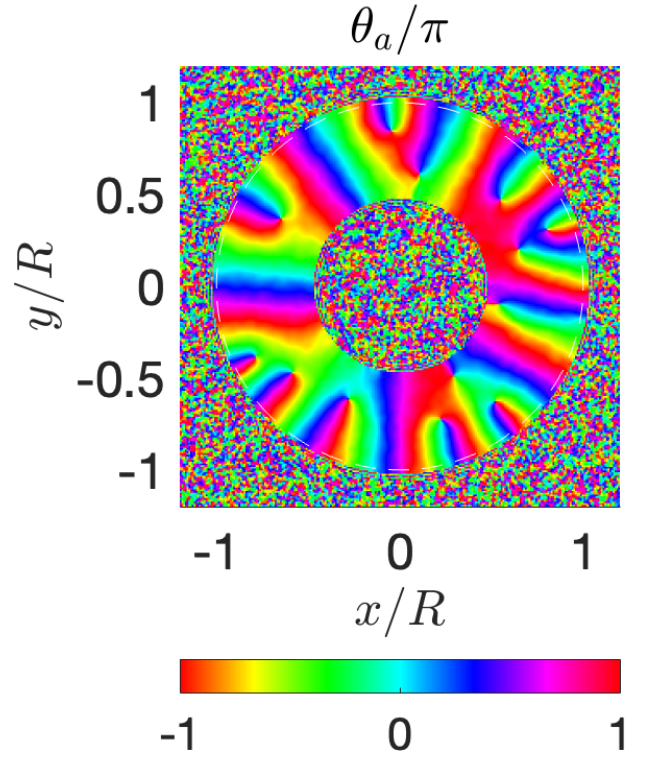}
\end{minipage}
\caption{Dynamical evolution of the two BECs in a trap with an annular geometry. Many qualitative features for the evolution of $16$ identical massive vortices initially arranged in a necklace are the same for the disk geometry represented in Figure \ref{fig:neckl_disk_16v}. The last panel shows the phase field $\theta_a$ at the time $t=3.40$ $s$ and corresponds to the last density snapshot in the second row. The inital precession direction of the necklace is anti-clockwise.
We have $N_b=10^3$ and $q=0.5$.
}
\label{fig:neckl_annulus_16v}
\end{figure}

In \textbf{Figure \ref{fig:neckl_disk_16v}} and \textbf{\ref{fig:neckl_annulus_16v}} we see the dynamics of a system of $16$ vortices initially arranged in a necklace with a uniform infilling-mass distribution, respectively in a 2D disk and a 2D annulus.
For both the trap geometries, this configuration shows a transient with an initial metastable behavior followed by the disruption of the necklace, disclosing the predominant unstable character of the system.
Note that this result is in agreement with the prediction given by the PL model illustrated in Fig. \ref{fig:sigma_star}, showing that a $16$-vortex necklace is always unstable for both the geometries.
In the case of the disk, the metastable transient consists of repeated radial contractions and expansions of the massive vortex necklace, whereas in the case of the annulus this transient is much shorter. Such necklace-breathing dynamics is found
in several instance of vortex necklaces at varying $N_v$.

During this breathing effect, the vortex necklace rotates, as expected, due to the effect of the global vortex-generated current \cite{Helmholtz1858, Kirchhoff1876}.

The breathing of necklace is followed by its rupture, occurring, in the example in Figure \ref{fig:neckl_disk_16v} of a 2D-disk geometry, at $t\simeq 0.45$ $s$, and in the example in Figure \ref{fig:neckl_annulus_16v} of a 2D-annulus geometry, at $t\simeq 0.43$ $s$.
In both the geometries, after the necklace rupture, the vortices arrange themselves in a ring of pairs, each one exhibiting bosonic tunneling. These pairs make up a set of BJJs at any instant of time, where the vortices in the BJJs swap partners as the dynamics goes on.
The clear pairwise bosonic exchange lasts, in the case of the disk, up to
$t\simeq 1.30$ $s$, whereas in the annulus $t\simeq 0.7$ $s$. 
Subsequently, the systems evolve in more complex configurations. These are as well characterized by the tunneling of the vortices' infilling population, and the tunneling rate between two vortices is the higher the lower the distance between the vortices is, so that the tunneling events predominantly occur between nearest neighbors. Over time, the vortex dynamics becomes disordered, suggesting the onset of a fully chaotic regime. In passing, note that in both Figure \ref{fig:neckl_disk_16v} and \ref{fig:neckl_annulus_16v} it is visible how the vortices' cores expand or contract depending on the amount of mass they host.

It is worth observing how the annulus geometry, exhibiting a double boundary, adds complexity to the vortex dynamics. For example, it makes it possible for a single vortex to have two different precession directions, depending on its position with respect to the boundaries, an effect previously observed in Ref. \cite{Caldara2024}.
Also, the vortex absorption at the inner boundary can take place, triggering a superfluid background current. Figure \ref{fig:neckl_annulus_16v} shows this effect. In the figure, the last panel of the lower row shows a snapshot of the phase field $\theta_a$ of the order parameter $\psi_a$. Here one can identify the vortex positions as the phase singularities around which $\theta_a$ has a winding of $2\pi$. As visible in the figure, some of the vortices initially present in the trap have been absorbed by the central hole of the annulus, giving rise to a background-current contribution.  
This background superfluid current, 
is caused by its non trivial topology, and allows for a phase winding around the inner hole, an effect not visible in simpler disk geometry.

In conclusion, we note that in both Figure \ref{fig:neckl_disk_16v} and \ref{fig:neckl_annulus_16v},
preceding the transition to a disordered configuration, a series of ``ordered'' configurations, displaying some discrete symmetries are well visible.

\subsection{Component-\texorpdfstring{$b$}{b} diffusion in initial states with a
strong population imbalance}
\label{sec:peak_propagation}

In this section we examine the diffusion process for some significant initial filling distribution of the necklace.
We study how a necklace configuration with a single peak in the $b$-component evolves within the disk geometry,  
observing how the peak diffuses and modifies the filling of the adjacent vortex wells.

\begin{figure}[ht]
\centering
\begin{minipage}{0.24\textwidth}
  \centering
  \includegraphics[width=1\linewidth]{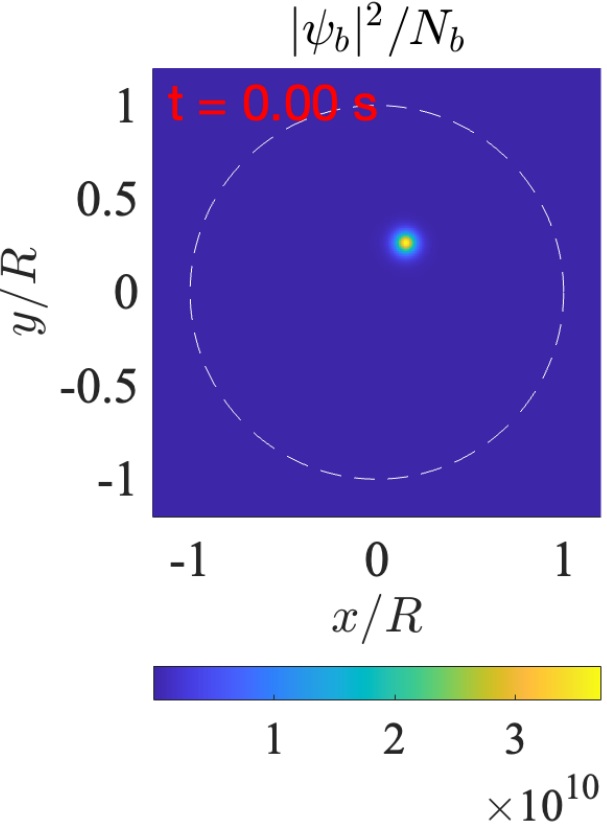}
\end{minipage}
\begin{minipage}{0.24\textwidth}
  \centering
  \includegraphics[width=1\linewidth]{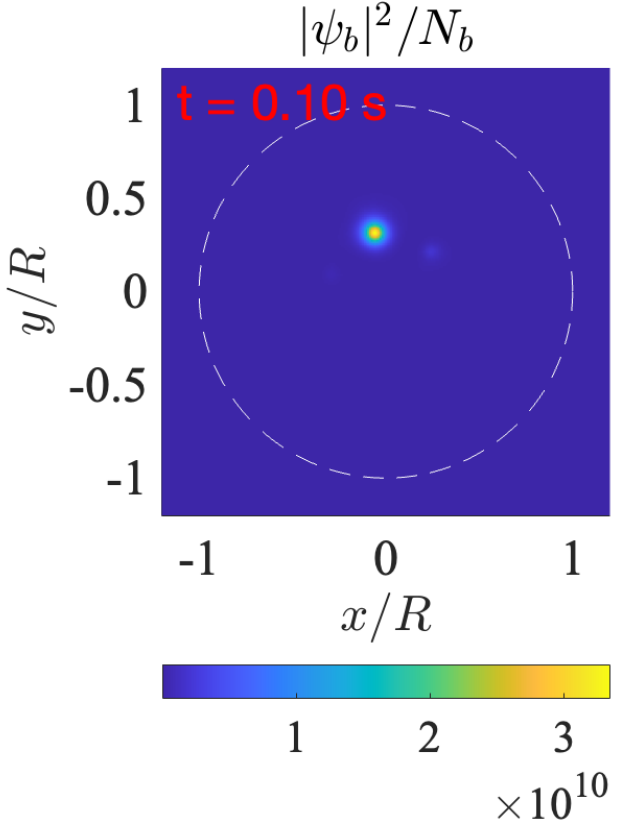}
\end{minipage}
\begin{minipage}{0.24\textwidth}
  \centering
  \includegraphics[width=1\linewidth]{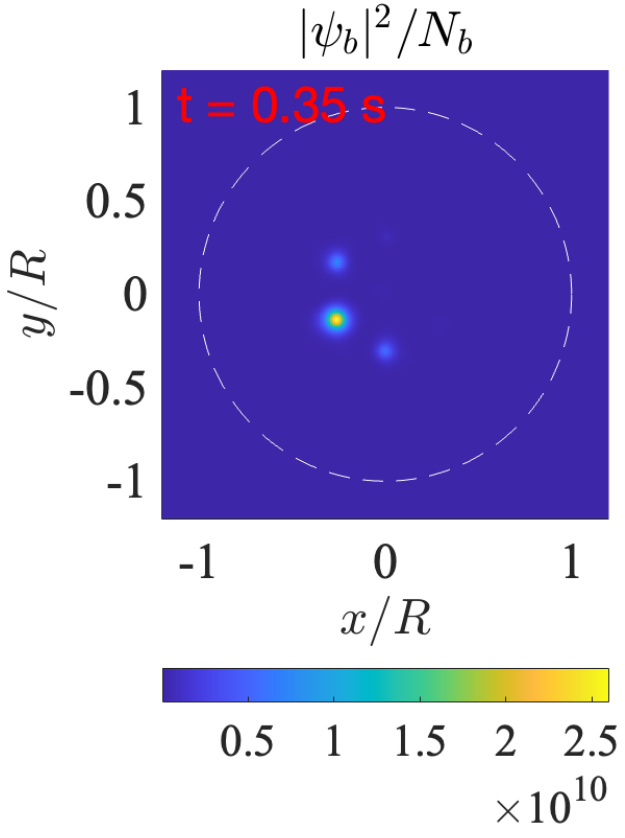}
\end{minipage}
\begin{minipage}{0.24\textwidth}
  \centering
  \includegraphics[width=1\linewidth]{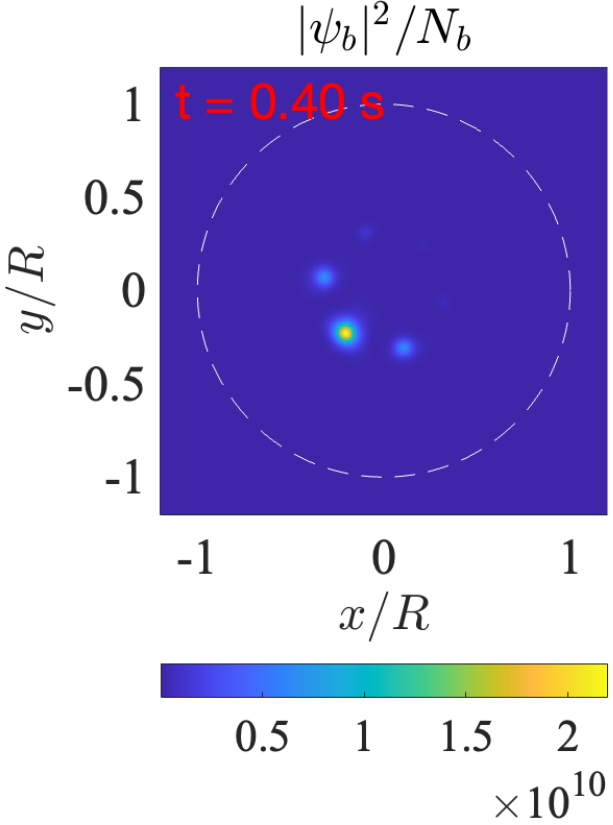}
\end{minipage}
\caption{Time evolution of the density field $|\psi_b|^2$ for a six-vortex necklace starting with a single $b$-peak located at a single vortex-well. Here $N_b= 10^2$, and $g_{ab}/\sqrt{g_a g_b}=1.26$.
}
\label{fig:sol_prop_disk}
\end{figure}

The first example concerns six vortices with $10^2$ infilling particles,
in a stable vortex-necklace, i.e., a configuration which persists for the total observation time (here: $10$ $s$). 
Here, the vortices are close enough to observe a tunneling between nearest neighbors, and $g_{ab}$ is a bit lower than the standard value (we have $g_{ab}/\sqrt{g_a g_b}=1.26$) to allow for such a configuration.
This system is 
reported in \textbf{Figure \ref{fig:sol_prop_disk}},
where we see how the effect of the rotation (anti-clockwise) induces a non-symmetric peak diffusion, namely the nearest well on the right of the $b$-peak is populated first, followed by the nearest well on the left. Subsequently, the bosonic current within the necklace acquires a more complex character (see Section \ref{sec:current_in_necklace}).
Interestingly, the peak diffusion takes place in a similar way in the annular geometry, as represented in \textbf{Figure \ref{fig:sol_prop_annulus}} in Appendix \ref{sec:app}.

Necklaces with more vortices and a smaller distance between nearest neighbors, or at larger values of $N_b$, feature
the same type of peak-diffusion dynamics shown in Figure \ref{fig:sol_prop_disk} and \ref{fig:sol_prop_annulus}.
In both geometries when the vortex number is increased, however, 
the necklace is not stable and gets destroyed after some time. 
Here, due to the presence of more vortices, the initial transient of the peak-diffusion
is, after the population of the two nearest neighbors, followed by the population
of the second-nearest well against the precession direction, and then  by the second-nearest well in the precession direction.
On the other hand, with $N_b=5\times 10^2$ or $N_b=10^3$ atoms concentrated in the initial peak, we see stable necklaces only in the case of the disk, but not for the annulus.

On the other hand, for fewer-vortex necklaces ($N_v=3,4,5$)
that feature a larger distance between nearest-neighbor vortices, we observe in both geometries the trapping of the peak within the initial vortex' site, with no filling of the other vortex wells for $N_v=3,4$ and with a negligible filling for $N_V=5$. In these examples, the vortex necklaces are also stable throughout the observation time of $10$ $s$.
Interestingly, the PL model predicts in fact stable necklace solutions for $N_v=3,4,5,6$ 
in the disk and in an annulus at $q=0.1$, with $N_b=10^2$. These are at an appropriate $\Omega$ given $r_0$, with $r_0$ as approximately in the initial configurations of the GPEs simulations. Namely, the survival of some necklaces throughout the whole simulation time could be explained by a similarity, up to the vortex-mass distribution, to stable fixed-point solutions in the PL framework.

\subsection{Supercurrents in a few-vortex necklaces}
\label{sec:current_in_necklace} 

In Section \ref{sec:peak_propagation} we found an example, in the disk trap, of a vortex necklace supporting 
bosonic currents between the vortex wells for a very long observation time, i.e. $10$ $s$. In this section,  we examine in more detail the tunneling phenomena occurring in the six-vortex necklace, within the disk-shaped trap, at varying $N_b$. We always consider an initial configuration where one vortex is filled and the others are approximately empty, i.e. $|\psi_b|^2$ has a single peak.
Furthermore, as in Section \ref{sec:peak_propagation}, we have $g_{ab}/\sqrt{g_a g_b}=1.26$, i.e. a higher miscibility with respect to the standard case. Thanks to this condition we can create an enough small-sized necklace and observe appreciable variations of the vortex-wells' populations.

\begin{figure}
\centering
\begin{minipage}{0.138\textwidth}
  \centering
  \includegraphics[width=1\linewidth]{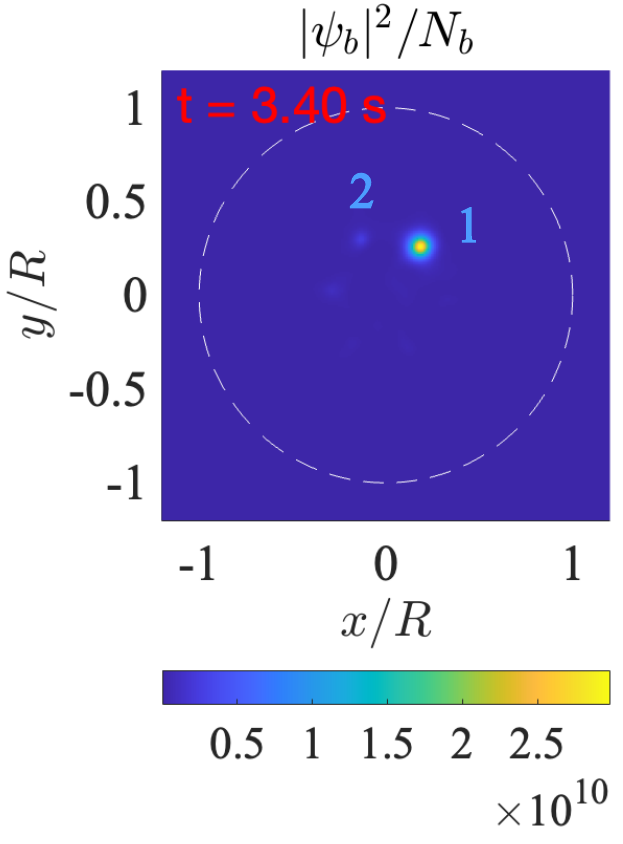}
\end{minipage}
\begin{minipage}{0.138\textwidth}
  \centering
  \includegraphics[width=1\linewidth]{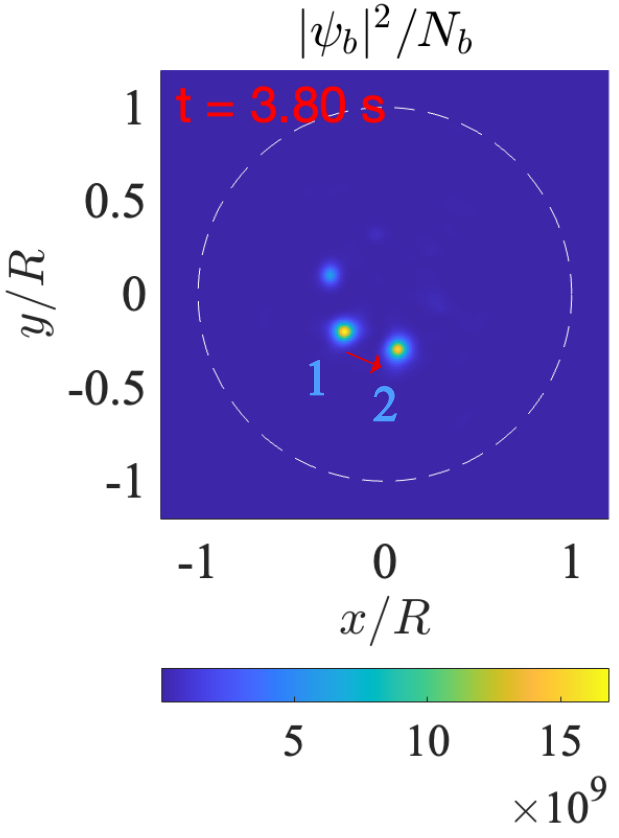}
\end{minipage}
\begin{minipage}{0.138\textwidth}
  \centering
  \includegraphics[width=1\linewidth]{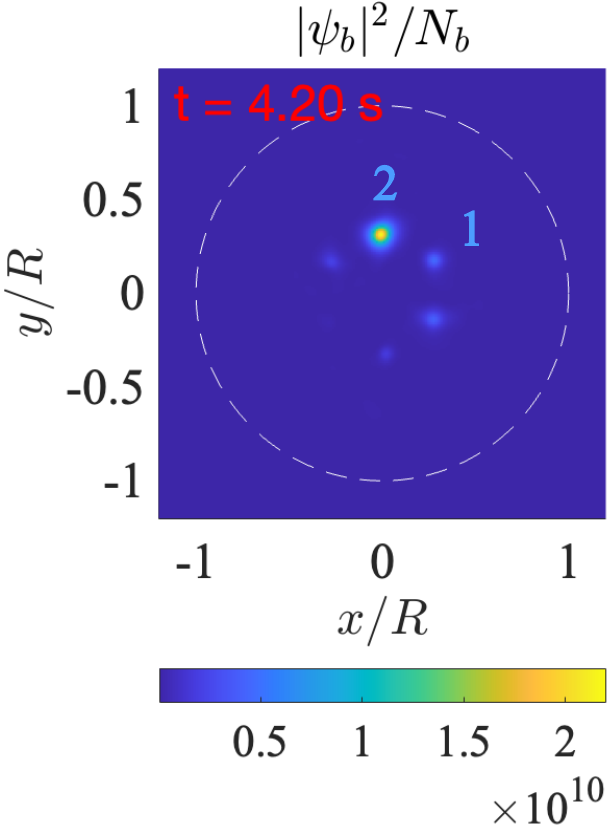}
\end{minipage}
\begin{minipage}{0.138\textwidth}
  \centering
  \includegraphics[width=1\linewidth]{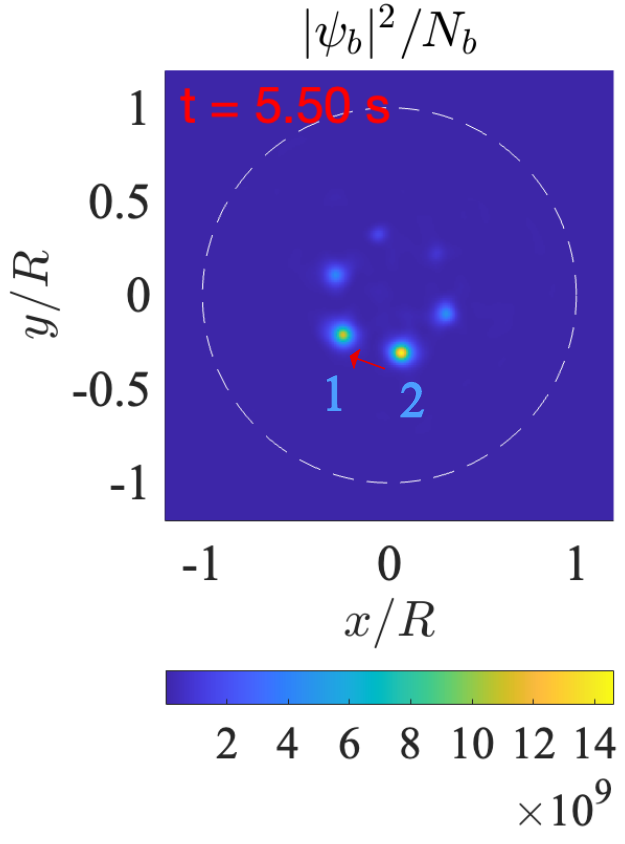}
\end{minipage}
\begin{minipage}{0.138\textwidth}
  \centering
  \includegraphics[width=1\linewidth]{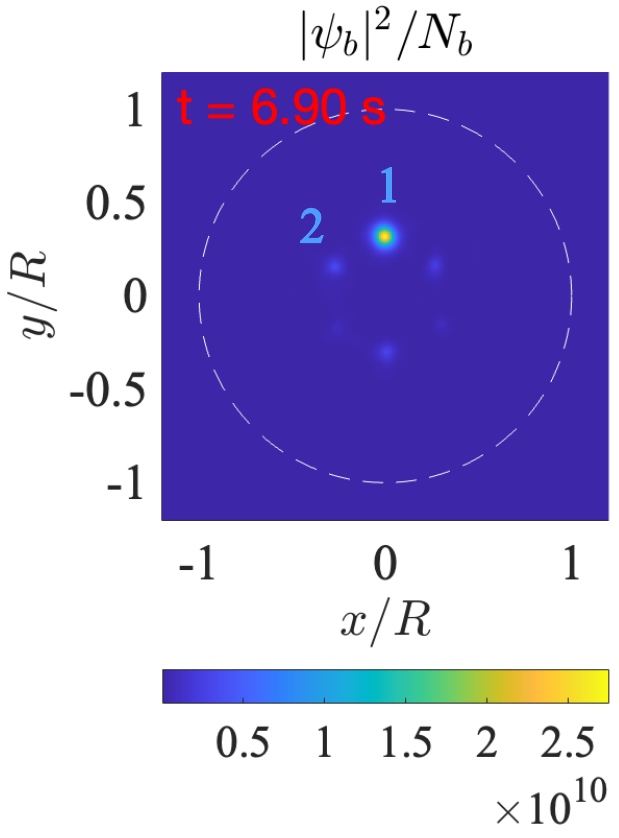}
\end{minipage}
\begin{minipage}{0.138\textwidth}
  \centering
  \includegraphics[width=1\linewidth]{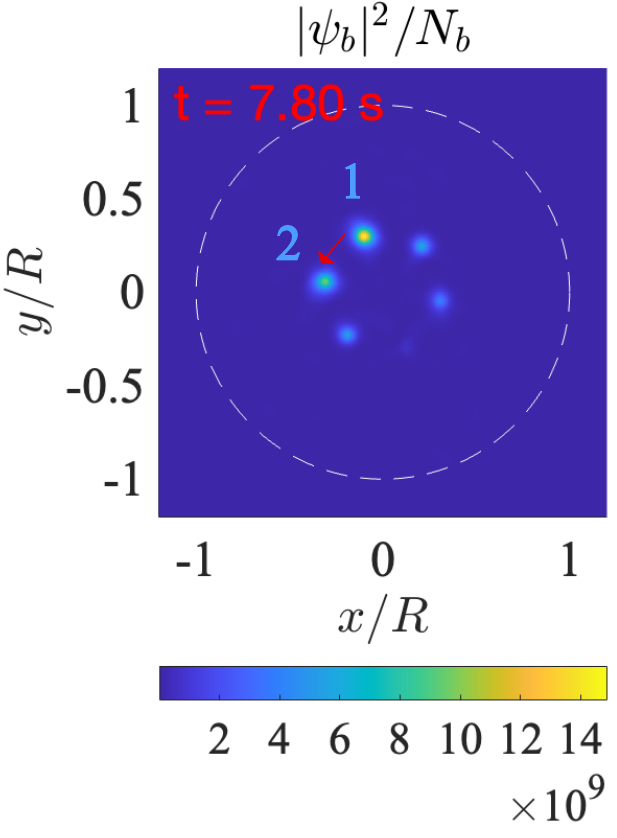}
\end{minipage}
\begin{minipage}{0.138\textwidth}
  \centering
  \includegraphics[width=1\linewidth]{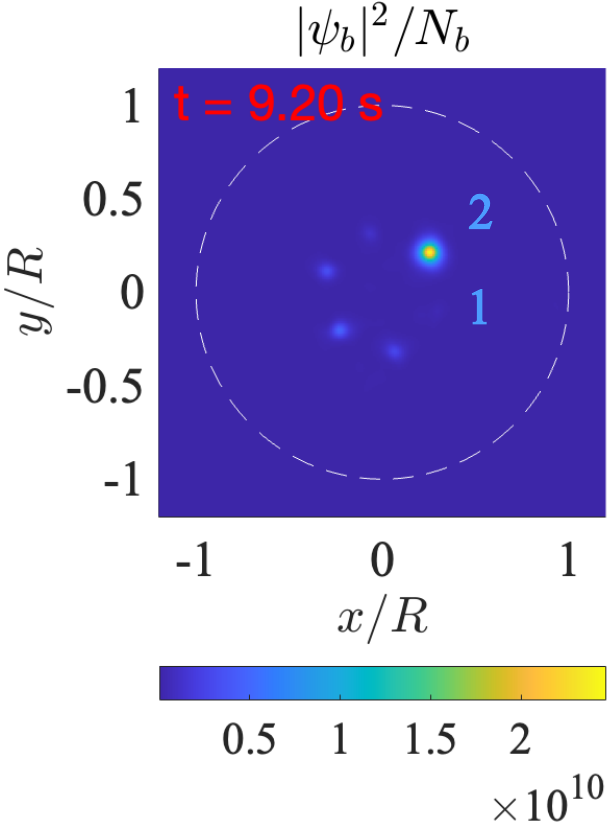}
\end{minipage}
\caption{Bosonic tunneling in a necklace of $6$ vortices. Within the complex nonlinear dynamics we recognize some major population's exchange between two given vortices, labeled by ``$1$'' and ``$2$''. In the panels the tunneling events leading to such population's exchange (at: $t\simeq3.80$ $s$, $t\simeq5.50$ $s$, and $t\simeq7.80$ $s$, occurring respectively in anti-clockwise, clockwise and anti-clockwise direction) are shown together with the configurations preceding the exchange. 
The red arrows highlight the direction of the peak transfer.
Here, $N_b=10^2$, $g_{ab}/\sqrt{g_a g_b}=1.26$.
}
\label{fig:6v_diskCurrent_100}
\end{figure}

The first example is the same of Figure \ref{fig:sol_prop_disk} (see the initial configuration in the first panel of Figure \ref{fig:sol_prop_disk}) involving $N_b=10^2$ tunneling atoms. After the initial transient shown in the figure, we observe other interesting phenomena occurring during the complex tunneling dynamics which takes place at longer times. Specifically, we see a transfer of the $b$-peak, over longer time scales, between the same two nearest-neighbor vortices. This dynamics is reminiscent of an open-BJJ behavior where the two wells are co-rotating.
As a matter of fact, it involves an oscillating population imbalance between two adjacent vortices, but the number of infilling particles in two-vortex sub-system is not fixed. In point of fact, the other four vortices get also significantly populated throughout the system's dynamics.
\textbf{Figure \ref{fig:6v_diskCurrent_100}} illustrates the tunneling events between the two nearest neighbors (labeled as ``$1$'' and ``$2$''), and the configurations preceding the tunneling, where the tunneling events in the BJJ-like subsystem are highlighted by the arrows. We have that the first population inversion occurs at $t\simeq3.80$ $s$ in the anti-clockwise direction, i.e. in the same direction of the necklace's rotation, whereas the second tunneling process leading to the peak's transfer occurs at $t\simeq5.50$ $s$ against the rotation direction, and the third peak's displacement occurs at $t\simeq 7.80$ $s$ again in the direction of the necklace rotation. In general, the wells not belonging to the subsystem feature appreciable population's variations during the dynamics, however
for a time lapse
before the peak's transfer they get very lowly populated or empty.

Examining a similar system but at $N_b=5\times 10^2$, we observe an analogous regime, where the peak swaps occur earlier, but only for an initial transient of time (up to $t\simeq 5$ $s$); 
subsequently, the peak is transferred to other vortices than the two that hosted the largest peak's exchanges before. As a consequence, the necklace undergoes severe deformation, e.g. resembling a triangle at some instants of time. Nonetheless, within the $10$ $s$ of our total observation it is not destroyed.
Moreover, we studied the analogous system at $N_b=10^3$, shown in Appendix \ref{sec:app} in \textbf{Figure \ref{fig:6v_diskCurrent_1000}}. We see here that several major peaks are created in the necklace, and several nearest-neighbors population's exchanges take place between varying pairs of vortices. Again, the presence of larger vortex masses with respect to the case at $N_b=10^2$ leads to strong deformations in the necklace, which however is not destroyed within our simulation time of $10$ $s$.

The emergence of stable Josephson supercurrents in the necklace of Figure \ref{fig:6v_diskCurrent_100} suggests that stable supercurrents could be supported also in presence of different initial conditions. We find that at uniform vortex fillings and $N_b=10^2$, the necklace persists stable over time, featuring periodic contractions and expansions,
and the vortex fillings stay uniform during the dynamics. 

\begin{figure}
\centering
\begin{minipage}{0.138\textwidth}
  \centering
  \includegraphics[width=1\linewidth]{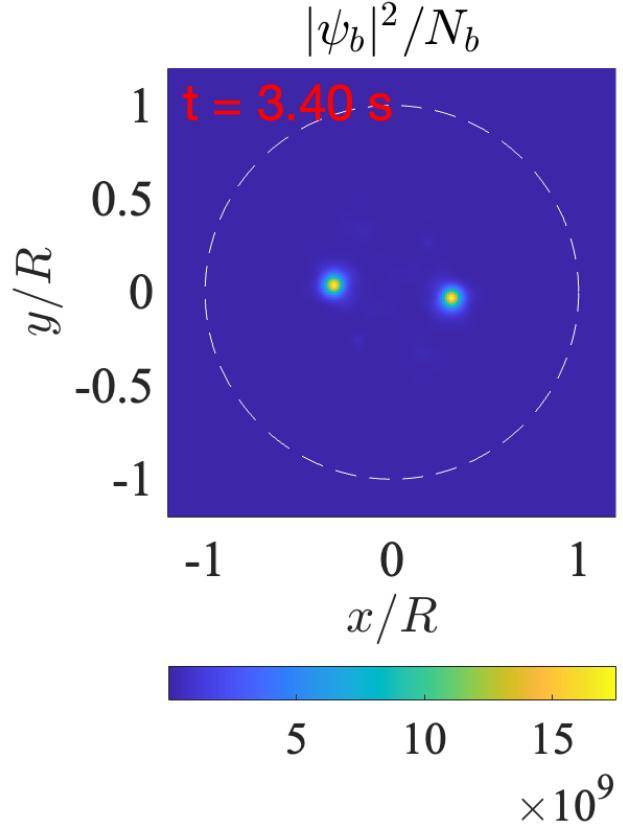}
\end{minipage}
\begin{minipage}{0.138\textwidth}
  \centering
  \includegraphics[width=1\linewidth]{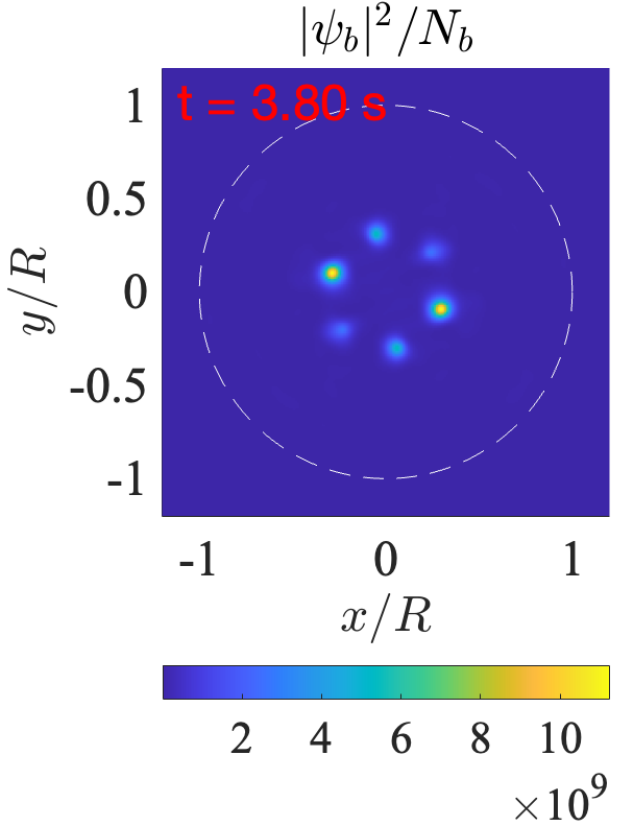}
\end{minipage}
\begin{minipage}{0.138\textwidth}
  \centering
  \includegraphics[width=1\linewidth]{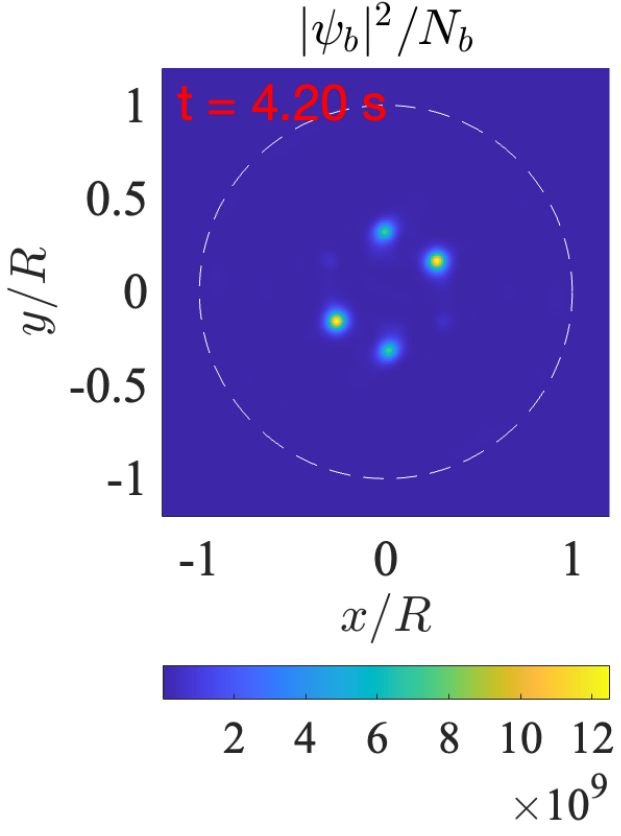}
\end{minipage}
\begin{minipage}{0.138\textwidth}
  \centering
  \includegraphics[width=1\linewidth]{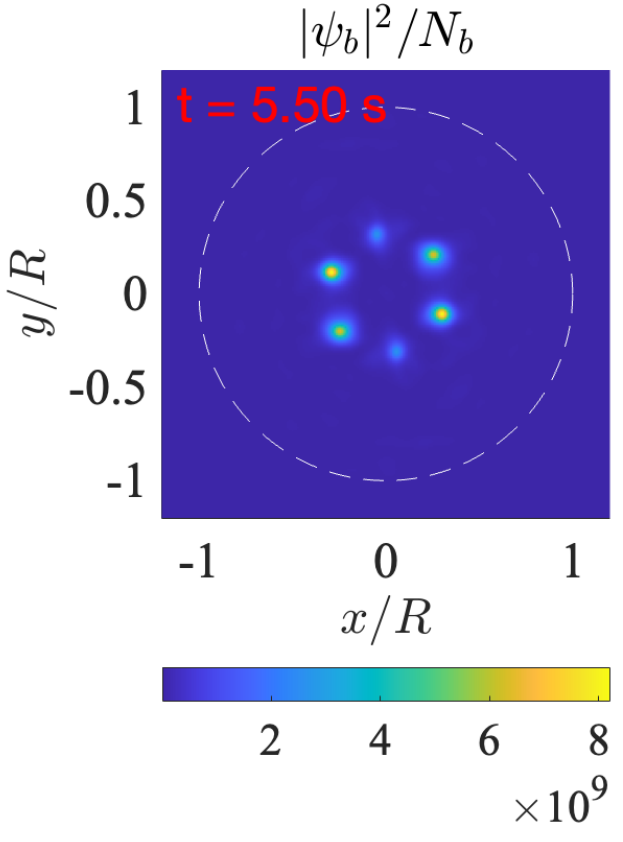}
\end{minipage}
\begin{minipage}{0.138\textwidth}
  \centering
  \includegraphics[width=1\linewidth]{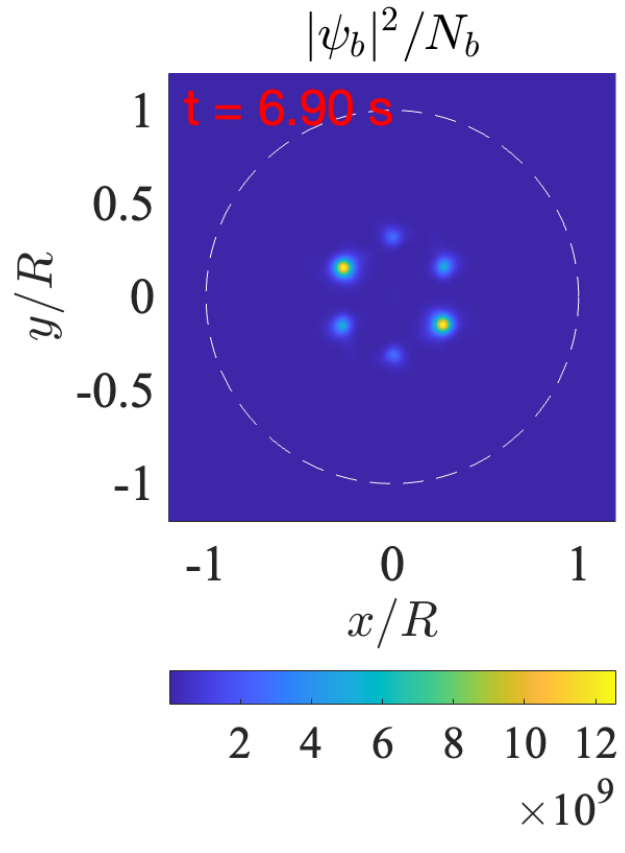}
\end{minipage}
\begin{minipage}{0.138\textwidth}
  \centering
  \includegraphics[width=1\linewidth]{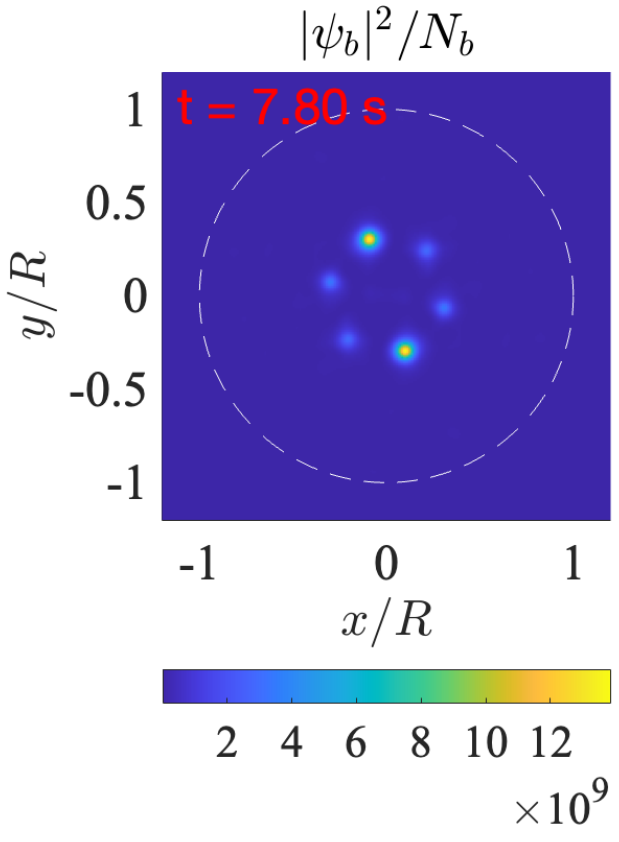}
\end{minipage}
\begin{minipage}{0.138\textwidth}
  \centering
  \includegraphics[width=1\linewidth]{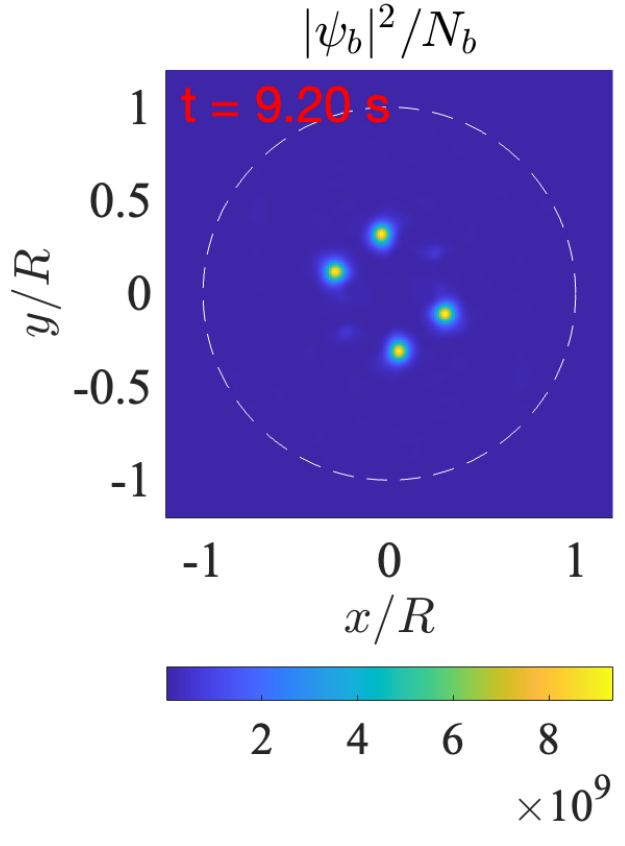}
\end{minipage}
\caption{Time evolution of $|\psi_b|^2$ in a six-vortex necklace, with an initial condition containing two specular $b$-peaks. In this case, $N_b=10^2$, $g_{ab}/\sqrt{g_a g_b}=1.26$.}
\label{fig:6v_diskCurrent_100_2peaks}
\end{figure}

At last, we consider the case with two specular peaks in the $b$-density, i.e. two filled vortices and four virtually empty ones.
Interestingly, in this case we have an ordered evolution of the $b$-supercurrents. This is illustrated in \textbf{Figure \ref{fig:6v_diskCurrent_100_2peaks}}, showing a central symmetry of the $b$-density distribution in the necklace at any instants of time.

We can quantify the Josephson supercurrents of the component $b$ within the vortex necklace by means of the average $x$- and $y$-currents $\bar j_x$ and $\bar j_y$,

\begin{equation}
    \bar  j_{k} =  \frac{-i\hbar}{2 m_b}\int  \, dx \, dy \,   \, (\psi_b^*   \nabla_k\psi_b  -\psi_b   \nabla_k\psi_b^* ),\;\;\;k=x,y.
\end{equation}

\begin{figure}[ht]
    \centering
    \includegraphics[width=0.5\linewidth]{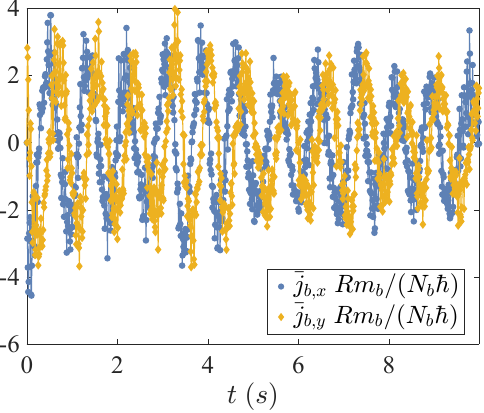}
    \caption{Normalized average currents in the $x$ and $y$ directions relative to the component $b$ for the case of a six-vortex necklace prepared with a single $b$-peak in a single vortex well. We have $N_b=10^2$ and $g_{ab}/\sqrt{g_a g_b}=1.26$.}
    \label{fig:currents}
\end{figure}

\medskip
\noindent
Remarkably, these average currents feature almost periodic oscillations in the case of the necklace prepared with a single-peak, as shown in \textbf{Figure \ref{fig:currents}}.
In this case, we recall that the latter was co-rotating with the necklace up to a few tunneling events.
On the other hand, in the case of a uniformly filled necklace and of the necklace with an initial (centrally symmetric) double-peak configuration, the net currents $\bar j_x$ and $\bar j_y$ turn out to  be zero.

To sum up, in this section we examined three cases at $N_v=6$ and $N_b=10^2$ (with three different initial conditions on the $b$-fillings), $5\times 10^2$, and $10^3$, and the GPEs' simulations showed that for very long times the necklaces did not get destroyed.
The initial configurations we took, in all the three cases, features a radial position of the vortices
$r_0\simeq 0.3\,R$.
Very interestingly, if we consider these initial configurations, up to the distribution of the vortex masses which is implicitly uniform, in the framework of PL model, we get that \textit{there exist stable} 
fixed-point solutions (see Equation \eqref{eq:necklace_sol}) 
in all the three cases. These first results are promising for a future more detailed comparison between the mean-field GPE dynamics and PL models.

\subsection{Collapse to fewer-vortex configurations with a background current}
\label{sec:collapse}

As already mentioned, the annular geometry allows for the absorption of quantum vortices at the inner boundary and the consequent creation of a background superflow. The closer to the inner boundary the vortices in the starting necklace configuration are, the quicker the system evolves in a fewer-vortex configuration with a higher background flow.
In a couple of examples, shown in \textbf{Figure \ref{fig:6v_collapse}} and in Appendix \ref{sec:app} in \textbf{\ref{fig:8v_collapse}}, that present 
necklaces with $6$ and $8$ vortices initially placed so to touch the inner boundary, we observe an evolution into particularly ordered final configurations. These respectively feature $2$ and $4$ massive vortices organized in a necklace of larger radius with respect to the initial one and a background flow carrying, as expected, four quanta of circulation in both cases. Such final
configurations 
are reached at $t\simeq 1.70$ $s$ and $t\simeq 0.83$ $s$ respectively for the two cases and
present, in good approximation, a discrete rotational symmetry and uniform, larger vortex masses. The transient dynamics that leads to the final configurations are complex and
involve the redistribution of the mass through tunneling events, and of the local vorticity.
In the last panel of Figure \ref{fig:6v_collapse} one can see the phase field $\theta_a$ of the vortex-hosting component, indicative of the collapse of four quantum vortices to the centre of the annulus
and the redistribution of their vorticities in a larger, background supercurrent circulating around the origin.

To explain these phenomena, we attempt at a rough comparison with the PL model for the annulus.
Remarkably, the GPEs' simulations are in agreement with the expectations based on the PL model. As a matter of fact, we find that there exist necklace configurations such as the initial and the final states of Figure \ref{fig:6v_collapse} and \ref{fig:8v_collapse}, at a given $\Omega$.
Secondly,
we find that the final configurations, involving two (four) massive vortices and a background superflow carrying four quanta of circulation, have a maximum instability growth rate that is approximately $0.07$ ($0.002$) that of the initial configuration. This indicates a much higher relative stability of the final configuration with respect to the initial one. The approximate character of these results is due to the approximation on the extracted value of $r_0$, then used for the \textit{Ansatz} \eqref{eq:necklace_sol}. As the values of $\sigma^*$ for systems similar to the final configurations in the GPE counterpart turn out to be extremely small, it is not excluded that a more precise comparison might show that such final configurations are in fact stable.

\begin{figure}[ht]
\centering
\begin{minipage}{0.38\textwidth}
  \centering
  \includegraphics[width=1\linewidth]{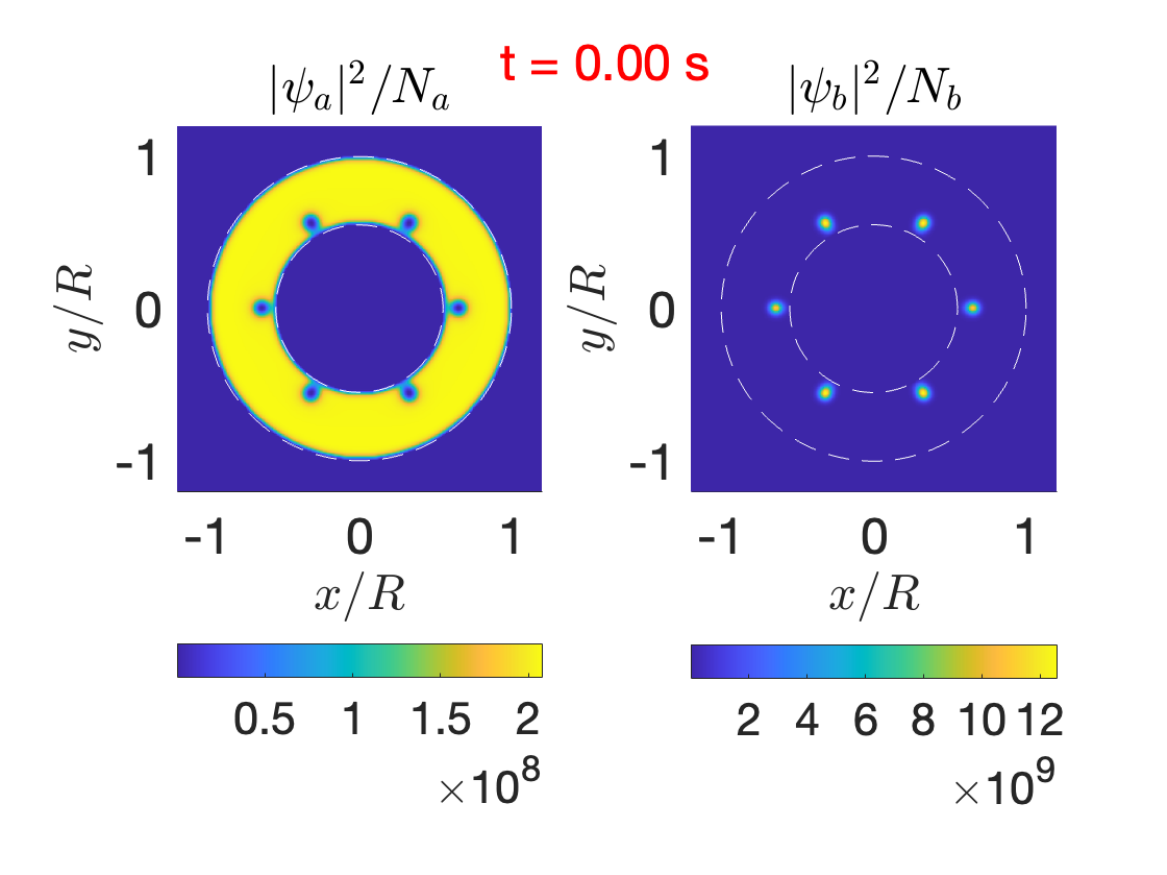}
\end{minipage}
\begin{minipage}{0.38\textwidth}
  \centering
  \includegraphics[width=1\linewidth]{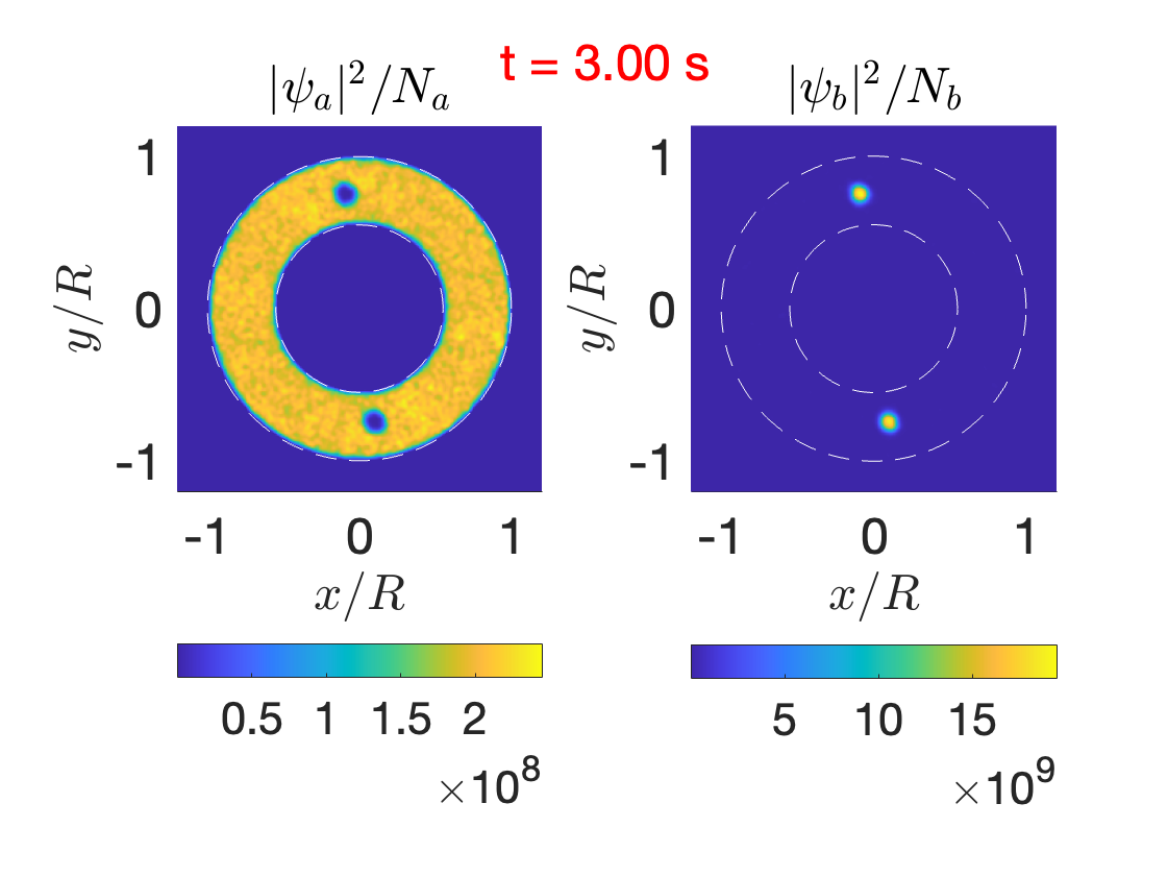}
\end{minipage}
\begin{minipage}{0.22\textwidth}
  \centering
  \includegraphics[width=1\linewidth]{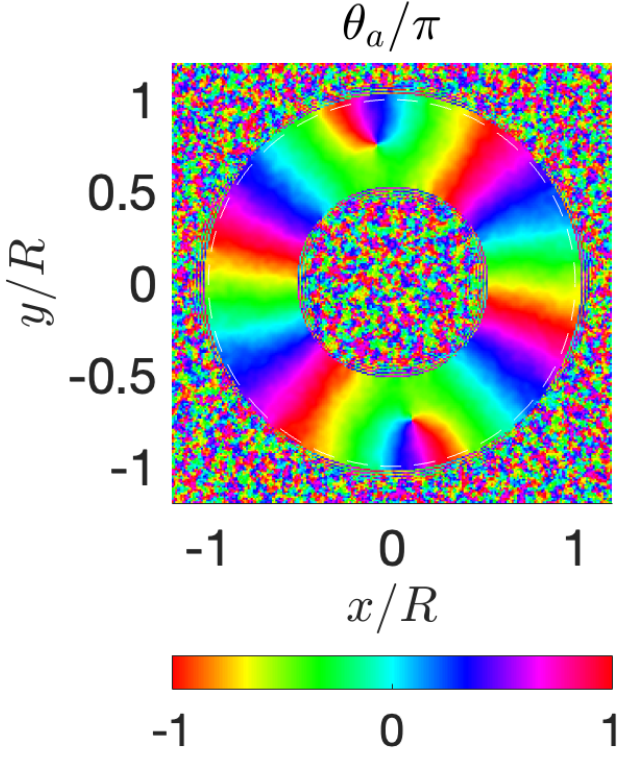}
\end{minipage}
\caption{Evolution of a massive six-vortex necklace placed so to touch the inner boundary of an annulus: the core-masses are redistributed and the system evolves into a massive two-vortex necklace and a background flow. The last panel is relative to the phase of $\psi_a$ at the time $t=3.00$ $s$. The two-vortex necklace persists until the final observation time (not shown here) of $10$ $s$.
The final precession direction is anti-clockwise.
Here, $N_b=10^3$, $q=0.55$.
}
\label{fig:6v_collapse}
\end{figure}

\section{Supercurrents in vortex lattices}
\label{sec:lattices}

In the previous sections, we saw how the distance between two vortices determined the time scale of any tunneling processes. 
This result is expected in the view of the effective vortices' potential wells, and it is in agreement with our results of Ref. \cite{Bellettini2024}, where we estimated the trend of the tunneling amplitude $J$ in a single Bosonic Josephson Junction formed by two massive vortices as $J \in \mathcal{O}(d^4)$, with $d$ the distance between the two vortices.

After exploring the necklace dynamics in the disk and the annulus geometries, it is quite natural to consider a third configuration still characterized by the same discrete rotational symmetry 
but with internal structure exhibiting a different topology.
We analyze what kind of tunneling processes can be supported by vortex lattices arranged in star structures (i.e. a vortex crown surrounding a central vortex), known to exist in stable configurations
\cite{Campbell1979, Abo-Shaeer2001, Engels2004}. We focus on cases where the distance between any pairs of nearest-neighbor vortices is comparable, so to ensure a similar tunneling probability.
During the dynamics the
vortex positions have very small variations, compared to the vortex size, around an average values.
Hence, we study such systems via the GPEs simulations, and present a few interesting examples
of their phenomenology, which open the path for future more detailed characterizations.

We always consider lattices consisting of a central vortex and a surrounding necklace of vortices.
The PL analogous, up to deformations of the necklace, is described by the solution in Equation \eqref{eq:lattice_sol}. 
Again, we study the Gross-Pitaevskii dynamics up to $10$ $s$ of time, and a ``stable'' vortex lattice, in the context of the GPEs simulation, is a lattice that survives until this final simulation time.

These systems make up some complex effective BH models, thanks to the $a$-density wells at the vortex sites and to the infilling condensate $b$.
We have that our vortex lattices always feature at least some small vibrations of the vortices around their average positions. These could in the future be in the first instance neglected, to then develop a mapping of the massive many-vortex system with a rotating BH model \cite{Kasamatsu2008, Jason2016, Arwas2015, \cite{Capuzzi2025}}, proceeding similarly as we did in Ref. \cite{Bellettini2024} for a single BJJ.
An interesting dynamical feature maintained by the vortex lattices is, similar to what observed for the necklace in Section \ref{sec:necklaces},
the contractions and expansions of the vortex crown over time. 
These seem to depend on the amount of mass filling the cores of the crown vortices, and to persist as long as the lattice structure is present.

\begin{figure}[ht]
\centering
\begin{minipage}{0.49\textwidth}
  \centering
  \includegraphics[width=1\linewidth]{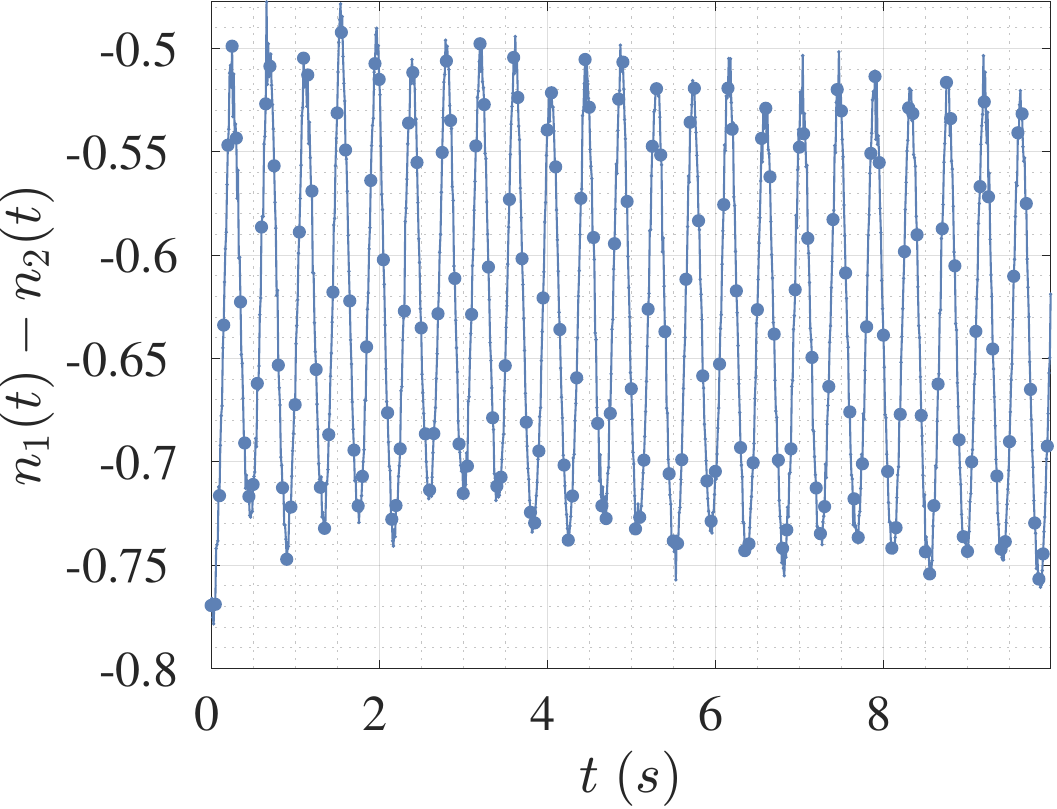}
\end{minipage}
\hfill
\begin{minipage}{0.49\textwidth}
  \centering
  \begin{minipage}{0.8\textwidth}
        \centering
        \includegraphics[width=1\linewidth]{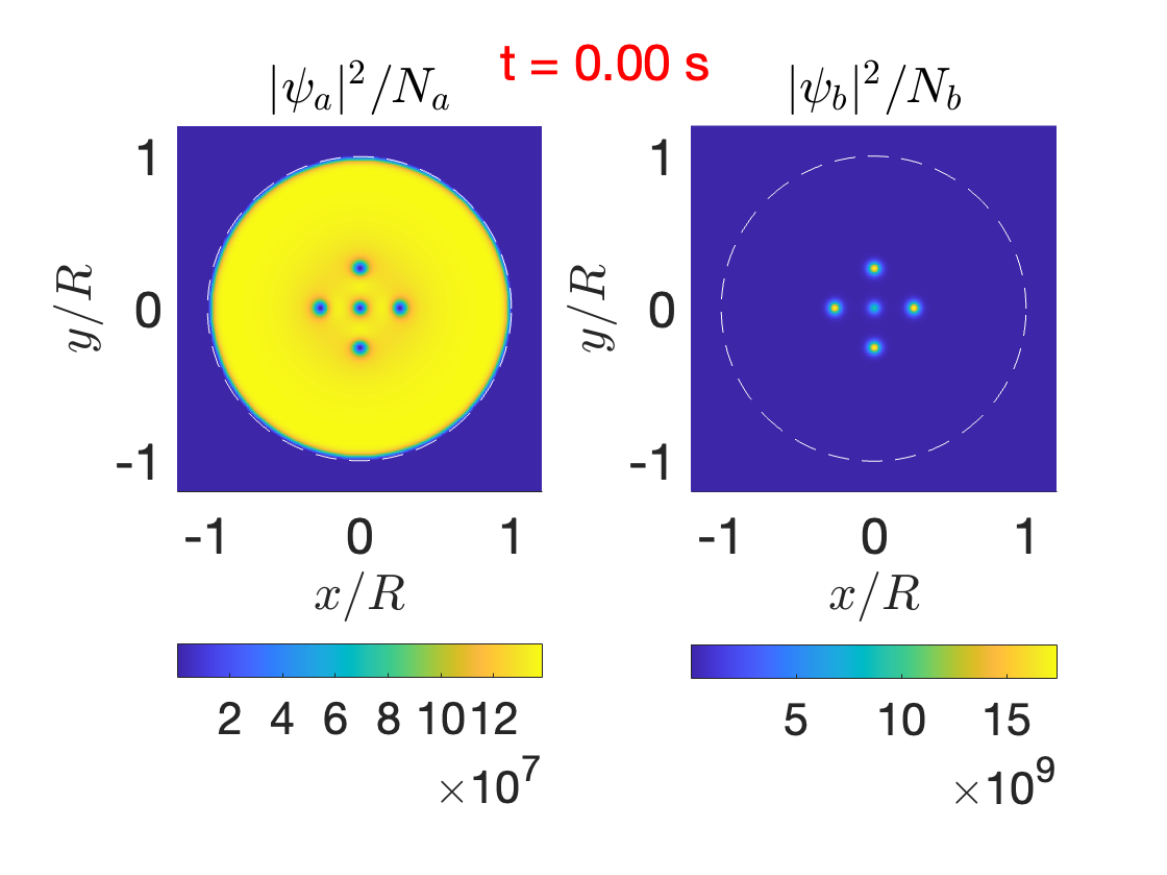}
    \end{minipage}
    \begin{minipage}{0.8\textwidth}
        \centering
        \includegraphics[width=1\linewidth]{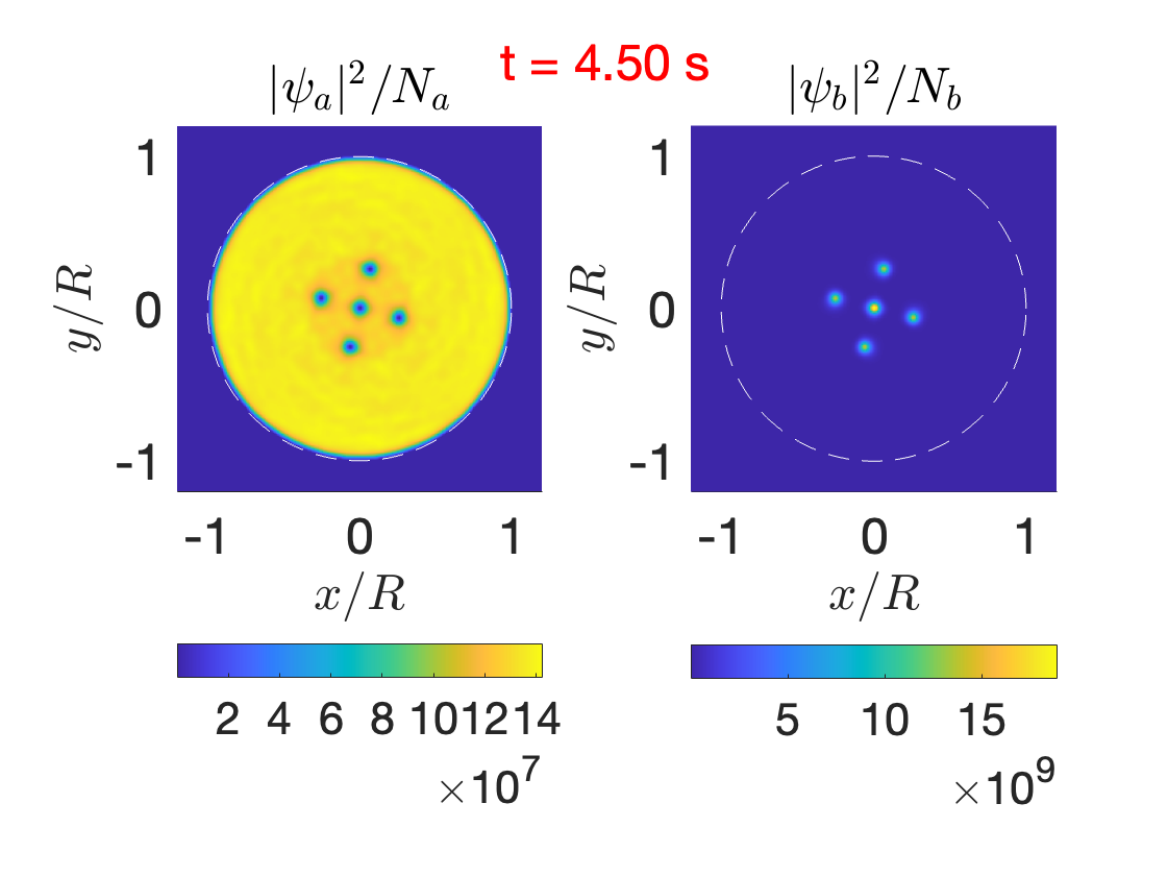}
   \end{minipage}
\end{minipage}
\caption{Oscillations, in the radial direction, of the $b$-atoms population in a five-vortex lattice. These represent a very ordered supercurrent with approximately uniform oscillation's amplitude and period, and a slightly varying average. As in the snapshots, the external vortex crown approximately features a uniform filling through the time evolution of the system. We have $N_b=10^2$.}
\label{fig:5v_100}
\end{figure}

A first fundamental 
aspect characterizing the dynamics of star-like arrays
is the macroscopic tunneling of the infilling component between the central vortex well and the vortex ``crown''. We start with $N_b=10^2$,
and an approximately uniform distribution of the $b$-atoms along the external necklace, with a negative population imbalance between the central vortex and the surrounding crown.
In this case, the dynamics is
extraordinarily ordered
if the lattice is composed of $5$ vortices, as shown in \textbf{Figure \ref{fig:5v_100}}. 
The latter shows a radial tunneling dynamics of the $b$-component from the central well to the vortex crown. This is quantified by the evolution of population imbalance $n_1(t)-n_2(t)$, where $n_1(t)$ and $n_2(t)$ are respectively the number of $b$-atoms in the central vortices and in the surrounding vortex crown. Such populations are extracted by integrating $|\psi_b|^2(t)$, in the case of $n_1(t)$, on the circle $\mathcal{C}$ centered in the origin and of radius approximately half the average distance of the vortices in the necklace from the origin, and in the case of $n_2(t)$ on the disk minus $\mathcal{C}$. 
In the case of Figure \ref{fig:5v_100} the extracted population imbalance features oscillations of similar amplitude and period,
and this radial tunneling suggests that, for some specific configurations, the population oscillations between the central vortex and the crown could be mapped to an effective (two-site) BJJ in presence of some symmetry conditions on $\psi_b$ (see e.g. Ref. \cite{Capuzzi2025}). 
The snapshots of Figure \ref{fig:5v_100} seem to confirm this scenario as the vortex crown always features an approximately uniform filling.

Increasing the number of vortices, and going to $N_v=6$, $7$, and $8$, as illustrated respectively in \textbf{Figure \ref{fig:6v_100}, \ref{fig:7v_100}}, and in Appendix \ref{sec:app} in \textbf{Figure \ref{fig:8v_100}}, 
the mean-field Gross-Pitaevskii model highlights a progressively ``more disordered'' dynamics, 
where the vortex crown develops a visibly non-uniform filling over the evolution time.
For $N_v=6$, we observe relatively regular supercurrents only restricted to some time windows, while the overall trend features a shift of the population imbalance towards the external vortex crown. 
For $N_v=7$, Figure \ref{fig:7v_100}, features globally an opposite trend of $n_1(t)-n_2(2)$, showing an increasing concentration of the $b$-atoms within the central well, with respect to the beginning. Interestingly, at increasing $N_v$ the radial tunneling seems to get faster, and in the case of seven vortices (Figure \ref{fig:7v_100}) it becomes more difficult to capture some periodic behaviors, while with eight vortices it is almost impossible (Figure \ref{fig:8v_100}). 
It is worth noticing how at increasing $N_v$, in our examples, a lower global variation of the imbalance $n_1(t)-n_2(2)$ takes place. The cases at $N_v=9$ and $10$ are very similar to 
the eight-vortex case, whereas at larger values of $N_v$ the lattice gets destroyed at some critical time. 

On the other hand, upon an increase on $N_b$, i.e. at $N_b=10^3$, the lattice at $5$ vortices, unlike the cases at $N_v=6,7,8$, gets destroyed before the end of the simulation of ten seconds. We have that the lattice-destruction process takes place via a depinning of the central vortex, which first oscillates around the origin until its displacement is too large and the lattice gets destroyed.

\begin{figure}[ht]
\centering
\begin{minipage}{0.49\textwidth}
  \centering
  \includegraphics[width=1\linewidth]{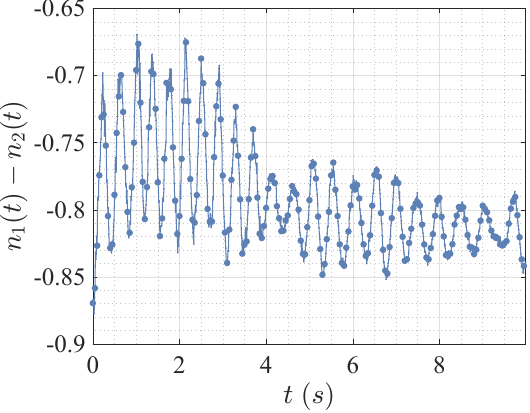}
\end{minipage}
\hfill
\begin{minipage}{0.49\textwidth}
  \centering
  \begin{minipage}{0.8\textwidth}
        \centering
        \includegraphics[width=1\linewidth]{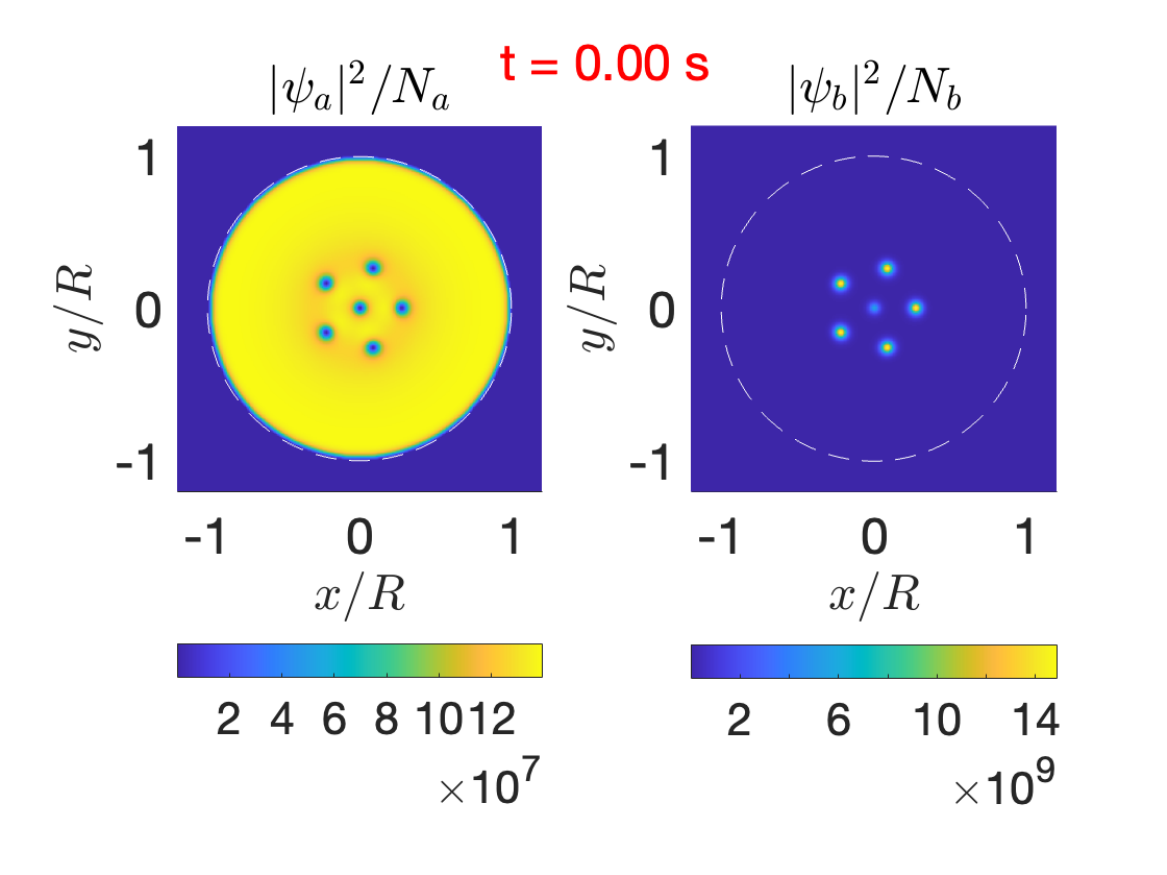}
    \end{minipage}
    \begin{minipage}{0.8\textwidth}
        \centering
        \includegraphics[width=1\linewidth]{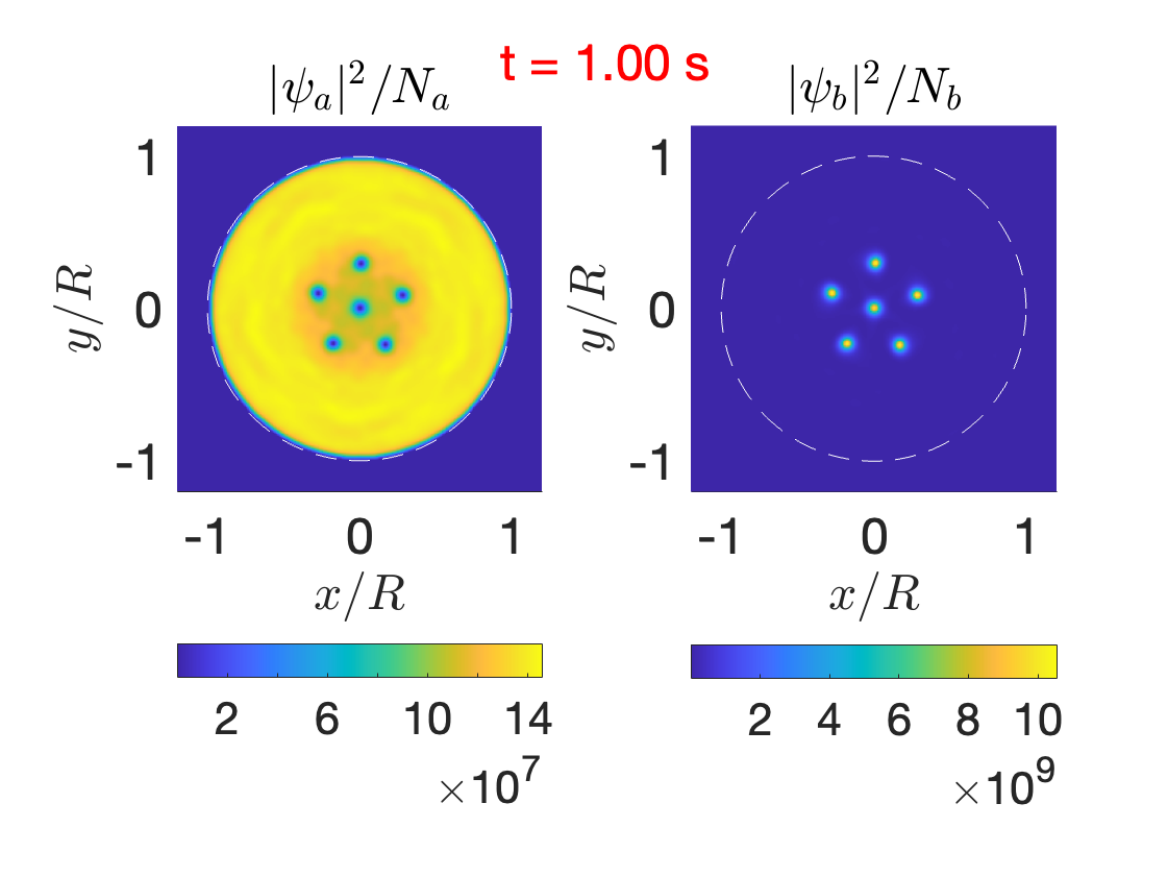}
   \end{minipage}
\end{minipage}
\caption{Dynamical evolution of a six-massive-vortex lattice according to the GPEs. The configuration, characterized by a central vortex and a five-vortex crown, persists throughout the simulation time ($10$ $s$). We estimate the amount of radial tunneling by the population imbalance $n_1(t)-n_2(t)$ between the central well and the surrounding necklace. In the right panels we see some snapshots of the condensates densities $|\psi_a|^2$ and $|\psi_b|^2$ at different instants. In this case, $N_b=10^2$.}
\label{fig:6v_100}
\end{figure}

\begin{figure}[ht]
\centering
\begin{minipage}{0.49\textwidth}
  \centering
  \includegraphics[width=1\linewidth]{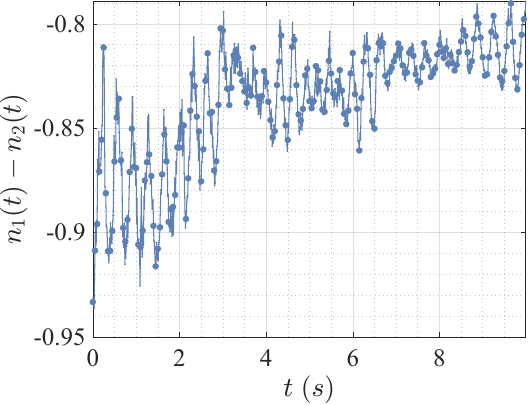}
\end{minipage}
\hfill
\begin{minipage}{0.49\textwidth}
  \centering
  \begin{minipage}{0.8\textwidth}
        \centering
        \includegraphics[width=1\linewidth]{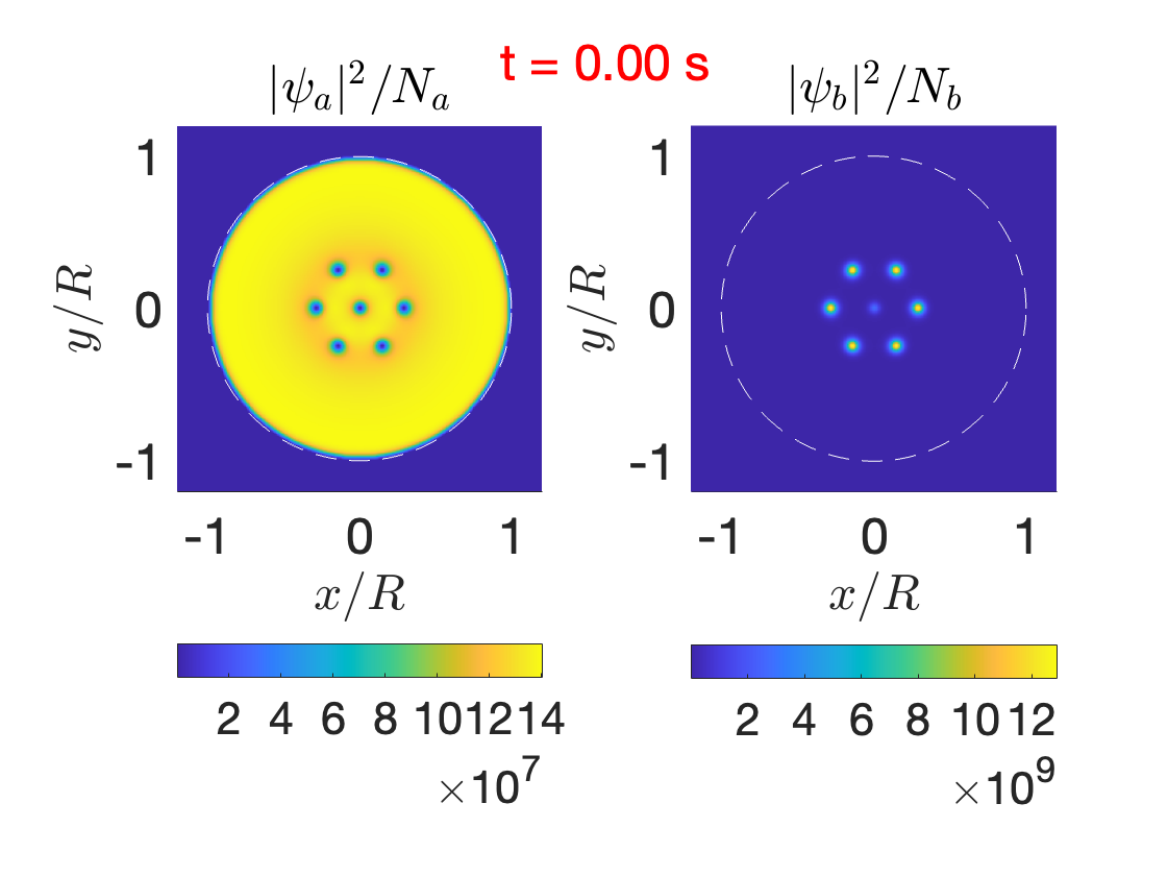}
    \end{minipage}
    \begin{minipage}{0.8\textwidth}
        \centering
        \includegraphics[width=1\linewidth]{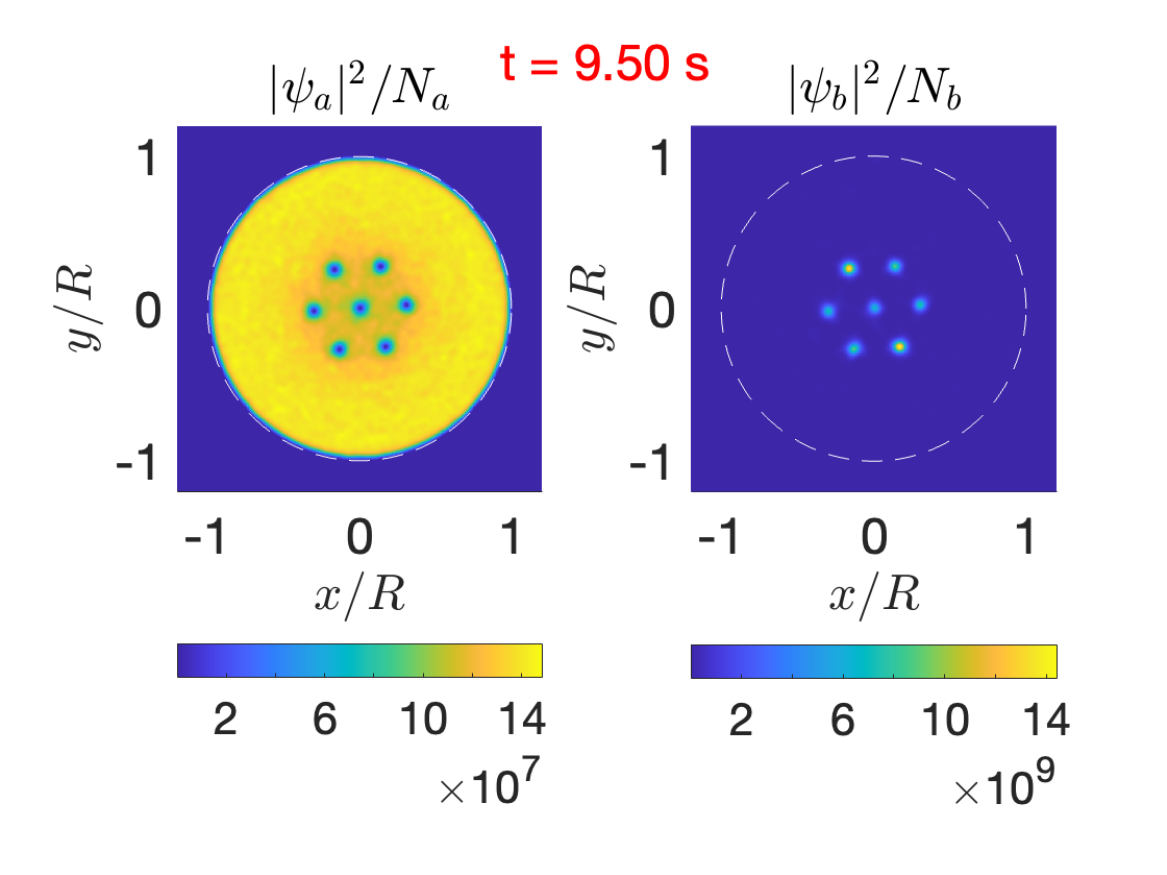}
   \end{minipage}
\end{minipage}
\caption{Population imbalance and density snapshots for a massive-vortex lattice of $7$ vortices. The same considerations of Figure \ref{fig:6v_100} hold. We also have $N_b=10^2$.}
\label{fig:7v_100}
\end{figure}

The emergence of sound waves within the $a$-component, shown in the different snapshots of Figure \ref{fig:6v_100} and \ref{fig:7v_100} is worth noticing.
These are related to the vortices' accelerations \cite{Barenghi2005} and to the repulsive interaction between the two components, governed by $g_{ab}$. Possibly, such sound waves could have an effect on the tunneling dynamics, a matter which of interest for a future study.
On the other hand, it would be interesting to examine how the vortex accelerations directly affect the dynamics of the Josephson supercurrents, as done in Ref. \cite{Roy2025} for a single BJJ.

\subsection{Effect of \texorpdfstring{$g_{ab}$}{gab}}

\begin{figure}
    \centering
    \includegraphics[width=0.5\linewidth]{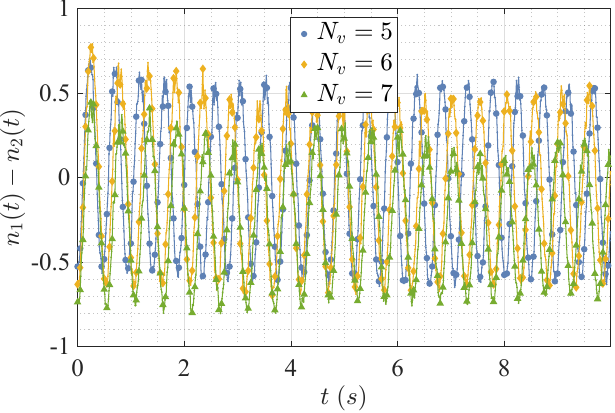}
    \caption{Population imbalance between the central vortex and the surrounding vortex crown for lattices of $5$, $6$, and $7$ vortices. We have $N_b=10^2$ and $g_{ab}/\sqrt{g_a g_b}=1.26$ in all cases.}
    \label{fig:imbalance_comp}
\end{figure}

We now investigate the effect of the coupling parameter $g_{ab}$, expected to play a crucial role in the tunneling dynamics, as a factor appearing in the vortex-wells effective potential (see also Secs. \ref{sec:peak_propagation} and \ref{sec:current_in_necklace}). Therefore, we set here $g_{ab}/\sqrt{g_a g_b}$, i.e. a \textit{lower immiscibility} with respect to the standard case. Interestingly, we see that this condition leads to much stronger oscillations in the imbalance $n_1(t)-n_2(t)$
for lattices of $5$, $6$ and $7$ vortices, as shown in \textbf{Figure
\ref{fig:imbalance_comp}},
Here we see that the imbalance varies from significantly negative to significantly positive values, meaning a strong filling of the internal vortex with respect to the whole collection of external vortices. These oscillations are approximately regular in all the three cases, unlike the corresponding systems at $g_{ab}/\sqrt{g_a g_b}\simeq 2.2$ (see Figure \ref{fig:5v_100}, \ref{fig:6v_100}, and \ref{fig:7v_100}), while for higher $N_v$ values the shift of the average imbalance towards negative values reveals a relatively more filled crown with respect to the internal vortex.
At $N_v=7$ we observe some leakages of the $b$-atoms out of the lattice at long times, phenomenon that increases at increasing $N_v$.
We also studied the same systems at $N_b=10^3$, and saw that the lattices, at $N_v \geq 5$ got destroyed before the final simulation time of $10$ $s$.

\section{Conclusions}
\label{sec:conclusions}

To sum up, we investigated mainly numerically some of the tunneling processes that the vortex-infilling component can undergo in a system of many vortices, and the stability properties of some ordered configurations at varying physical parameters. Our work was prompted by our recent finding, in Ref. \cite{Bellettini2024} of sustained Josephson supercurrents in two-vortex systems, and aimed at exploring some of the features occurring at the level of the GPEs simulations when many vortices exchanging mass among each other are involved.
We started, in Section \ref{sec:PL}, by outlining the PL models present in literature for approximating the dynamics of massive vortices with a frozen mass, which were shown to work well with a negligible quantum tunneling of the minority component. We then explored in Section \ref{sec:GPE} and \ref{sec:lattices} the tunneling phenomenology displayed by the GPEs simulations in regimes where the tunneling in the infilling component was triggered. Surprisingly, we still found some qualitative agreement on the stability character of necklace vortex configurations,
even in presence of a varying vortex mass. In essence, we found that persisting necklaces were in fact configurations very close to stable configurations at the level of the PL model, whereas necklaces that got destroyed according to the mean-field dynamics were very close to unstable configurations in the PL framework. This first qualitative considerations foster future more detailed comparisons, e.g. on the rotation frequency or the instability rate, and the development of a PL model with varying masses.
Specifically, in Section \ref{sec:necklaces} we presented the GPEs dynamics of many-vortex necklaces in a disk and an annular geometry, illustrating an example of their destruction mechanism and the onset of a chaotic motion accompanied by the vortices' exchange of mass.

Subsequently, in Section \ref{sec:peak_propagation} we examined the initial diffusion dynamics of a single $b$-peak in a stable few-vortex necklace ($N_v=6$). The effect of the necklace rotation, in both the disk and the annular geometry, was retrieved in an asymmetric diffusion of the initial peak.

In Section \ref{sec:current_in_necklace} we then started from the system of Section \ref{sec:peak_propagation}, for the disk trap, and illustrated the longer-times dynamics of the $b$-supercurrent in the six-vortex necklaces, also at varying $N_b$. 
Here we saw how the asymmetry in the vortices' populations led to asymmetrical lattice deformations. We found several instances of Josephson supercurrents supported by the vortex necklace, and we identified a BJJ-like regime in the case of a necklace initially prepared with a single $b$-peak in a single vortex well. Furthermore, we observed that the case of a centrally symmetric vortex-filling evolves so that $|\psi_b|^2$ approximatly preserves this symmetry.

Further on, in Section \ref{sec:collapse} we illustrated two examples where massive vortex-necklaces in an annulus, with an initially uniform vortex-mass distribution, collapsed to fewer-massive-vortex necklaces with a frozen mass and a background superflow of the majority component. This behavior, involving mass exchanges, could
be explained at the level of the PL model by comparing the stability character of the initial and final configurations.

Finally, in Section \ref{sec:lattices} we studied the radial bosonic Josephson current between a vortex placed in the origin of a disk trap and a surrounding necklace of vortices. We selected vortex configurations that are characterize by comparable distances between any two pairs of nearest neighbors, to favor there the tunneling effects.
We found a substantial current for a lattice of $5$ vortices, with periodic oscillations of the population imbalance. Increasing the number of vortices, we observed a decrease of this radial current and a mass imbalance $n_1(t)-n_2(t)$ between the central vortex and the external vortex crown trapped around very low values.
On the other hand, by decreasing the inter-species repulsive coupling $g_{ab}$, we observed a relative increase in the radial current, observing larger oscillations in the imbalance $n_1(t)-n_2(t)$ with respect to previous case, and a periodicity in the imbalance oscillations for larger time windows.

In general, in all the sections relevant to the Gross-Pitaevskii results, e.g. Section \ref{sec:GPE} and \ref{sec:lattices}, we individuate some metastable dynamical behaviors exhibited by the many-vortex systems, and study our configurations' stability properties at varying physical parameters.

Our work prompts numerous future outlooks. We found that vortices can support BJJ arrays in place of optical lattices; however, they are all but defect free, as they make up vibrating, time dependent potentials.
As an example, it would be most intriguing to study the coupling between the vortices' dynamics and that of the tunneling component. Here the Tkachenko modes could be involved \cite{Schweikhard2004Tkachenko, Kim2004}, as well as the breathing of the vortices, changing their core size to accommodate more or less $b$-atoms. It would be interesting to also investigate how this vortex breathing affects in turn the sound waves in the component $a$, and finally how the latter affect the tunneling dynamics. 
On the other hand, one could proceed as we did in Ref. \cite{Bellettini2024} to study the Josephson supercurrents within stable vortex structures.
This phenomenology could be extended to vortex configurations with particular symmetry conditions, as the periodic distribution of the $b$ population among the wells. In this case, the systems could be reduced to few-mode models within the BH picture \cite{Capuzzi2025}. 
Finally, our ongoing work is focused on systems of three and four vortices, which represent the building blocks to explore chaotic and semi-periodic 
regimes \cite{Aref1983, Franzosi2003}, and so to 2D quantum turbulence \cite{Griffin2017}.

\medskip
\textbf{Acknowledgments} \par   
We warmly thank Dr. Andrea Richaud for the helpful discussions, and for his time and precious advices.
Computational resources were provided by HPC@POLITO (http://hpc.polito.it).

\begin{appendix}

\section{Additional figures}
\label{sec:app}

\begin{figure}[ht]
\centering
\begin{minipage}{0.24\textwidth}
  \centering
  \includegraphics[width=1\linewidth]{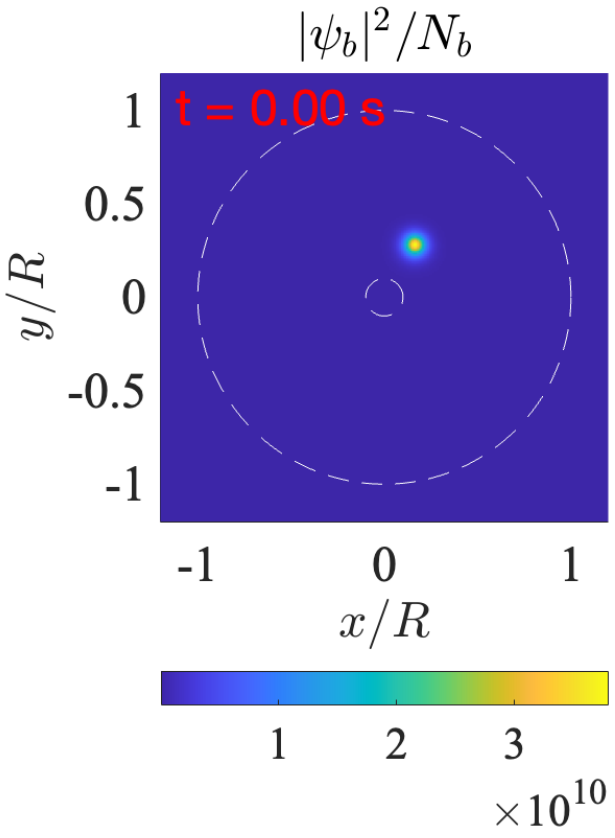}
\end{minipage}
\begin{minipage}{0.24\textwidth}
  \centering
  \includegraphics[width=1\linewidth]{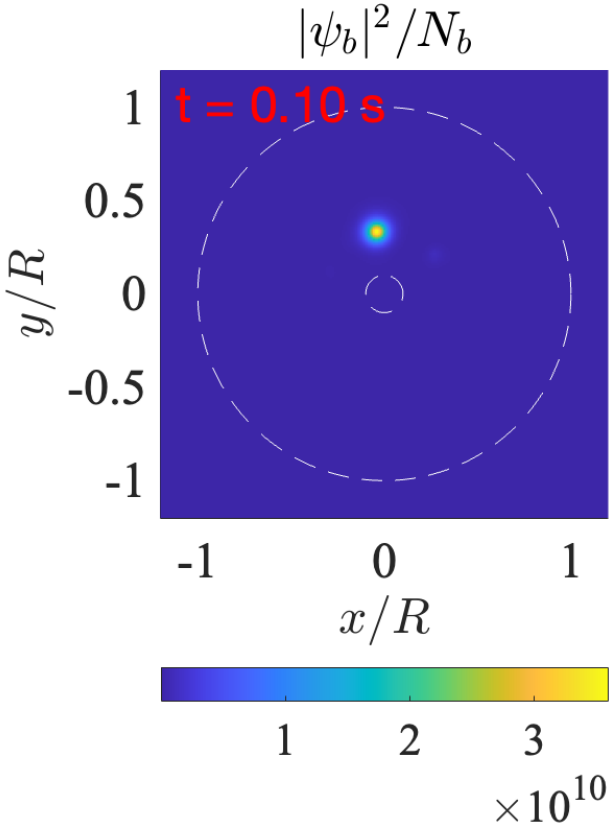}
\end{minipage}
\begin{minipage}{0.24\textwidth}
  \centering
  \includegraphics[width=1\linewidth]{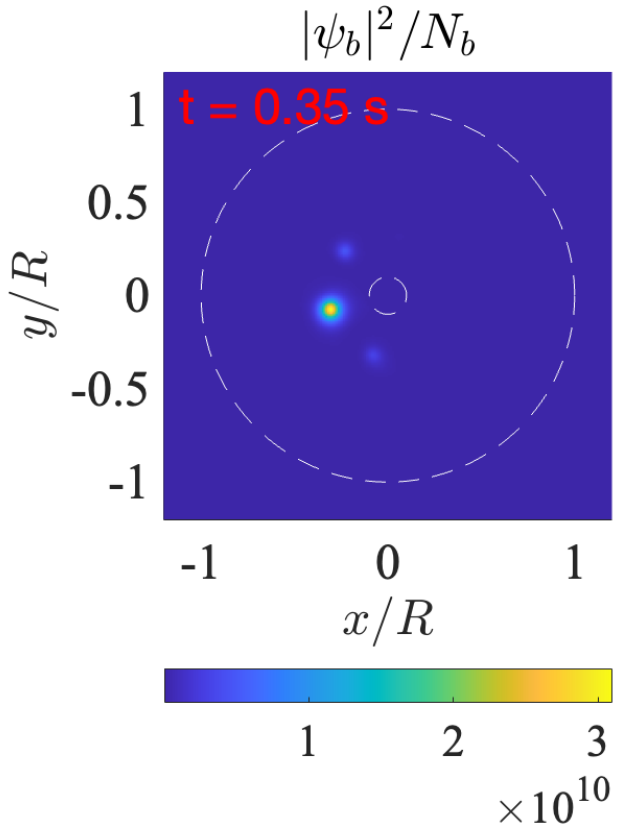}
\end{minipage}
\begin{minipage}{0.24\textwidth}
  \centering
  \includegraphics[width=1\linewidth]{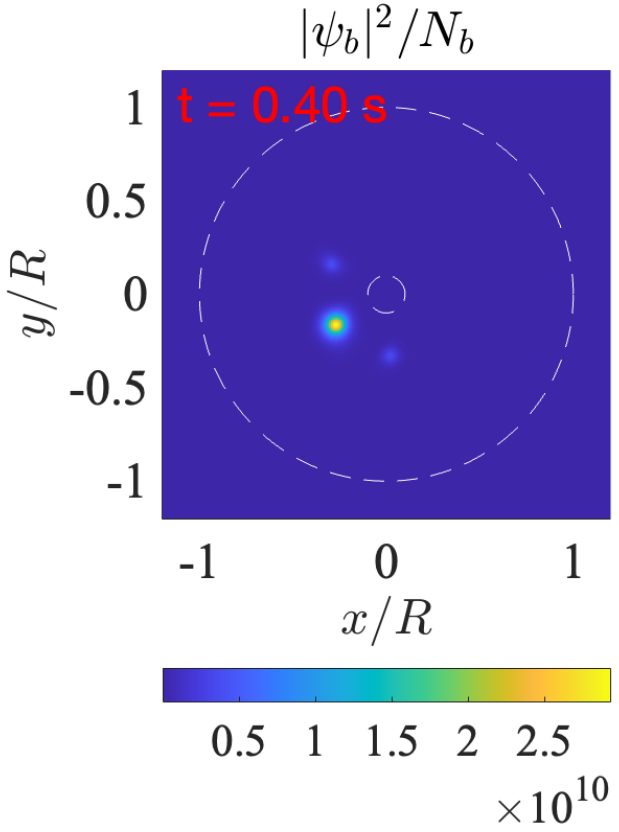}
\end{minipage}
\caption{Propagation of a single $b$-peak in a six-vortex necklace in a planar annulus, whose boundaries are represented by the white dotted 
lines. The initial population of the empty sites occurs in a similar way to the case of the disk in Figure \ref{fig:sol_prop_disk}, but slower. The precession direction is anti-clockwise.
We have $N_b= 10^2$, $q=0.1$, and $g_{ab}/\sqrt{g_a g_b}=1.26$.
}
\label{fig:sol_prop_annulus}
\end{figure}

\begin{figure}[ht]
\centering
\begin{minipage}{0.38\textwidth}
  \centering
  \includegraphics[width=1\linewidth]{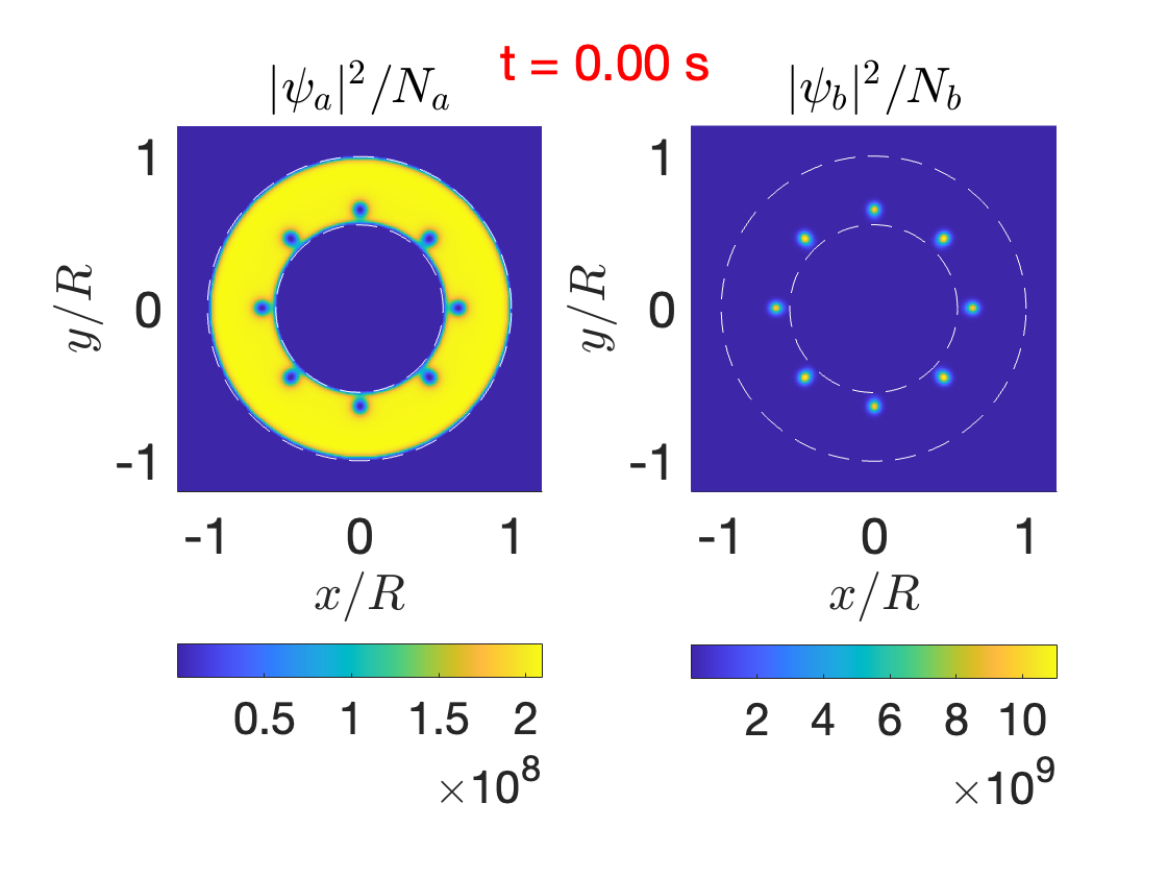}
\end{minipage}
\begin{minipage}{0.38\textwidth}
  \centering
  \includegraphics[width=1\linewidth]{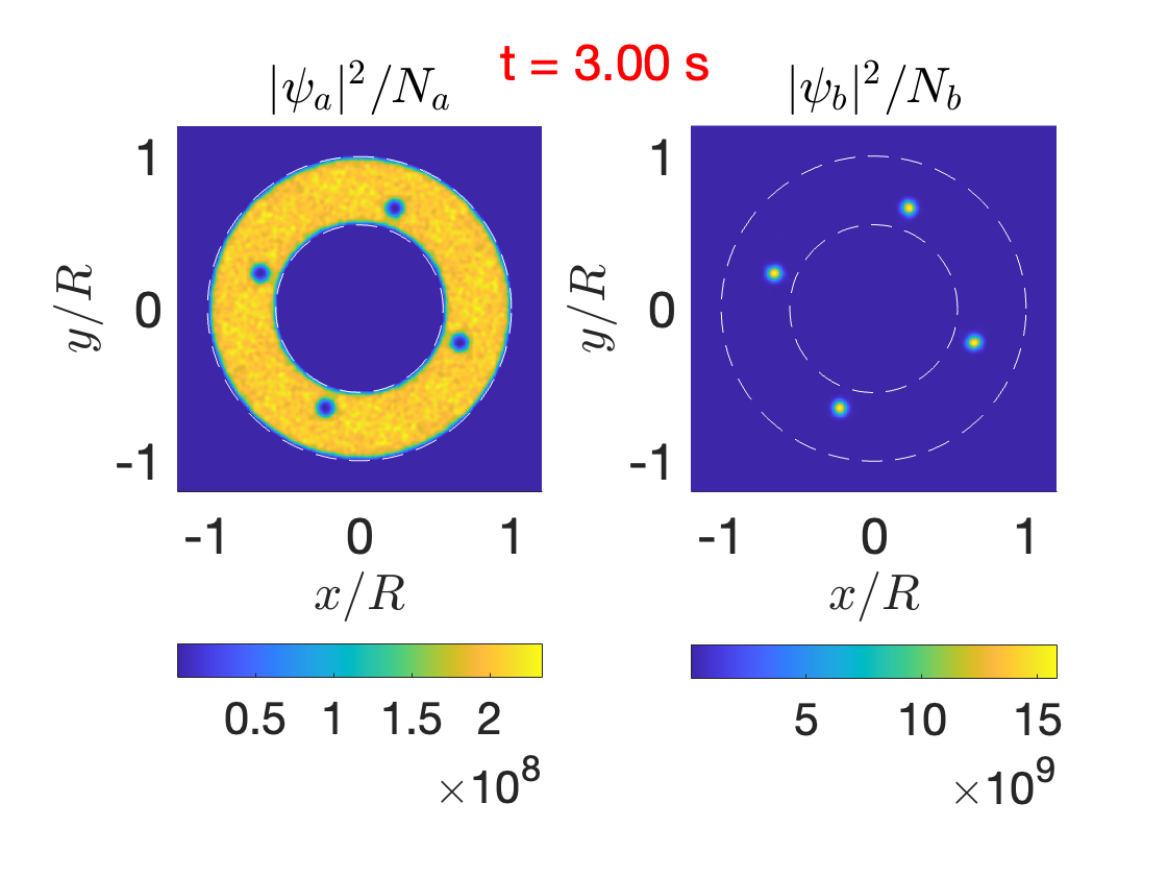}
\end{minipage}
\begin{minipage}{0.22\textwidth}
  \centering
  \includegraphics[width=1\linewidth]{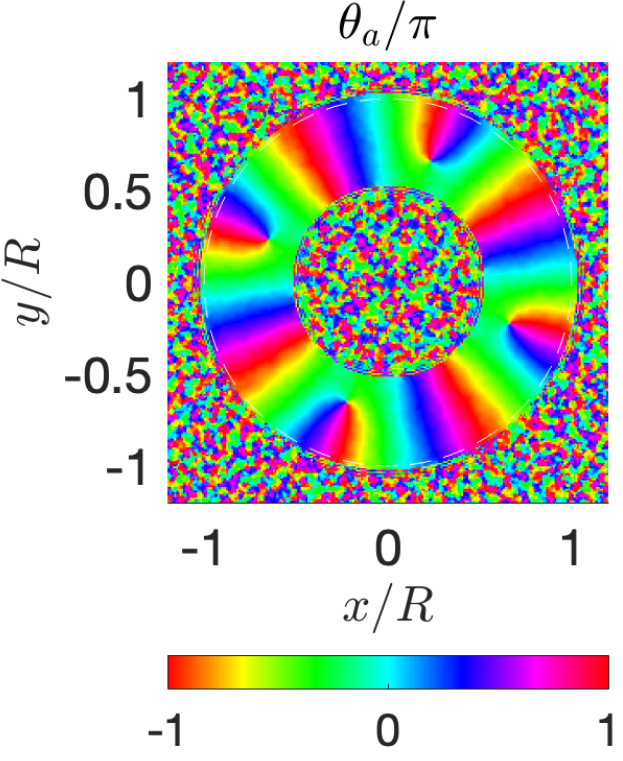}
\end{minipage}
\caption{Evolution, likewise to the case of Figure \ref{fig:6v_collapse}, of a massive eight-vortex necklace. After three seconds the system's configuration turns into an annular background supercurrent and $4$ massive vortices arranged in a necklace that is stable for a long observation time (i.e. until $10$ $s$). The last panel represents $\theta_a$ after three seconds of time. The final precession direction of the necklace is anti-clockwise.
We have: $N_b=10^3$, and $q=0.55$.}
\label{fig:8v_collapse}
\end{figure}

\begin{figure}[ht]
\centering
\begin{minipage}{0.24\textwidth}
  \centering
  \includegraphics[width=1\linewidth]{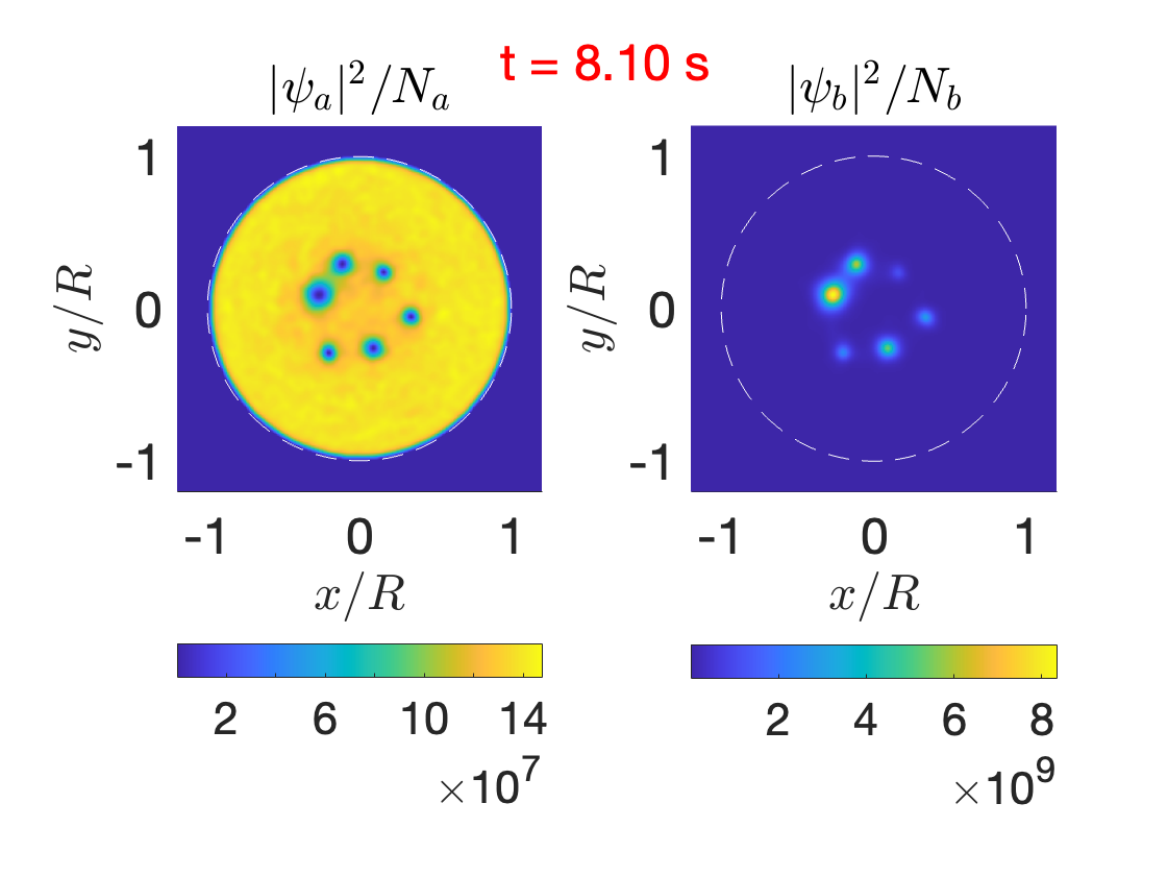}
\end{minipage}
\begin{minipage}{0.24\textwidth}
  \centering
  \includegraphics[width=1\linewidth]{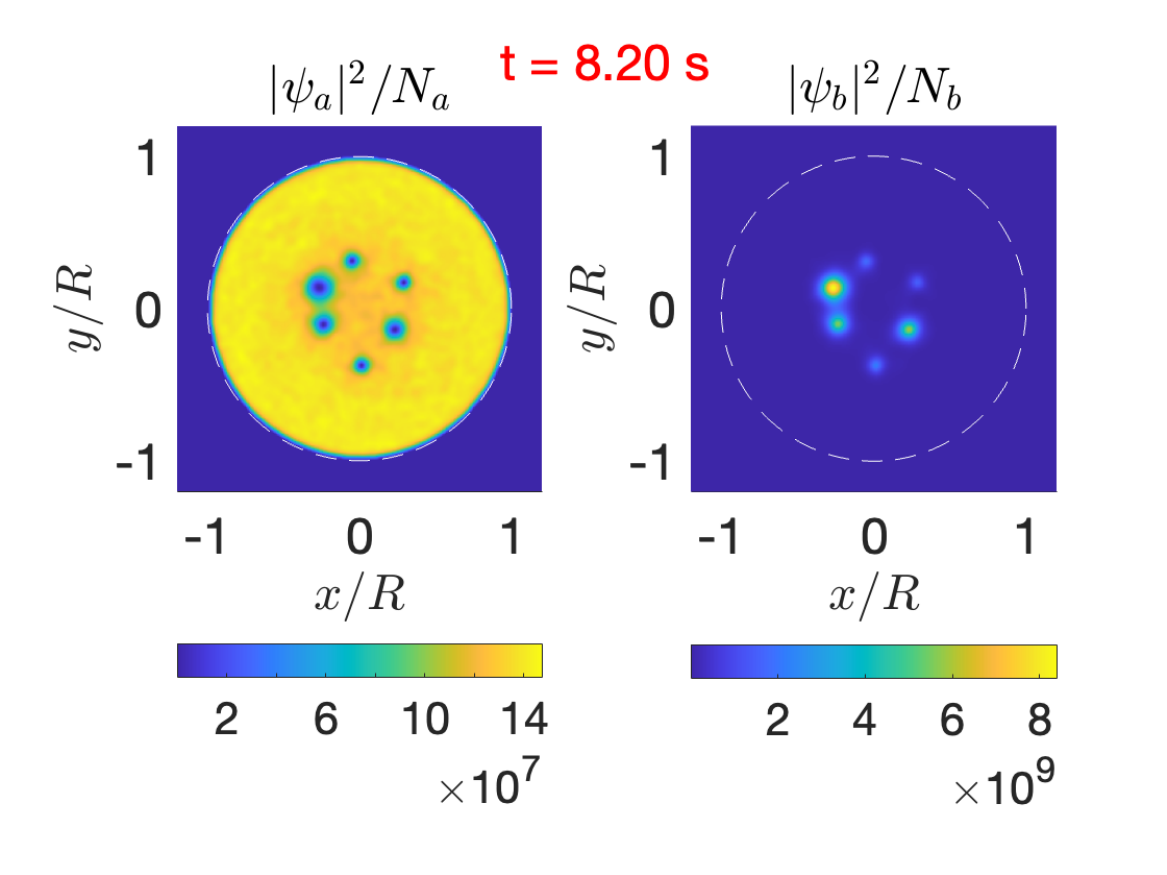}
\end{minipage}
\begin{minipage}{0.24\textwidth}
  \centering
  \includegraphics[width=1\linewidth]{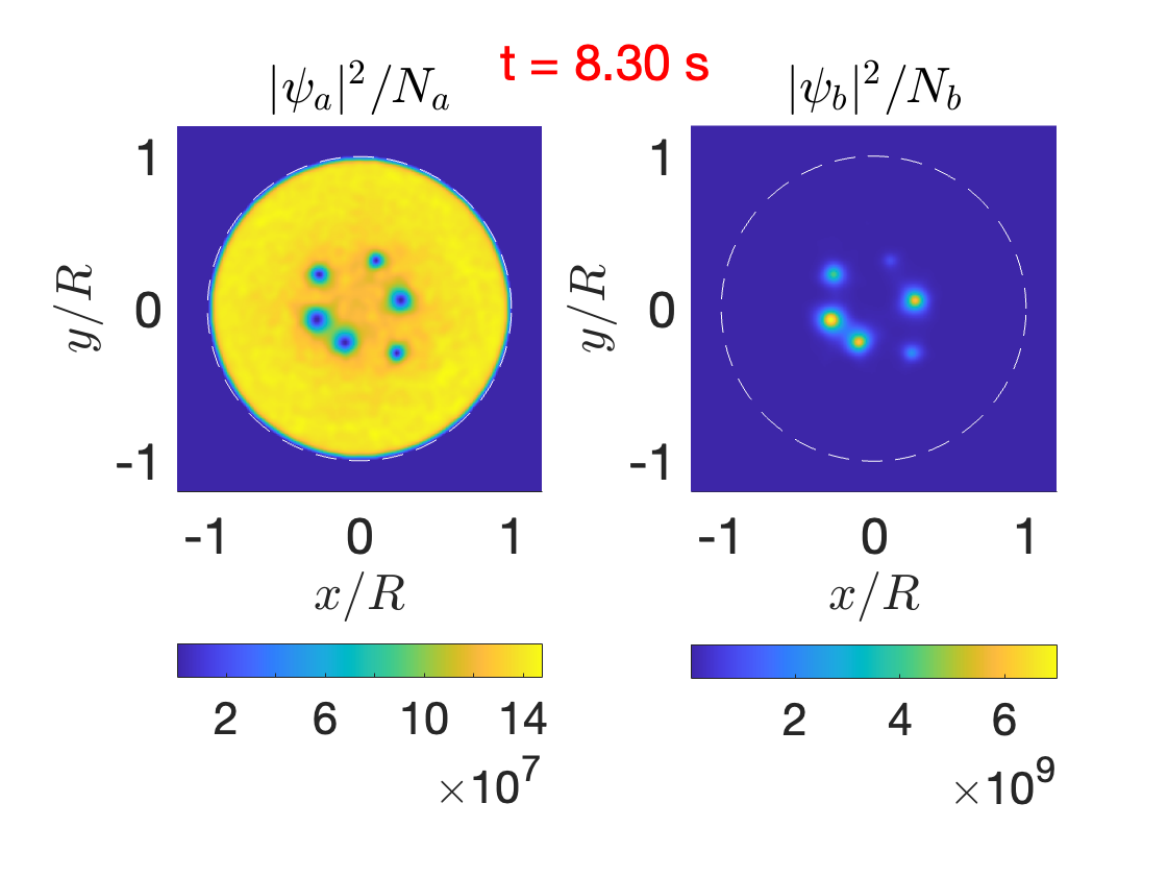}
\end{minipage}
\begin{minipage}{0.24\textwidth}
  \centering
  \includegraphics[width=1\linewidth]{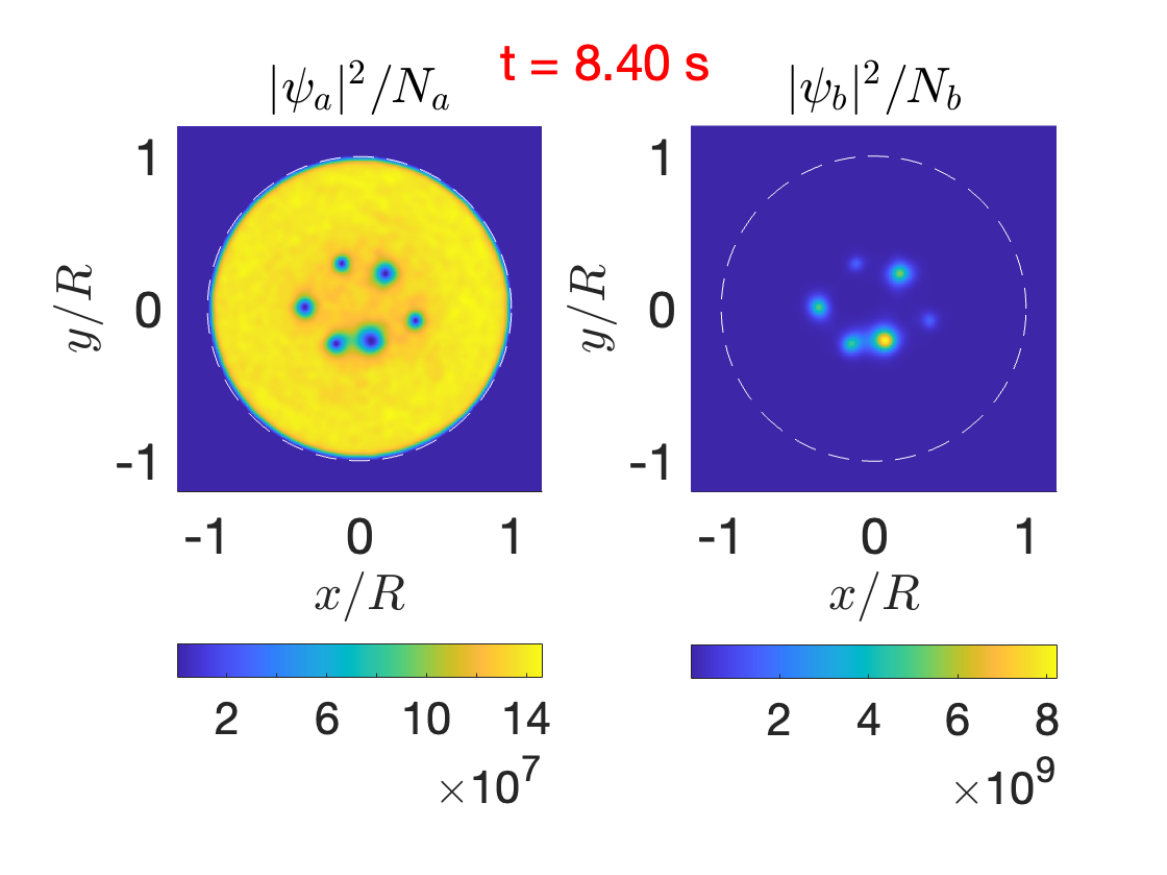}
\end{minipage}
\caption{During the complex tunneling dynamics occurring at relatively large $N_b$ values in a six-vortex necklace, we see several appreciable tunneling events between nearest neighbors, as in the exemplary sequence above. We have $N_b=10^3$ and $g_{ab}/\sqrt{g_a g_b}=1.26$.
}
\label{fig:6v_diskCurrent_1000}
\end{figure}

\begin{figure}[ht]
\centering
\begin{minipage}{0.49\textwidth}
  \centering
  \includegraphics[width=1\linewidth]{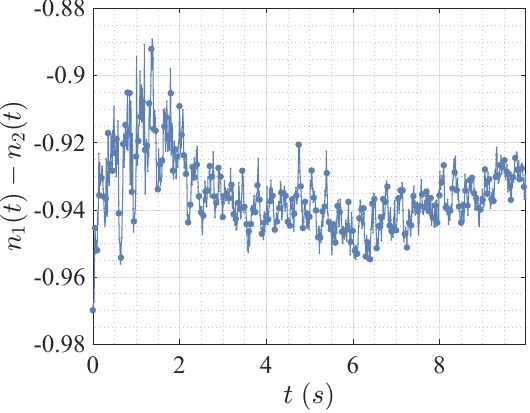}
\end{minipage}
\hfill
\begin{minipage}{0.49\textwidth}
  \centering
  \begin{minipage}{0.8\textwidth}
        \centering
        \includegraphics[width=1\linewidth]{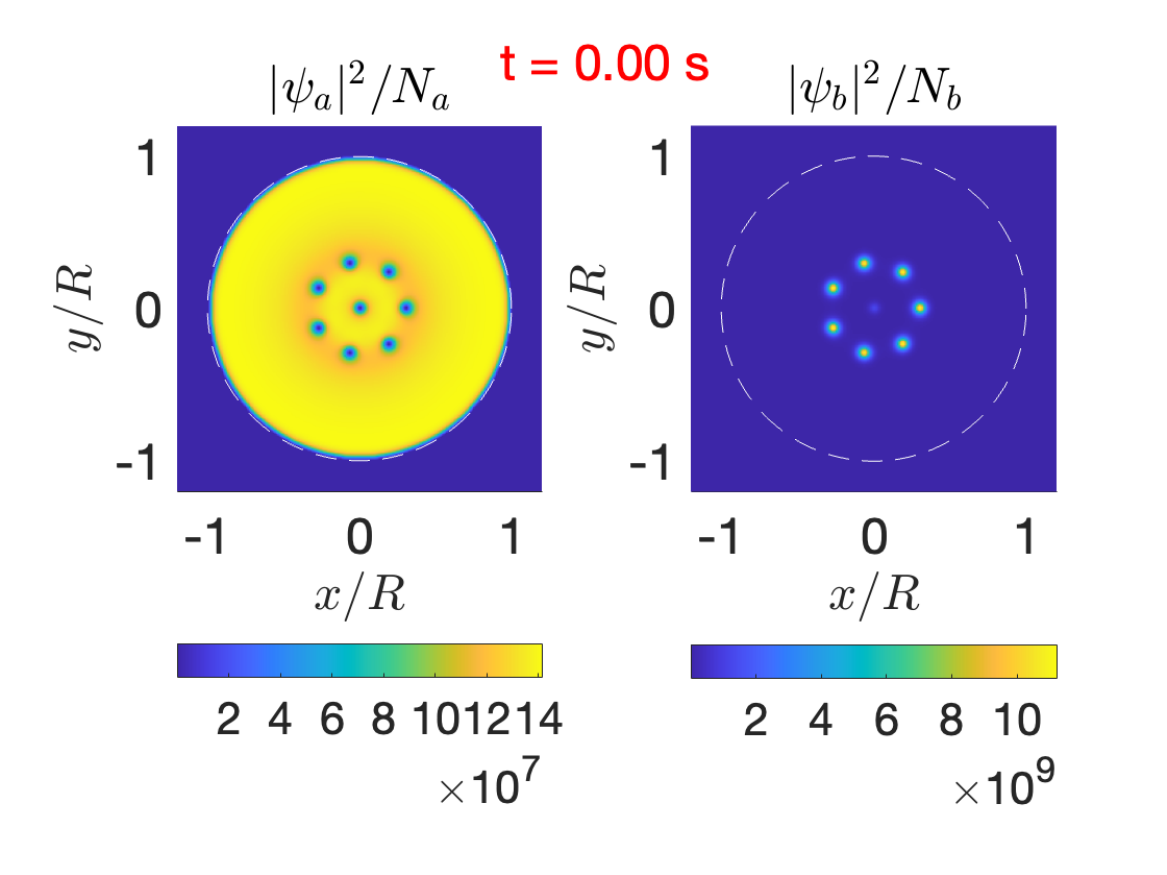}
    \end{minipage}
    \begin{minipage}{0.8\textwidth}
        \centering
        \includegraphics[width=1\linewidth]{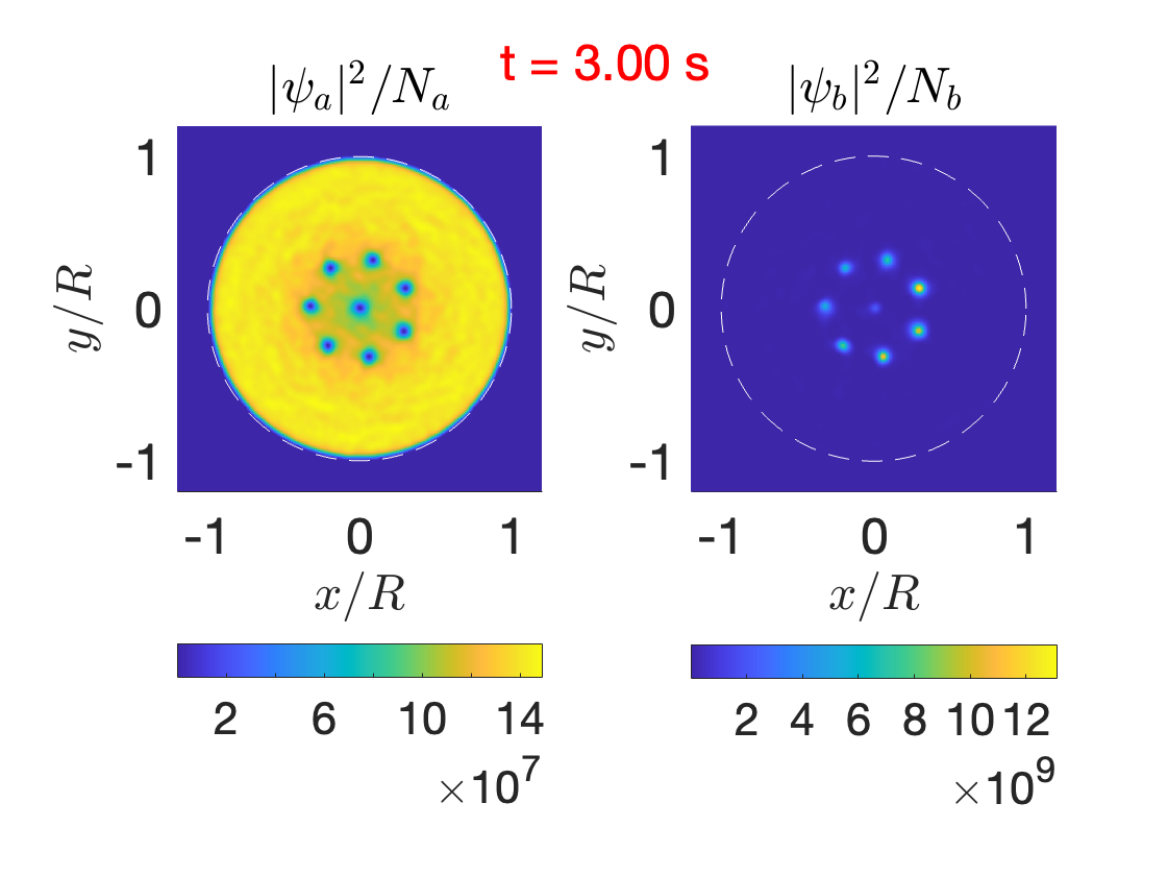}
   \end{minipage}
\end{minipage}
\caption{Dynamical evolution and tunneling of a system characterized by $8$ massive vortices. We have that the same considerations of Figure \ref{fig:6v_100} apply and $N_b=10^2$.}
\label{fig:8v_100}
\end{figure}

\end{appendix}

\end{document}